\documentclass[12pt,a4paper]{article} 
\pdfoutput=1
\usepackage{jheppub} 
\usepackage{graphicx} 
\usepackage{amsmath} 
\usepackage{amssymb} 
\usepackage{slashed} 
\usepackage{bigstrut} 
\usepackage{mathtools} 
\usepackage{multirow} 

\newcommand{\beq}{\begin{equation}} 
\newcommand{\eeq}{\end{equation}} 
\def\bsp#1\esp{\begin{split}#1\end{split}} 
\def\bal#1\eal{\begin{align}#1\end{align}} 
\newcommand{\beeq}{\begin{eqnarray}} 
\newcommand{\eeeq}{\end{eqnarray}}

\newcommand{\gsim}{\raisebox{-0.07cm}{$\:\:\stackrel{>}{{\scriptstyle \sim}}\:\: $} }
\newcommand{\lsim}{\raisebox{-0.07cm}{$\:\stackrel{<}{{\scriptstyle \sim}}\: $} }
\newcommand{\msbar}{$\overline{\text{MS}}\, $}

 \title{ 
Lepton fluxes from atmospheric charm revisited
} 
\author[a]{M. V. Garzelli} 
\author[a]{\!,~~~S. Moch} 
\author[a]{\!,~~~G. Sigl} 
\affiliation[a]{II.~Institute for Theoretical Physics, Hamburg University
\\
Luruper Chaussee 149, D--22761 Hamburg, Germany}

\abstract{ 
We update predictions for lepton fluxes from the hadroproduction of charm
quarks in the scattering of primary cosmic rays with the Earth's atmosphere. 
The calculation of charm-pair hadroproduction 
applies the latest results from perturbative QCD through next-to-next-to-leading order  
and modern parton distributions, together with estimates on various sources of uncertainties. 
Our predictions for the lepton fluxes turn out to be compatible, within the uncertainty band, 
with recent results in the literature. 
However, by taking into account contributions neglected in previous works, 
our total uncertainties are much larger.
The predictions are crucial for the interpretation of results
from neutrino experiments like IceCube, 
when disentangling signals of neutrinos of astrophysical origin from the atmospheric background.  
}  
 

\preprint{DESY 15-107,  MITP/15-049}
\begin{document} 
\maketitle 

\section{Introduction} 
\label{sec:intro} 
Atmospheric lepton fluxes are important backgrounds in the search of neutrinos of astrophysical origin~\cite{Gaisser:2014eaa}. 
In particular recent claims from the IceCube experiment, which has detected a statistically significant sample 
of leptonic events at very high energies~\cite{Aartsen:2013jdh,Aartsen:2014gkd}, whose interpretation is still under
debate~\cite{Fong:2014bsa, Chen:2014gxa}, require an estimate of the
background as accurately as possible. 
One of the most uncertain components of
this background is the prompt contribution due to the hadroproduction of charm quarks in the hard scattering
of primary cosmic rays with the Earth's atmosphere, the so-called atmospheric charm. 
In this paper we will concentrate on the contribution to the lepton fluxes that
can be ascribed to atmospheric charm. 

After initial studies on the basis of phenomenological models (see e.g. Ref.~\cite{Battistoni:1995yv,Bugaev:1998bi} and references therein), 
previous predictions for lepton fluxes from atmospheric charm have been obtained  within the framework of perturbative Quantum Chromodynamics (QCD) 
for proton-proton collisions according to the standard QCD collinear 
factorization formalism, with hard-scattering evaluated 
at leading order (LO) in Ref.~\cite{Gondolo:1995fq}
and including radiative corrections at next-to-leading order (NLO) in Ref.~\cite{Pasquali:1998ji}, respectively.
As an alternative description motivated by the high collision energies of the underlying hard scattering, 
Ref.~\cite{Enberg:2008te} has used the so-called color dipole approach as an
effective model for the production of colored particles at high energies, in
order to compute the production rates for atmospheric charm.

All these predictions, however, are subject to very large theoretical uncertainties. While the results of Ref.~\cite{Enberg:2008te} are very sensitive to the 
parameters of the phenomenological model for the color dipole, which are poorly constrained by experimental data, 
also the standard perturbative QCD predictions 
for the hadroproduction of charm quarks acquire big uncertainties, of the order of several ten percents, 
in the kinematical regions of interest for astrophysical applications. 
In the latter case, these uncertainties are due to estimates of the missing radiative corrections at higher orders, 
the knowledge of parton distribution functions (PDFs), especially the gluon PDF
at small fractions $x$ of the momenta of the colliding protons,
as well as the precision on the charm quark mass. 

Since the start of the Large Hadron Collider (LHC) and thanks to both theoretical and experimental progress, 
our understanding of charm-pair hadroproduction at high energies has significantly improved.
It is, therefore, the aim of the present paper to provide new predictions for atmospheric charm 
and its contribution to lepton fluxes, on the basis of standard perturbative QCD, 
taking into account the most recent developments in this field.

For inclusive charm-pair hadroproduction we use QCD predictions
up to next-to-next-to-leading order (NNLO) in order to establish the apparent
convergence of the perturbative expansion and the stability under variation
of the renormalization and factorization scales, together with recent
determinations of PDFs compatible with constraints from LHC measurements.
We also investigate the dependence on the renormalization scheme and the value 
for the charm quark mass and discuss differences between the running mass and the pole mass schemes.
A consistency check is performed by a comparison of the theory predictions to available 
LHC data from ALICE \cite{Abelev:2012vra}, ATLAS~\cite{ATLAS:2011fea}
and the LHCb~\cite{Aaij:2013mga} experiments obtained in the runs at 
$\sqrt{S}=7$ and $8$~TeV center-of-mass energy, 
because the data are within the kinematic region of interest for atmospheric charm.
The differential distributions for charmed hadron production which are
necessary in order to compute the lepton fluxes are obtained in
perturbative QCD with a consistent matching between NLO QCD corrections and
parton showers, 
as the respective predictions at NNLO are currently not available.
The interface to the {\texttt{PYTHIA}} event generator~\cite{Sjostrand:2006za} accounts
for the full effect of parton showers and the hadronization.
Our study features a detailed discussion of the different sources of  
theoretical uncertainties affecting predictions, 
which are propagated to the computation of the lepton fluxes.
In this way, a total uncertainty band for the final predictions of the prompt
lepton fluxes is established and compared to previous results in the literature. 
The effects on the fluxes due to modifications in primary cosmic ray spectra are also
shown by making use of the latest spectra available in an analytic form.

Recently, the authors of Ref.~\cite{Pasquali:1998ji} have proposed an update of that work in Ref.~\cite{Bhattacharya:2015jpa}.
Our computation is independent and differs from Ref.~\cite{Bhattacharya:2015jpa} 
because of the up-to-date perturbative QCD results and methods used in the computation of charm and $D$-hadron production
cross-sections and because of the choice and the variation of the input parameters. 
In summary, this leads to a more comprehensive estimate of the related
uncertainties for the prompt lepton fluxes.
 
The paper is organized as follows: 
In Sec.~\ref{sec:method} we present the method, the input, and the tools we have used
for the calculation, together with examples of results from its intermediate steps. Sec.~\ref{sec:predictions} contains our predictions for the lepton fluxes
along with a discussion of the related uncertainties. 
In Sec.~\ref{sec:astroimpli} we sketch the astrophysical implications of these predictions, after comparing them
to those used so far by the astrophysical community, and discuss their potential
implications for the IceCube experiment. 
Finally, in Sec.~\ref{sec:conclusions} we draw our conclusions, 
with reference to future theory progress and measurements
which could help to decrease the uncertainties on the predictions presented.

\section{Method: cascade equations and their solution} 
\label{sec:method} 

The particle evolution through an air column 
of depth $X$ in the Earth's atmosphere can be
obtained by solving a set of coupled differential equations, so-called cascade equations.
Following Ref.~\cite{Lipari:1993hd, Pasquali:1998ji} one has
\begin{eqnarray}
\frac{d \phi_j}{dX} = - \frac{\phi_j}{\lambda_{j,int}} - \frac{\phi_j}{\lambda_{j,dec}} + \sum_{k \ne j} S_{prod} (k \rightarrow j) + \sum_{k \ne j} S_{decay} (k \rightarrow j)
+S_{reg} (j \rightarrow j)\, .\nonumber\\ 
\label{cascade}
\end{eqnarray}
A dependence on the energy $E_j$ is understood in all terms of eq.~(\ref{cascade}), $j$ labels a
particle species, $\lambda_{j,int}$ and $\lambda_{j,dec}$ its interaction and
decay lengths, respectively, while $S_{prod}$ and $S_{decay}$ denote the generation
functions for production and decay: 
\begin{eqnarray}
S_{prod}(k\rightarrow j) = \int_{E_j}^\infty dE_k \frac{\phi_k(E_k, X)}{\lambda_k (E_k)}
\frac{1}{\sigma_k} \frac{d\sigma_{k\rightarrow j}(E_k, E_j)}{dE_j} \sim
\frac{\phi_k(E_j,X)}{\lambda_k(E_j)} Z_{kj}(E_j) \,,
\label{sprod}
\end{eqnarray}
\begin{eqnarray}
S_{decay}(j\rightarrow l) = \int_{E_l}^\infty dE_j \frac{\phi_j(E_j, X)}{\lambda_j (E_j)}
\frac{1}{\Gamma_j} \frac{d\Gamma_{j\rightarrow l}(E_j, E_l)}{dE_l} \sim
\frac{\phi_j(E_l,X)}{\lambda_j(E_l)} Z_{jl}(E_l) \, .
\label{sdecay}
\end{eqnarray}
Here, $\phi_k(E_k,X)$ is the flux of particle $k$, $\sigma_k$ is the total
inelastic cross-section for the interaction of particle $k$ in the atmosphere,
${d\sigma_{k\rightarrow j}}/{dE_j}$ is the energy distribution of particle $j$ produced by $k$, 
$\Gamma_j$ is the total decay width of particle $j$ and ${d\Gamma_{j\rightarrow l}}/{dE_l}$ is the energy
distribution of particle $l$ produced by the decay of $j$. 
Regeneration functions in eq.~(\ref{cascade}), 
i.e., $S_{reg}(j \rightarrow j)$, can be viewed as a particular case
of $S_{prod}(k \rightarrow j)$ when $k$~=~$j$. According to the nature of
particle $j$ (nucleon, heavy-hadron, neutrino), some of the terms in
eq.~(\ref{cascade}) may be absent.

The right hand sides of eqs.~(\ref{sprod}) and~(\ref{sdecay}), defining
the so-called $Z$-moments for the production and decay of particle $j$, respectively, 
are obtained after noticing that the $X$ dependence of fluxes approximately factorizes
from their $E$ dependence. 
In this approximation, analytic solutions exist for eq.~(\ref{cascade}) 
in the limit where the energy of intermediate particles leading to final
leptons is either very small or very large with respect to their critical energy,
the latter being proportional to the particle mass $m$ and to the inverse of its proper life-time $\tau_{0}$. 
In the vertical direction, $E_{crit}~=~m~c^2~h_0/(c\tau_{0})$, 
where $h_0$ is the vertical depth of the atmosphere, for which
an isothermal model is assumed with the density of the atmosphere evolving
with depth as $\rho (h)~=~\rho_0~\exp(-h/h_0)$.

In fact, the competition between hadron interaction and hadron decay 
is crucial in determining the final lepton fluxes and $E_{crit}$ represents an
approximate energy above which the hadron decay probabilities are suppressed with
respect to their interaction probabilities. 
In particular, one can distinguish between the conventional neutrino flux and
the prompt neutrino flux, according to the nature of the intermediate hadrons. 
The conventional flux is produced by the decays of charged kaons and pions which
dominate over their interaction rates at relatively low energies, as the critical energies for these particles are smaller than $1$~TeV. 
On the other hand, for larger energies, the probability of secondary interactions overcomes the pro\-ba\-bi\-lity that these mesons decay, 
thereby progressively suppressing the flux of neutrinos from them. 
At energies above $10^5 - 10^6$~GeV, neutrinos are thus mainly produced by the decay of other particles. 
In the framework of the Standard Model, these are, in particular, 
charmed and bottomed heavy-hadrons, 
which are characterized by a larger critical energy ($E_{crit} > 10^7$~GeV)
than pions and kaons~\footnote{
More precisely, the critical energies in vertical direction for the charmed hadrons considered in this work amount to: 
$E_{D^0}^{crit}$ = 9.71 $\cdot 10^7$~GeV,$\,\,\,$ 
$E_{D^+}^{crit}$ = 3.84 $\cdot 10^7$~GeV,$\,\,\,$
$E_{D_s^+}^{crit}$ = 8.40 $\cdot 10^7$~GeV,$\,\,\,$
$E_{\Lambda_c}^{crit}$ = 24.4 $\cdot 10^7$~GeV.}. 
These immediately decaying particles ($\tau \sim 10^{-12}$~s) give rise to the so called prompt flux, that is the object of study of this paper. 

In case of hadrons decaying into leptons, the flux of leptons coming from
low energy hadrons, i.e., from hadrons with $E \ll E_{crit}$, can be approximated by 
\begin{eqnarray}
\phi_{h\rightarrow l}^{low} = Z_{h\,l}^{low}\, \frac{Z_{p\,h}}{1-Z_{p\,p}}\, \phi^0_p \, ,
\label{philow}
\end{eqnarray}
whereas the flux of leptons from hadrons with $E \gg E_{crit}$
is approximated by
\begin{eqnarray}
\phi_{h \rightarrow l}^{high} = Z_{h\,l}^{high}\, \frac{Z_{p\,h}}{1-Z_{p\,p}}\, \frac{E_{crit,h}}{E_h} \, \frac{\ln(\Lambda_h/\Lambda_p)}{1-\frac{\Lambda_p}{\Lambda_h}} \, f(\theta) \, \phi^0_p \, ,
\label{phihigh}
\end{eqnarray}
with $\Lambda_j (E_j)$ defined as $\Lambda_j (E_j) = \lambda_j (E_j)/(1-Z_{jj} (E_j))$. 
In the approximated solutions to the cascade equations outlined
above, an energy dependence is understood in all fluxes $\phi$, all $Z$-moments
and all attenuation lengths $\Lambda$. 
Note that the low energy lepton flux is isotropic, whereas the high
energy lepton flux is characterized by an angular dependence 
$f(\theta)\sim 1/\cos(\theta)$ for $\theta$ $< 60^{o}$, 
where $\theta$ is the angle with respect to the zenith, and by a more
complex angular dependence close to the horizon.

The solution in the intermediate energy range $E \sim E_{crit}$ is obtained
by the geometric approximation
\begin{eqnarray}
\phi_{h \rightarrow l} (E_l) = \frac{\phi_{h \rightarrow l}^{low} (E_l)
  \phi_{h \rightarrow l}^{high} (E_l)}{(\phi_{h \rightarrow l}^{low}(E_l) +
  \phi_{h \rightarrow l}^{high}(E_l))}
\, ,
\label{geometric}
\end{eqnarray}
whose quality and validity depend on the particular shapes 
of $\phi_{h \rightarrow l}^{low}$ and $\phi_{h \rightarrow l}^{high}$ as a
function of the lepton energy, see, e.g., 
Fig.~\ref{fig:interpola} in Sec.~\ref{sec:predictions} below, 
for an example of this interpolation.
In this way one gets the contribution $\phi_{h \rightarrow l}$ 
to the lepton flux from each hadron species $h$. 
Summation over all hadron species finally provides 
the total lepton flux $\phi_l$ for each lepton species $l$, that is 
$\phi_l = \sum_h \phi_{h \rightarrow l}$. 
Alternatively, the system of differential equations in eq.~(\ref{cascade}) can
also be solved numerically.

\subsection{Input: cosmic ray primary spectrum}
\label{subsec:cosmic}

The primary cosmic ray (CR) spectrum is an important input of our calculation as it enters the solution 
of the cascade equations~(\ref{philow}) and~(\ref{phihigh}) both implicitly through $Z$-moments and explicitly.

\begin{figure}[ht!]
\begin{center}
\includegraphics[width=0.75\textwidth]{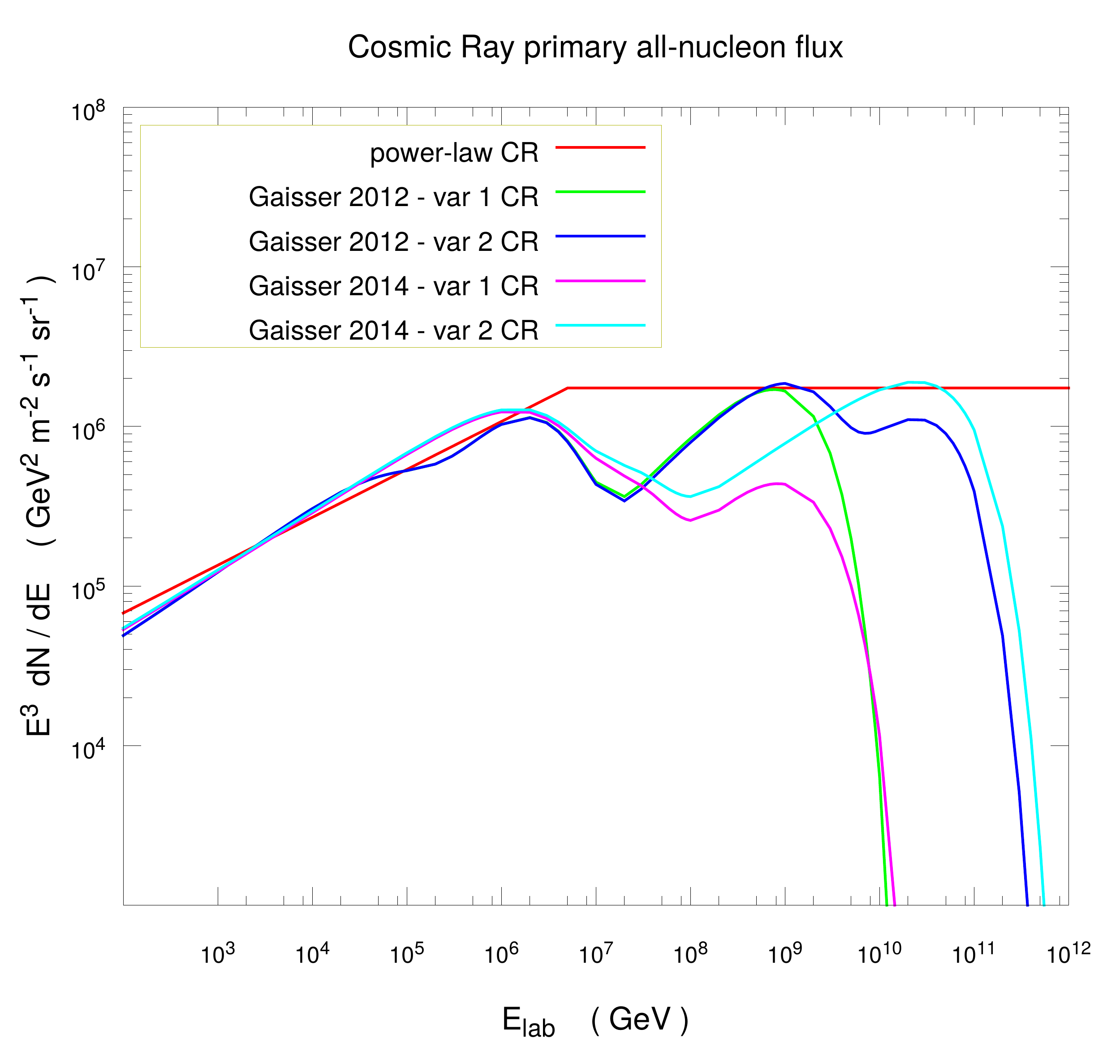}
\caption{\label{crspettri} 
  The all-nucleon primary cosmic ray spectra used as input in this work.
See text for more detail.}
\end{center}
\end{figure}

CR spectra in the upper layer of our atmosphere
are very well constrained at low energies by many direct measurements.
Balloon-borne and space experiments, like AMS and CREAM, 
are able to discriminate with high precision the individual elements included in
the cosmic ray composition up to energies around $E_{lab} = 10^5$~GeV~\cite{Peixoto:2015ava}. 
On the other hand, the high energy tail of cosmic rays is subject to significant uncertainties, 
in particular related to the different possible CR compositions
(protons or heavier ions, up to iron) and the CR origin (galactic or extra-galactic). 
The high energy region is investigated by ground-based experiments, like KASCADE, KASCADE-Grande
and the Pierre Auger Observatory, which glo\-bal\-ly cover the energy range between $E_{lab} = 10^6$~GeV and up to several $10^{11}$ GeV.
In this context it is important to note that the lepton fluxes at a given energy are affected by
CR spectra at energies even larger by a factor of order $O(100-1000)$, 
due to the integration over primary energies in the expressions of the generation functions, eqs.~(\ref{sprod}) and~(\ref{sdecay}).
Therefore, in order to parametrize the uncertainty on our knowledge of cosmic ray spectra at high energies 
we consider the following possibilities, i.e., 
we evaluate lepton fluxes separately for each of the following primary cosmic ray
spectra, which are available in literature~\footnote{Given the fact,
  that we parameterize the $p$-$A$ cross-sections
  in terms of $p$-$p$ ones, cf. Sec.~\ref{subsec:charm},
  we consider the all-nucleon version of each spectrum.}:
\begin{itemize}
\item[1)] Power-law spectrum, composed by two parts:
\begin{eqnarray}
\phi^0_p(E) & = & 1.7\, E^{-2.7} 
\mathrm{cm}^{-2}\,\mathrm{s}^{-1}\,\mathrm{sr}^{-1}\,\mathrm{GeV}^{-1} 
\qquad
{\mathrm{for}}\,\, E < 5 \cdot 10^6 \, \mathrm{GeV} 
\, ,
\nonumber \\
& & 174\, E^{-3}
\mathrm{cm}^{-2}\,\mathrm{s}^{-1}\,\mathrm{sr}^{-1}\,\mathrm{GeV}^{-1} 
\, \,\qquad
{\mathrm{for}}\,\, E > 5 \cdot 10^6 \, \mathrm{GeV}
\, .
\end{eqnarray}
This is a reference spectrum used in earlier works on 
prompt lepton fluxes, and, although recent measurements have shown
that it basically overestimates nucleon fluxes at the highest energies,
we consider it for reference and comparison with older
works~\cite{Pasquali:1998ji, Enberg:2008te, Bhattacharya:2015jpa}.
\item[2)] Gaisser 2012 (variant 1 and 2)~\cite{Gaisser:2012zz}: \\
The first variant of the Gaisser spectrum, fitting available experimental data
of different origin to an analytic expression with a number of
parameters, is based on the hypothesis that three populations, one including
CR particles accelerated by SuperNova remnants in our galaxy, a second one
still of galactic origin but with an higher energy, and a third one of
particles accelerated at extra-galactic sources, contribute to the measured CR
spectrum. The three populations are characterized by different
rigidities~\footnote{The rigidity of each population multiplied by the atomic
  number of each nuclear group, determines the characteristic energy where the
  corresponding all-particle CR spectrum exponentially cuts off. The larger
  the rigidity is, more extended is the spectrum at high energy.}, they all include protons and nuclear groups (He, CNO, Mg-Si, Fe) with different spectral indices.
The second variant of the Gaisser spectrum provides a special treatment of
the third population, which is assumed to be composed of protons only, with large rigidity.

\item[3)] Gaisser 2014 (variant 1 and 2)~\cite{Stanev:2014mla, Gaisser:2013bla}: \\
 This uses the same functional form as in Gaisser 2012, but with updated parameters for an alternative fit of experimental data. In particular, the first variant of the spectrum involves three populations, two of galactic and one of extra-galactic origin, involving the p, He, C, O, Fe nuclear groups, with dif\-fe\-rent rigidities with respect to the Gaisser 2012 case. The second variant differs from the first one because it includes an additional component from heavier nuclei, plus a fourth population, characterized by extra-galactic protons only, with large rigidity. This affects the ultra-high-energy part of the spectrum and improves the agreement with Auger data on cosmic ray composition at high-energy~\cite{Kampert:2012mx}.

\end{itemize}
The all nucleon spectra corresponding to these cases are shown in Fig.~\ref{crspettri}. The effect of the different options on the shape of lepton fluxes is extensively discussed in Sec.~\ref{sec:uncQCD}. 

\subsection{Input: p-Air total inelastic cross-section}
\label{subsec:pair}

\begin{figure}[ht!]
\begin{center}
\includegraphics[width=0.75\textwidth]{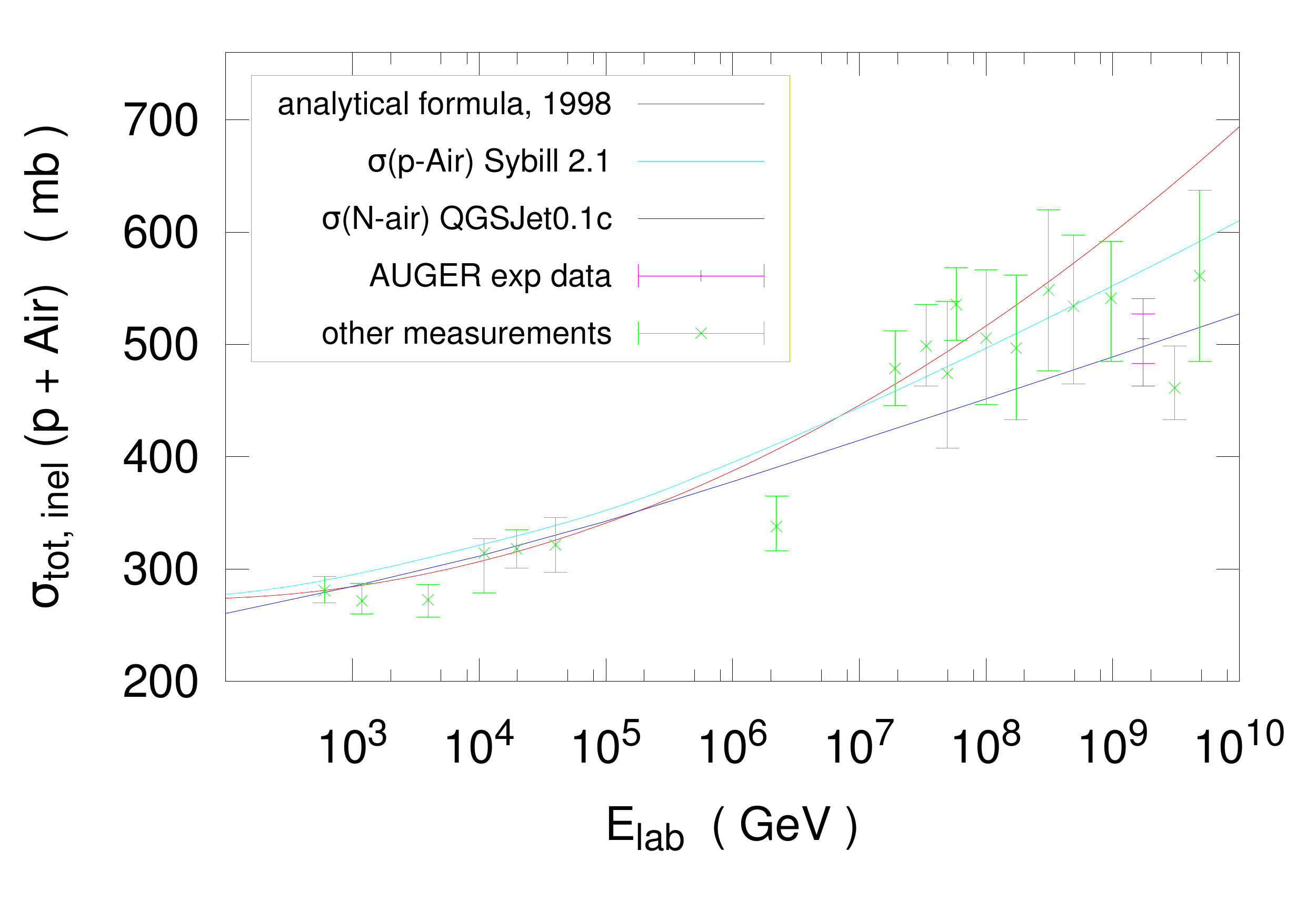}
\caption{\label{fig:pair-a} 
  The $p$-Air total inelastic cross-section according
  to different models ({\texttt{QGSJet0.1c}}, {\texttt{SIBYLL2.1}}, analytical) as compared to
  measurements from Auger and older experiments. The experimental data are
  taken from Ref.~\cite{Collaboration:2012wt} and references therein.
}
\end{center}
\end{figure}

The total inelastic proton-Air cross-section as a function of the laboratory
energy is an input in the denominator of the integrand of the generation
function $S_{prod}$ in eq.~(\ref{sprod}), and as a consequence, for the $Z$-moments for
heavy-hadron hadroproduction. 
Several measurements exist for this quantity, performed by different experiments 
(for a collection of results see Ref.~\cite{Collaboration:2012wt} and references therein), 
together with theoretical predictions, on the basis of phenomenological models. 
In this paper, for com\-pa\-ti\-bi\-li\-ty with previous works, we consider both the analytical formula~\cite{Mielke:1994un}, 
already used in old estimates of prompt neutrino fluxes (see, e.g.,~\cite{Pasquali:1998ji}), 
\begin{equation}
\label{eq:sigmapAir}
\sigma_{p-Air}^{inel} (E) = 290 - 8.7\, \mathrm{ln} (E/\mathrm{GeV}) + 1.14\, \mathrm{ln}^2 (E/\mathrm{GeV}) \,\,\,  \mathrm{mb} \, ,
\end{equation}
and predictions from the {\texttt{SIBYLL2.1}}~\cite{Ahn:2009wx}  
and the  
{\texttt{QGSJet0.1c}}~\cite{Kalmykov:1993qe}
models for hadronic interactions
included in the {\texttt{CORSIKA}} package~\cite{Heck:1998vt}. 
Cross-sections corresponding to those possible options are shown in Fig.~\ref{fig:pair-a} together with presently available expe\-ri\-men\-tal data. 
We also point out that predictions from other {\texttt{CORSIKA}} models, like {\texttt{EPOS~1.99}}~\cite{Werner:2005jf} 
lie within the band that one can draw from these two choices, as shown in Fig.~2 of Ref.~\cite{Collaboration:2012wt}, 
so that we consider them as upper and lower limits. 
The recent measurement from the Auger collaboration at $\sqrt{s}$ = 57~TeV, reported in
Ref.~\cite{Collaboration:2012wt} turns out to be in agreement, within the
error bands, with the predictions from {\texttt{QGSJet0.1c}}.  

Discussions on the effects of the different assumptions for the $p$-Air
cross-section on our final results of lepton fluxes are reported in Sec.~\ref{subsec:other}.

\subsection{Charm hadroproduction cross-section}
\label{subsec:charm}

Heavy-quark hadroproduction has been extensively studied in perturbative QCD.
The QCD corrections at NLO have first been obtained in Refs.~\cite{Nason:1987xz,Beenakker:1988bq,Beenakker:1990maa}
and are a\-vai\-la\-ble in public tools, like
{\texttt{hvqmnr}}~\cite{Mangano:1991jk}, 
{\texttt{MCFM}}~\cite{Campbell:2010ff} or 
{\texttt{HELAC-NLO}}~\cite{Bevilacqua:2011xh}
for the automatic computation of fully differential observables.
For the inclusive cross-section, the QCD corrections
are complete to NNLO~\cite{Baernreuther:2012ws,Czakon:2012zr,Czakon:2012pz,Czakon:2013goa} 
and, thus far, have been applied to top-quark pair production.
All these theory predictions have adopted the on-shell renormalization scheme for the heavy-quark mass.
The conversion to the \msbar scheme for the heavy-quark mass has been
discussed in Refs.~\cite{Langenfeld:2009wd,Aliev:2010zk,Dowling:2013baa}.

Beyond the perturbative expansion at fixed order, the resummation of large
lo\-ga\-rithms features important improvements, 
cf., e.g., the review in Ref.~\cite{Cacciari:2012ny} for charm and bottom production at the LHC.
In dynamical regimes where the transverse momentum $p_T$ of the heavy quark is much larger
than its mass $m$ the semi-analytical resummmation of logarithms in $p_T/m$ has been performed 
in Ref.~\cite{Cacciari:1998it} in the so-called FONLL approach.
On the other hand, when the NLO corrections are consistently matched with parton showers (PS) as in the {\texttt{POWHEG}}~\cite{Nason:2004rx,Frixione:2007vw} or MC@NLO~\cite{Frixione:2002ik}
approaches using the frameworks of {\texttt{POWHEG-BOX}}~\cite{Alioli:2010xd}
or {\texttt{(a)MC@NLO}}~\cite{Frixione:2010wd}, respectively, 
the leading logarithms are effectively resummed through 
the Monte Carlo (NLO~+~PS) event generators.

In summary, there exists a robust theoretical framework with a set of well tested tools 
for the computation of top, bottom and charm-pair hadroproduction at high energies, 
which has been developed for and used extensively in the LHC environment.
For estimating lepton fluxes from atmospheric charm, the core of our calculation 
is an updated estimate of the $NN$ $\rightarrow$ $c\bar{c}$ production cross-section 
in perturbative QCD, where $N$ indicates a nucleon~\footnote{We use the
  approximation $p \simeq n \simeq N$, neglecting mass differences between
  protons and neutrons by approximating all masses as $m_p$. At high energies
  differences in the partonic content of $p$ and $n$ may also be safely
  neglected, whereas at low energies differences in the PDFs of these partons
  imply differences between $pp$ and $NN$ cross-sections up to a factor,
  depending on the specific PDF set, of a few percent at $E_{lab}$ = 50~GeV,
  reducing to a few per mil above $E_{lab}$ = 500~GeV.}. 
We use the QCD predictions for the inclusive cross-section $pp$ $\rightarrow$ $c\bar{c}$ at NNLO, 
and their comparison to those at NLO, to study the stability of the perturbative expansion 
as a guide for fixing parameters and inputs to be adopted in the application to atmospheric charm.
In detail, this includes the PDF dependence, the choice of the central values for renormalization and
factorization scales $\mu_R$ and $\mu_F$ and the charm mass, as well as plausible intervals for their variations, 
given the fact that the global uncertainty bands at NLO due to scale, PDF and charm mass variation are large. 
Our findings are summarized in the sequel.

\begin{figure}[ht!]
\begin{center}
\includegraphics[width=0.475\textwidth]{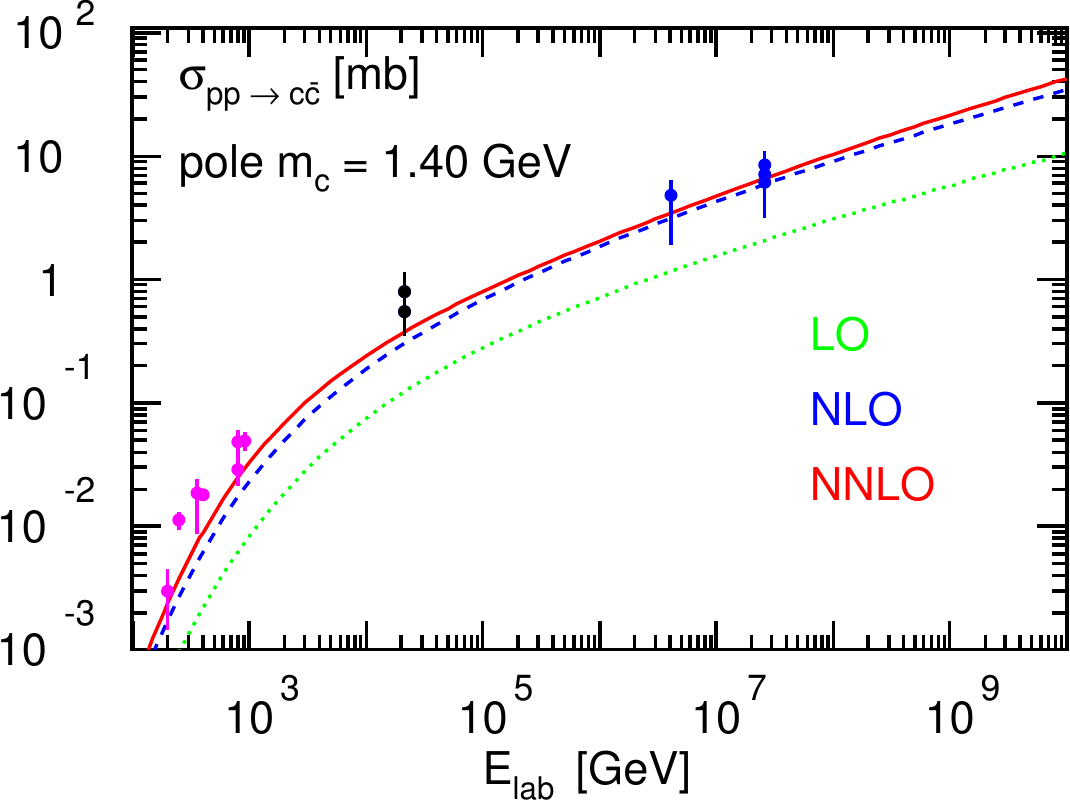}
\hspace*{5mm}
\includegraphics[width=0.475\textwidth]{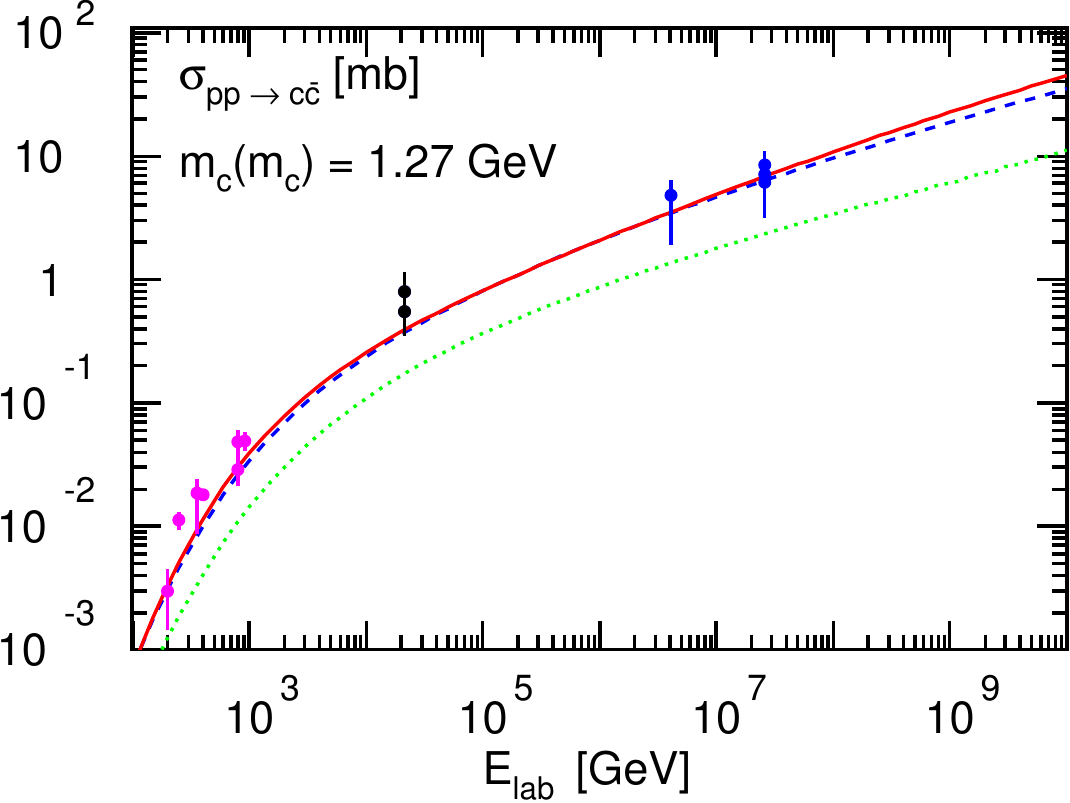}
\caption{\label{fig:sigmatot-elab} 
  Theoretical predictions for the total $pp \rightarrow c\bar{c}$  cross-section
  as a function of the laboratory energy $E_{lab}$ 
  at LO (dotted), NLO (dashed), NNLO (solid) QCD accuracy in the pole mass (left) and in the \msbar\
  mass scheme (right) 
  using the central set of the ABM11 PDFs in the FFNS with $n_f = 3$.
  The scales were chosen as $\mu_R = \mu_F = 2m_c^{\rm pole}$ 
  with $m_c^{\rm pole}~=~1.4$~GeV in the on-shell scheme 
  and as $\mu_R = \mu_F = 2m_c(m_c)$ with $m_c(m_c)~=~1.27$~GeV
  in the \msbar\ mass scheme, respectively. 
  See text for details and references on the experimental data from fixed target experiments and colliders (STAR, PHENIX, ALICE, ATLAS,
  LHCb).
}
\end{center}
\begin{center}
\includegraphics[width=0.475\textwidth]{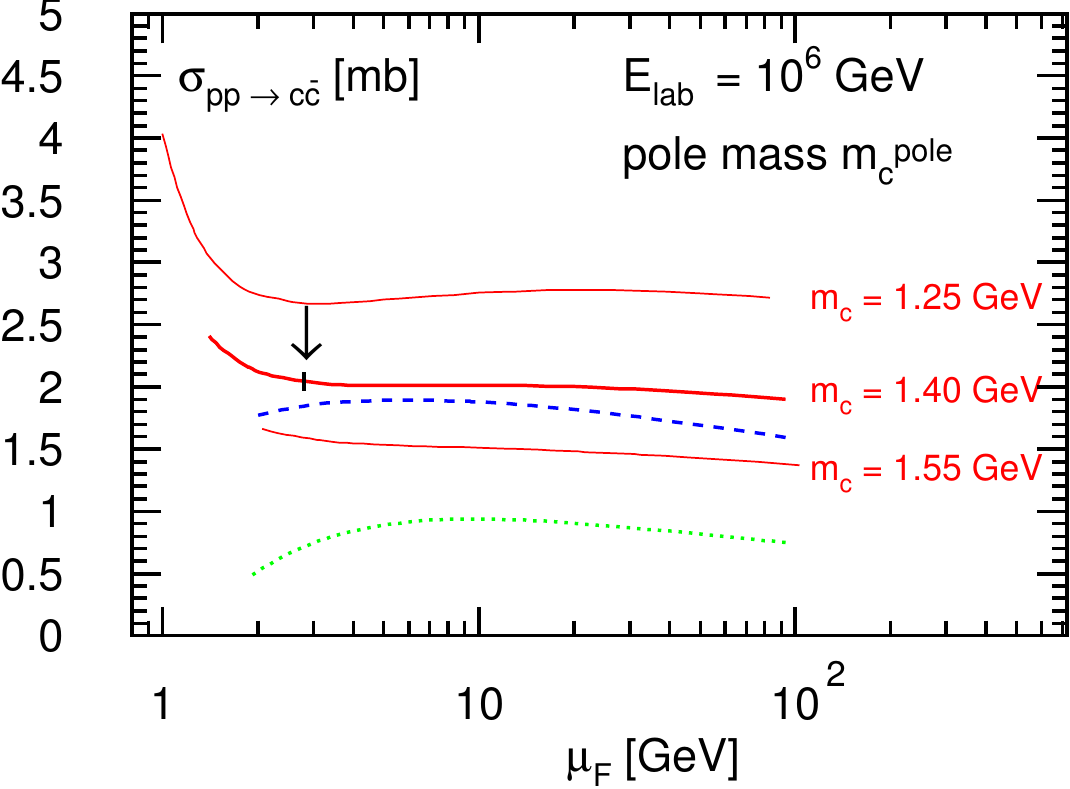}
\hspace*{5mm}
\includegraphics[width=0.475\textwidth]{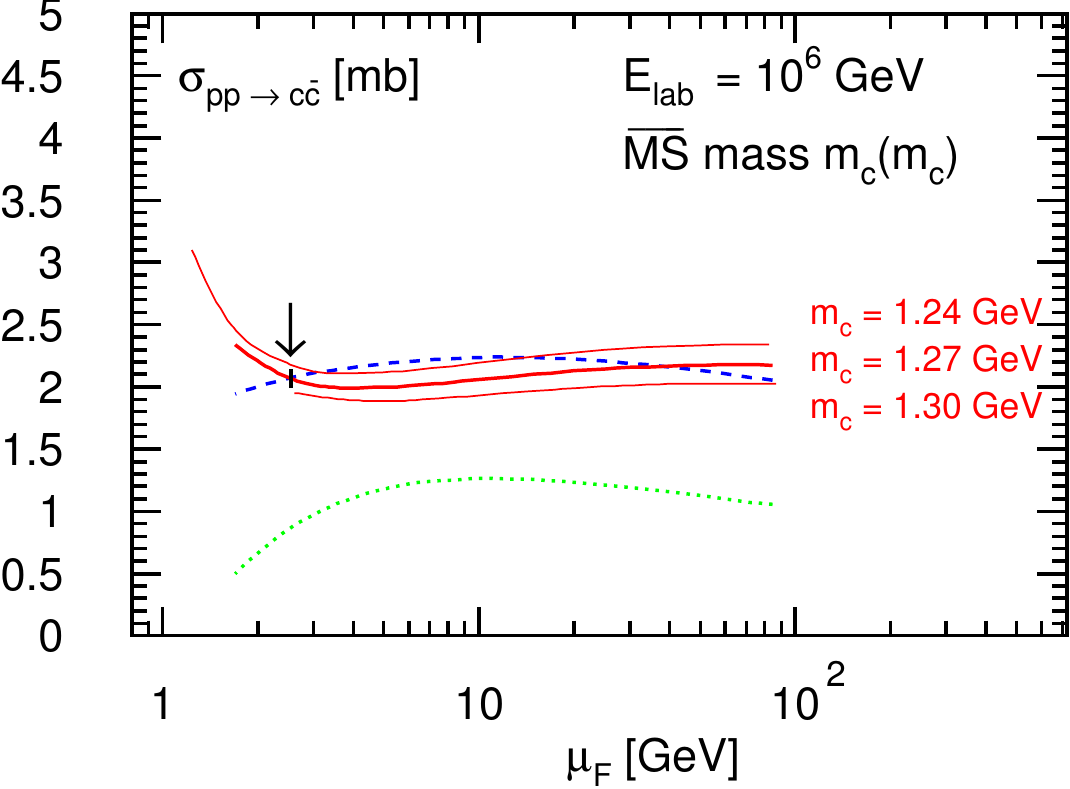}
\caption{\label{fig:sigmatot-mass} 
  Sensitivity of the total cross-section for $pp \rightarrow c\bar{c}$ to the factorization scale $\mu_F$ 
  at LO (dotted), NLO (dashed), NNLO (solid) QCD accuracy, in the pole mass (left) and in the \msbar\
  mass scheme (right). 
  The charm mass and PDFs were fixed as in Fig.~\ref{fig:sigmatot-elab}. 
  The central line at each order denotes the choice $\mu_R = \mu_F$.
  The upper and the lower lines at NNLO denote the cross-sections from the
  mass variation $m_c^{\rm pole}=1.40 \pm 0.15$ GeV and $m_c(m_c)=1.27 \pm 0.03$ GeV, respectively.  
  The arrows indicate the scale $\mu_R = \mu_F$ equal to $2m_c^{\rm pole} $ (left) and $2m_c(m_c)$ (right), respectively.  
}
\end{center}
\end{figure}

\begin{figure}[ht!]
\begin{center}
\includegraphics[width=0.475\textwidth]{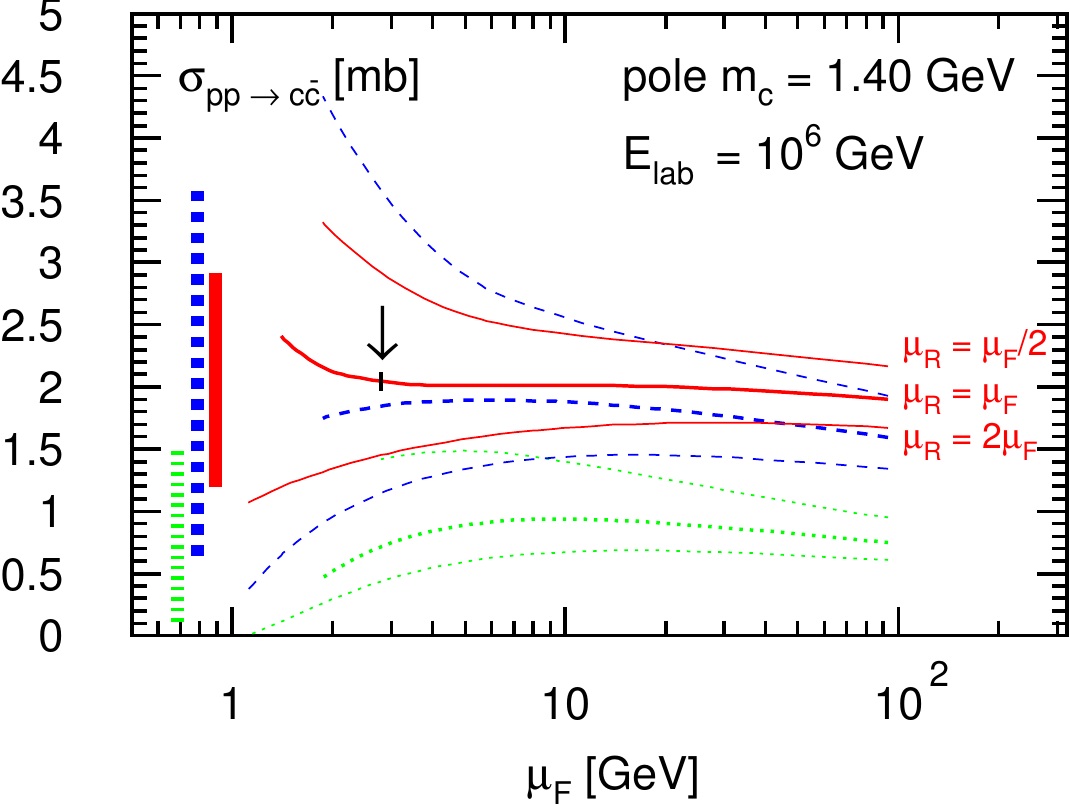}
\hspace*{5mm}
\includegraphics[width=0.475\textwidth]{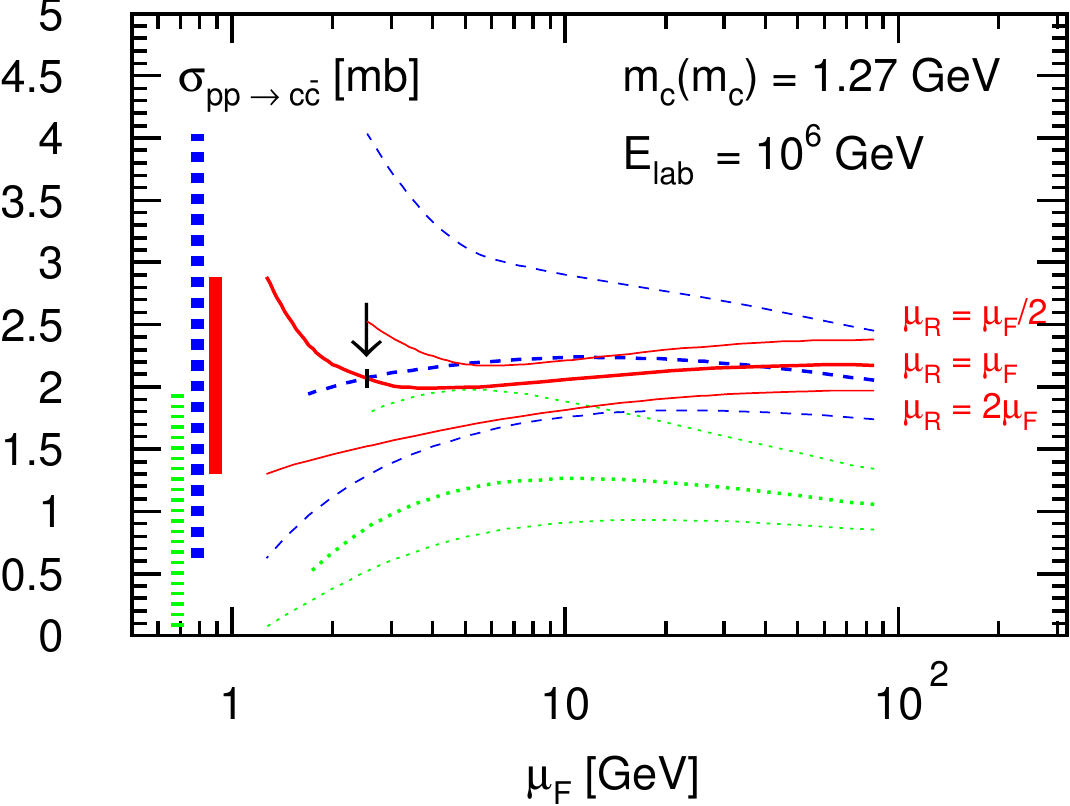}
\caption{\label{fig:sigmatot-scale} 
  Sensitivity of the total cross-section for $pp \rightarrow c\bar{c}$ to the factorization scale $\mu_F$ 
  with the same PDFs and charm mass central values as in Fig.~\ref{fig:sigmatot-mass}.
  The central line at each order denotes the choice $\mu_R = \mu_F$,
  the upper and the lower line the choices $\mu_R = \mu_F/2$ and 
  $\mu_R = 2 \mu_F$, respectively.
  The vertical bars give the size of the independent variation 
  of $\mu_R$ and $\mu_F$ in the standard range $m_c^{\rm pole}/2 \le \mu_R, \mu_F \le 2m_c^{\rm pole}$ 
  and $m_c(m_c)/2 \le \mu_R, \mu_F \le 2m_c(m_c)$, respectively, with the restriction that 
  $1/2 \le \mu_R/\mu_F \le 2$.
  Again, the arrows indicate the scale $\mu_R = \mu_F$ equal to $2m_c^{\rm pole} $  (left) and $2m_c(m_c)$ (right).  
}
\end{center}
\begin{center}
\includegraphics[width=0.475\textwidth]{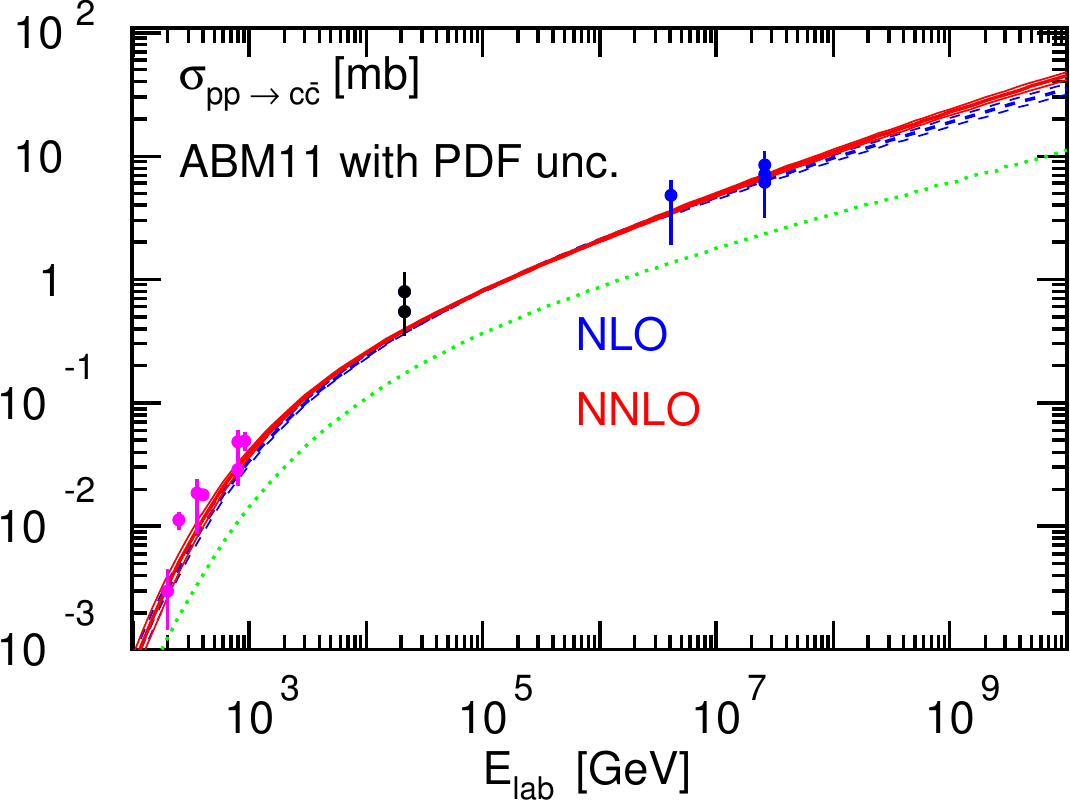}
\hspace*{5mm}
\includegraphics[width=0.475\textwidth]{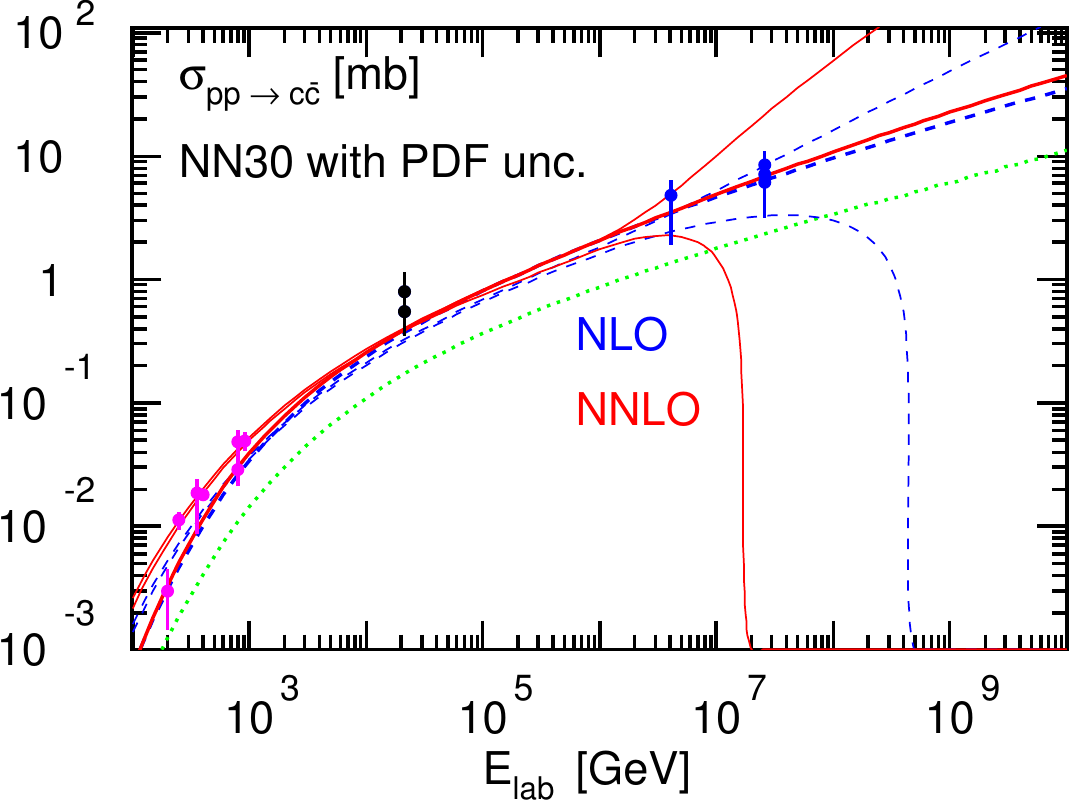}
\caption{\label{fig:sigmatot-pdf} 
  Dependence of the total cross-section for $pp \rightarrow c\bar{c}$ on the PDF choice 
  at LO (dotted), NLO (dashed), NNLO (solid) QCD accuracy in the \msbar\
  mass scheme.
  The charm mass and the scales were fixed as in Fig.~\ref{fig:sigmatot-elab}. 
  The upper and the lower lines at NLO and NNLO indicate 
  the total 1$\sigma$ PDF uncertainty band for ABM11 (left)
  and NNPDF3.0 PDF set (right) with $n_f$ = 3. Experimental data are the same as in Fig.~\ref{fig:sigmatot-elab}.
} 
\end{center}
\end{figure}

Fig.~\ref{fig:sigmatot-elab} displays the dependence of the total cross-section $\sigma_{pp \to c{\bar c}}$ on the laboratory energy $E_{lab}$.
The computation is performed in the fixed flavor number scheme (FFNS) with the number of flavors $n_f = 3$, 
implying that charm is considered as a heavy state consistently included in the matrix elements
with its mass different from zero and its presence excluded from the PDFs.
The computation is performed in the theoretical framework as implemented in the {\texttt{HATHOR}} code~\cite{Aliev:2010zk}.
Fig.~\ref{fig:sigmatot-elab} applies two different schemes for the heavy quark mass renormalization, 
the commonly chosen on-shell scheme with the pole mass $m_c^{\rm pole}$ and 
the \msbar scheme with the running mass $m_c(\mu_R)$, where the renormalization
scale for the evaluation of the mass has been fixed at $\mu_R=m_c$. 

The experimental data in Fig.~\ref{fig:sigmatot-elab} for the fixed target experiments with energies up to $E_{lab} = 10^3$~GeV
are taken from Ref.~\cite{Lourenco:2006vw} and for HERA-B from Ref.~\cite{Abt:2007zg} (purple points in Fig.~\ref{fig:sigmatot-elab}).
RHIC data from PHENIX and STAR have been published in Refs.~\cite{Adare:2006hc,Adamczyk:2012af}
(black points in Fig.~\ref{fig:sigmatot-elab}) and LHC data are available from ALICE~\cite{Abelev:2012vra},
ATLAS~\cite{ATLAS:2011fea} and LHCb~\cite{Aaij:2013mga} (blue points in Fig.~\ref{fig:sigmatot-elab}), 
see also Ref.~\cite{Bhattacharya:2015jpa}.
Fig.~\ref{fig:sigmatot-elab} demonstrates the stability of the perturbative expansion 
of the $\sigma_{pp \to c{\bar c}}$ cross-section through NNLO up to very high
energies and good consistency of the predictions with the experimental data.

Related to the heavy quark mass renormalization is the choice of the numerical value for the charm quark mass.
The Particle Data Group (PDG)~\cite{Agashe:2014kda} reports a very precise value
of $m_c(m_c) = 1.275 \pm 0.025$~GeV  in the \msbar\ scheme.
In case of charm, 
the conversion of the \msbar\ to the pole mass 
suffers from well-known convergence problems, see, e.g., Ref.~\cite{Marquard:2015qpa}.
In addition, the definition of the pole mass is based on the unphysical idea of quarks as
asymptotic states of the $S$-matrix, so the accuracy in the pole mass is limited to be of
the order ${\cal O}(\Lambda_{\rm QCD})$ by the renormalon ambiguity.
The comparison of the $\sigma_{pp \to c{\bar c}}$ cross-sections 
in the two mass renormalization schemes 
at the nominal scales $\mu_R = \mu_F$ equal to $2m_c^{\rm pole} $ and $2m_c(m_c)$ 
and taking the result with the precise PDG value as a reference, motivates
our choice for the charm pole mass $m_c^{\rm pole} = 1.40 \pm 0.15$~GeV, 
as illustrated in Fig.~\ref{fig:sigmatot-mass}.

The behavior of the total cross-section as a function of the factorization
scale $\mu_F$ is further explored in Fig.~\ref{fig:sigmatot-scale} 
by considering three different renormalization scales, $\mu_R = \mu_F/2, \mu_F$ and $2 \mu_F$, at LO, NLO and NNLO QCD. 
Fig.~\ref{fig:sigmatot-scale} demonstrates, that 
the choice of the mass renormalization scheme is important, because the \msbar scheme 
leads to predictions with slightly improved convergence.
Scale stability of the perturbative expansion at NNLO is reached in both
schemes for scales $\mu_R \sim \mu_F \gsim 2$~GeV.
As shown in Fig.~\ref{fig:sigmatot-scale} the use of the running mass
scheme leads to a somewhat reduced scale uncertainty band at NNLO 
for the independent variation of $\mu_R$ and $\mu_F$ in the standard range 
$\mu_R/m_c(m_c)$ and $\mu_F/m_c(m_c) \in [1/2, 2]$ 
and restricting the ratio $1/2 \le \mu_R/\mu_F \le 2$, as compared to the pole mass scheme.
Similar features, although much more pronounced, have been found already for the $t\bar{t}$ hadroproduction cross-sections and
differential distributions in Ref.~\cite{Dowling:2013baa}.
Both in the pole and in the running mass scheme the point of minimal sensitivity, 
i.e., the region where the cross-sections predictions at NLO and NNLO approximately coincide, 
turned out to be around scales $\mu_R = \mu_F \sim 2m_{c}$ and larger. 
This justifies scale choice adopted in Fig.~\ref{fig:sigmatot-elab}.
Translating this value into a dynamical scale, more suitable to describe the dynamics of
heavy-quarks in differential distributions, 
we will use in the following a dynamical central scale for our calculation fixed to $\mu_F = \mu_R = \sqrt{p_T^2 + 4 m_{c}^2}$, 
where $p_T$ is the transverse momentum of the emitted charm quark.  

Another important factor contributing to the theoretical uncertainties of the $c\bar{c}$ 
hadroproduction cross-section originates from the choice of the PDF set.
We have taken predictions at NNLO accuracy as the basis of our central PDF choice.
Among the different possibilities, currently available in the LHAPDF
interface~\cite{Buckley:2014ana}, we have chosen the ABM11 one~\cite{Alekhin:2012ig},
together with the respective value for the strong coupling constant $\alpha_s(M_Z)$, as a default.
In the FFNS with $n_f = 3$, this PDF features a central set complemented by 28 variations, allowing to
estimate a PDF uncertainty band at the $1\sigma$ level. 
The ABM11 PDFs are compatible with ABM12~\cite{Alekhin:2013nda}, where the
latter set has been tuned to LHC data.
Moreover, the predictions of the ABM11 and ABM12 family for gluon PDF at low Bjorken-$x$ values are in complete compatibility with the only PDF fit
available in literature so far including LHCb data on $c\bar{c}$ and $b\bar{b}$ hadroproduction, that has recently been performed by the PROSA collaboration~\cite{Zenaiev:2015rfa, Zenaiev:2015qea}~\footnote{See also footnote~\ref{others}.}. 

As an alternative, we have used the 3-flavor central PDF set at NLO available from CT10~\cite{Lai:2010vv}, 
which also provides results characterized by partial com\-pa\-ti\-bi\-li\-ty with the
PROSA fit (differences lie within about 2$\sigma$) and with ABM11.
At NNLO, both the CT10 and the ABM11 PDF sets give positive results 
for the total cross-section for $c\bar{c}$ hadroproduction 
in the highest energy range ranging up to $E_{lab} \sim 10^{10}$~GeV, 
together with an uncertainty band for the PDFs which always stays positive as well,
see Fig.~\ref{fig:sigmatot-pdf} (left) for predictions from ABM11.

On the other hand, PDF sets  with unconstrained gluons at small $x$ lead to very different results.
In particular, the NNPDF3.0 set~\cite{Ball:2014uwa}, characterized by a different
parameterization, leads in case of the highest energies, to a huge uncertainty band, 
even covering a range with negative cross-section values.
As shown in Fig.~\ref{fig:sigmatot-pdf} (right), 
the cross-sections obtained with NNPDF3.0 at NNLO do not remain positive anymore already for $E_{lab} \gsim 5 \cdot 10^7$~GeV, 
an energy well below the one so far covered in run 1 by the LHC~\footnote{\label{others}
  Data for the hadroproduction of heavy quarks at the LHC can therefore be used
  to further constrain these PDFs at small $x$.
  Very recently, following a research line non too different from the one
  already pointed out by the PROSA collaboration, 
  Ref.~\cite{Gauld:2015yia} has considered constraints on
  the small-$x$ gluon from charm hadroproduction at the LHC, working in
  a scheme with 5 flavors, though. In contrast, Refs.~\cite{Zenaiev:2015rfa, Zenaiev:2015qea}
  and our work make consistent use of the FFNS with 3 flavors.
}.
Thus, in the remainder of this paper we only consider the central value of the NNPDF3.0 set, 
with the purpose of quantifying differences with respect to the central values of the other families. 

The PDFs from MMHT~\cite{Harland-Lang:2014zoa}, and in particular their central fit, 
lead to negative $c\bar{c}$ hadroproduction cross-sections at NNLO for energies above $E_{lab} \gsim 5 \times 10^8$~GeV.
Such an unphysical feature also affects predictions for the longitudinal structure function
$F_L$ in deep inelastic scattering as noted, e.g., in Ref.~\cite{CooperSarkar:2011pa}. 
While the MMHT set seems to be valid for the description of the production of heavier
particles at LHC energies, its extrapolation to higher
energies characterizing several astroparticle physics problems is quite questionable.
Thus, we neglect this PDF family in the present study.

Finally, the small $x$ region is not only of importance for the gluon PDF, 
but also when considering the behavior of the perturbative hard parton scattering cross-section.
The high-energy factorization of the cross-section~\cite{Catani:1990eg,Ball:2001pq}
in the limit when the center-of-mass energy $S$ is much larger than the heavy-quark mass provides an effective theory for the description of the high-energy logarithms in $S/m^2$.
These behave as $\ln^0(S/m^2) \simeq {\rm const.}$ at NLO, as $\ln^1(S/m^2)$ at
NNLO, and so on, see, e.g., Ref.~\cite{Moch:2012mk} and studies of operator matrix elements in deep-inelastic scattering at three
loop order in the small-$x$ limit~\cite{Kawamura:2012cr,Ablinger:2014nga}.
At the energies currently considered, even up to $E_{lab} \sim 10^{10}$~GeV, 
their numerical importance is, however, strongly suppressed in the convolution integral of the hard partonic cross
section at small $x$ with the large $x$ part of the gluon PDF (and vice versa).
The apparent convergence of perturbative expansion for the 
$\sigma_{pp \to c{\bar c}}$ cross-section  through NNLO 
observed in Fig.~\ref{fig:sigmatot-elab} underpins this fact.

The $c\bar{c}$ differential distributions which we use in this paper are 
at parton level exact to NLO in QCD, because differential predictions 
for $c\bar{c}$ hadroproduction are not yet available at NNLO.
For generating these distributions we use the {\texttt{POWHEG-BOX}}~\cite{Alioli:2010xd}, 
complemented by the event generator {\texttt{PYTHIA-6.4.28}}~\cite{Sjostrand:2006za}, 
in a $p_T$-ordered tune belonging to the family of Perugia tunes~\cite{Skands:2010ak}, for describing
parton shower and hadronization. 
This provides us with differential distributions of $D$-hadrons at NLO
accuracy in QCD with NLO matching to parton showers according to the {\texttt{POWHEG}} method.
Beyond the resummation to leading logarithmic accuracy provided by the parton showers approaches, 
next-to-leading logarithmic (NLL) contributions of $p_T/m$, as obtainable by an
approach like FONLL,
are not considered here. 
This is justified because the computation of $Z$-moments requires an 
integration over the whole kinematically accessible range in $p_T$ and thus exhibits less sensitivity 
to the shape of the $p_T$ distribution at large $p_T$, which is mainly influenced by the NLL
corrections provided by the FONLL approach.

Moreover, following Ref.~\cite{Pasquali:1998ji}, 
we derive the total charm cross-section in 
the i\-ne\-la\-stic proton-Air ($pA$) collisions, from the $pN$ cross-section, by using the formula
\begin{eqnarray}
\sigma_{pA \rightarrow c\bar{c}} = A^\gamma \sigma_{pN \rightarrow c\bar{c}}
\end{eqnarray}
where $A=14.5$ for a nucleus of air. Here, we take $\gamma=1$ assuming a linear superposition since, 
for light nuclei, the effect of nuclear shadowing is expected to be small~\cite{Bhattacharya:2015jpa, BlancoCovarrubias:2009py}.

\subsection{$Z$-moments: $Z_{p\,h}$, $Z_{h\,l}$, $Z_{p\,p}$, $Z_{h\,h}$}

\subsubsection{$Z_{p\,h}$}
The inclusive $c\bar{c}$ cross-section is an essential ingredient for the estimate of the charm contribution to lepton fluxes. 
The latter ones, in fact, depend on the $Z$ production moments which can be 
expressed as integrals over the differential distribution $d\sigma_{p \rightarrow charm} (E/x_E)/dx_E$, through the formula,
\begin{eqnarray}
\label{eq:Z-moms}
Z_{p \,,\, charm} (E)  = \int_0^1 \frac{dx_E}{x_E}\,  
\frac{\phi_p(E/x_E)/\phi_p(E)}{\sigma_{p}(E)}\, 
\frac{d\sigma_{p\rightarrow charm}(E/x_E)}{d x_E} \,,
\end{eqnarray}
with the ratio $x_E=E/E_k$. 
Here, $E_k$ is the nucleon energy in the laboratory frame and $E$ the energy of the produced particle (charm).
The primary CR nucleon flux is $\phi_p$ 
and we have assumed that charm is all produced in $c\bar{c}$ pairs from
primary CR nucleons (denoted by $p$) interacting with the Earth atmosphere, i.e., 
\begin{eqnarray}
d\sigma_{p\rightarrow charm}/d x_E = 2\, d\sigma_{pA\rightarrow c\bar{c}} /d x_E \,,
\end{eqnarray}
and that $\sigma_{p}(E)$ in eq.~(\ref{eq:Z-moms}) coincides with the total
inelastic proton-Air cross-section $\sigma^{inel}_{p-Air} (E)$. 
The lower integration limit would ideally correspond to the case of 
$E_k~\rightarrow~+\infty$. We thus replace it with $\epsilon$, as we compute
the cross-sections for $E_k$ limited to a finite value, with $\epsilon$ small enough that the results for $Z$-moments are almost independent of its variations $\epsilon_{var} < \epsilon$. 

As discussed in Sec.~\ref{subsec:charm}, for the differential 
cross-sections $d\sigma_{pA\rightarrow c\bar{c}} /d x_E$ and $d\sigma_{pp\rightarrow c\bar{c}} /d x_E$ no predictions at NNLO are available at present. 
Thus we compute it at NLO through {\texttt{POWHEG-BOX}}, using PDFs and
parameters such as $m_c$ and scale choices as described above, 
so that we are close to the point of minimal sensitivity for the total
cross-sections, where the differences between NLO and NNLO QCD predictions are small.

The hadronic moments $Z_{p\,h}$  with $h = D^0$, $\bar{D}^0$, $D^\pm$,
$D_s^\pm$ and $\Lambda_c^\pm$ were calculated in Ref.~\cite{Pasquali:1998ji}
from the partonic moment $Z_{p, charm}$, through the relation 
$Z_{p\,h}~=~f_{charm, h}~\,~Z_{p, charm}$, including conversion factors $f_{charm, h}$
which represent the fraction of charm quarks emerging as specific hadrons after
fragmentation. 
In Ref.~\cite{Bhattacharya:2015jpa} a more refined approach was used: 
differential distributions for hadrons were obtained from those for quarks 
after convoluting the latter ones with fragmentation functions that were
assumed to depend on the energies through the ratio $E_{h}/E_{charm}$ and
to be independent of the beam energy. 
On the other hand, the use of {\texttt{POWHEG-BOX}} allows us to follow a
different path, i.e., we take into account parton shower and fragmentation
effects by means of the Monte Carlo {\texttt{PYTHIA}} event generator applied
to the Les Houches events at first radiation emission level obtained by
running {\texttt{POWHEG-BOX}}. 
This allows us to directly extract differential distributions $d\sigma_{pp\rightarrow h+X} /dx_E$ for $D$-hadrons ($x_E = E_{h}/E_k$), whose shape
may, in general, differ from those of the charm quarks 
$d\sigma_{pp\rightarrow c+X} /dx_E$ ($x_E = E_{charm}/E_k$), 
implying that a global rescaling factor 
is too naive an approximation for the translation of quark distributions at
parton level into the corresponding ones at hadron level.  
\begin{figure}[ht!]
\begin{center}
\includegraphics[width=0.495\textwidth]{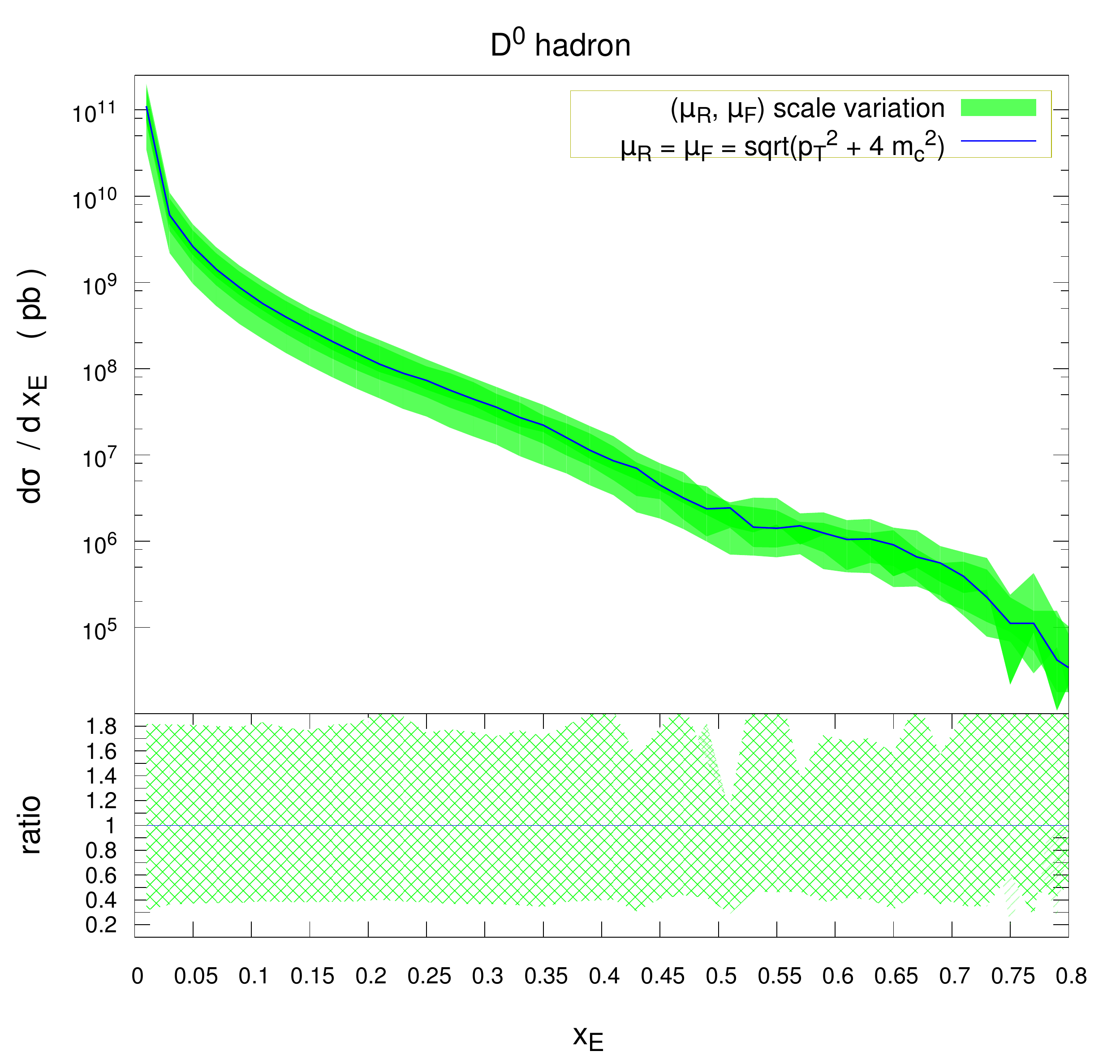}
\includegraphics[width=0.495\textwidth]{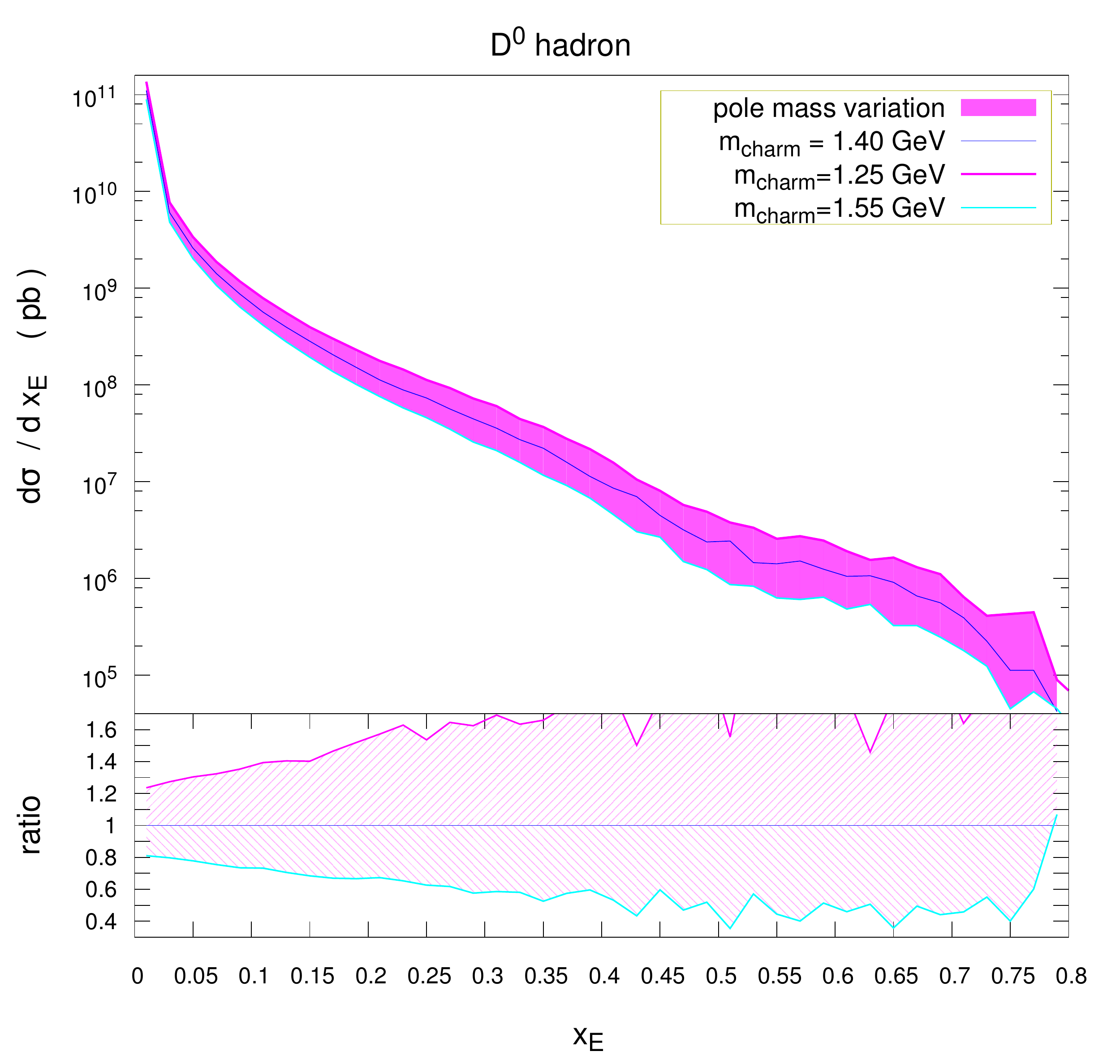}
\end{center}
\caption{\label{fig:dsigmadx} 
Differential distribution $d\sigma/d x_E$ 
for $pp \rightarrow c\bar{c} \rightarrow D^0 + X$ 
from {\texttt{POWHEG-BOX}}
interfaced to {\texttt{PYTHIA}} at $E_{lab}=10^7$~GeV. 
Central scales were fixed as  $\mu_R=\mu_F=\sqrt{p_{T,c}^2 + 4 m_c^2}$, 
central mass as $m_{c}^{pole} = 1.4$ GeV, and PDFs as the central set
of the ABM11 NLO family with $n_f$ = 3.   
The uncertainty bands related to scale variation (at fixed $m_c^{pole}$ and
PDFs) and mass variation (at fixed $\mu_R$, $\mu_F$ and PDFs) 
are shown in the left and right panels, respectively.
The lower subpanels display the band for the relative uncertainties 
when normalized with respect to the central prediction.
} 
\end{figure}
\begin{figure}[ht!]
\begin{center}
\includegraphics[width=0.495\textwidth]{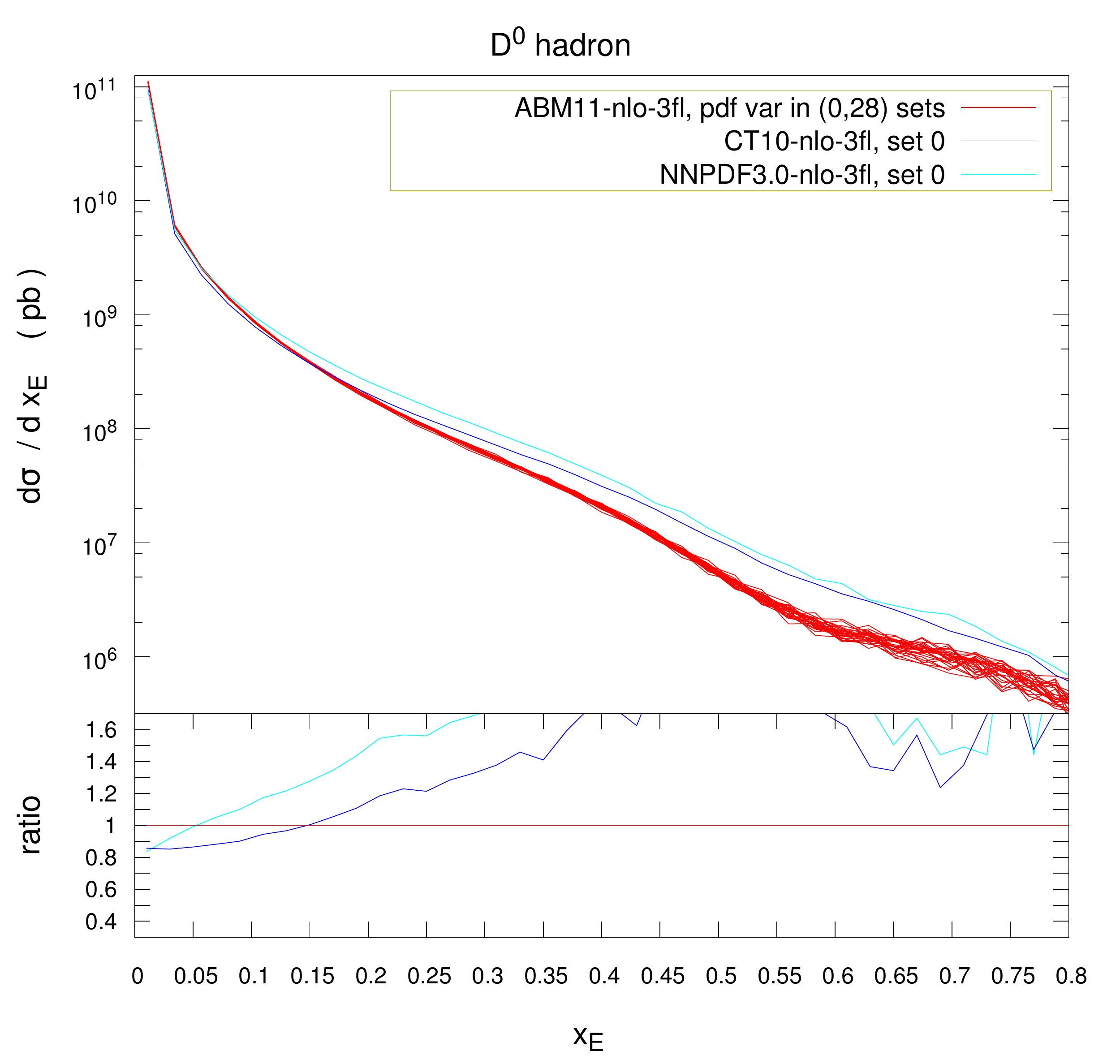}
\includegraphics[width=0.495\textwidth]{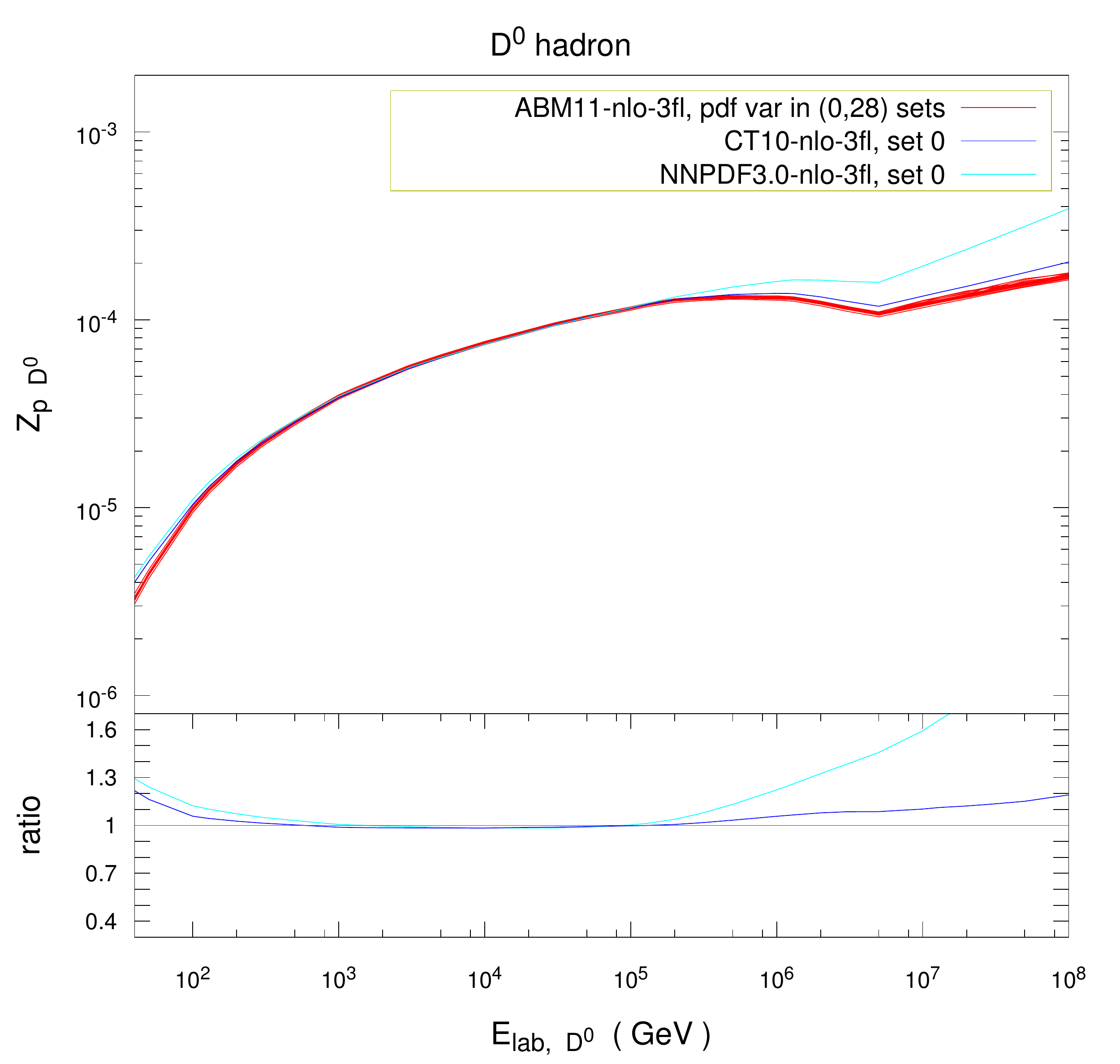}
\end{center}
\caption{\label{fig:dsigmadxpdf} 
Differential distribution $d\sigma/d x_E$ 
for $pp \rightarrow c\bar{c} \rightarrow D^0 + X$ from
{\texttt{POWHEG-BOX}} interfaced to {\texttt{PYTHIA}} 
at $E_{lab}=10^7$~GeV (left) 
and $Z$-moment for $D^0$ hadroproduction as a function of $E_{lab,\, D^0}$ (right). Scales were fixed as  $\mu_R=\mu_F=\sqrt{p_{T,c}^2 + 4 m_c^2}$, 
mass as $m_{c}^{pole} = 1.4$ GeV. 
The red lines correspond to the central fit and the 28 additional sets for
parametrization of the PDF and $\alpha_s$ uncertainties of the ABM11 PDFs at NLO with $n_f = 3$.
Predictions using central fits of the CT10 and NNPDF3.0 sets at NLO, 
with $n_f = 3$ each, are also shown (in blue and light-blue, respectively), together with their ratio to the predictions with the central set of ABM11 at NLO.
} 
\end{figure}

As an example, the differential distribution $d\sigma_{pp\rightarrow h+X} /dx_E$ 
is shown for the case of $D^0$ hadrons in Fig.~\ref{fig:dsigmadx} for a
$pp \rightarrow c\bar{c}$ collision characterized by $E_{lab} = 10^7$~GeV. 
The uncertainties due to scale and mass variation around the central predictions
are shown in the left and right panels of the figure, respectively. 
They are, in general, large. 
The scale uncertainties turn out to be almost constant with $x_E$,
whereas the uncertainties due to the variation of the mass increase with increasing $x_E$ on kinematical grounds.
Differential distributions for different $E_{lab}$ show a qualitatively
similar shape, a scaling property already pointed out in the literature. 

Uncertainties due to the PDF variation are shown in the left panel of Fig.~\ref{fig:dsigmadxpdf}. 
For illustration, the predictions for the central fit and the 28 additional
variations which parametrize the PDF and $\alpha_s$ uncertainties of the ABM11
PDFs with $n_f = 3$ in the FFNS, are displayed individually~\footnote{All differential results for PDF (scale) variations in this paper have been obtained after sho\-we\-ring with {\texttt{PYTHIA}} sets of Les Houches events generated in the {\texttt{POWHEG-BOX}} framework by explicitly fixing different PDFs (scales) in the input cards, without making use of reweighting facilities.
This ensures a fully consistent computation of the Sudakov form-factors,
including the specific PDF set and scale in the whole integrand,
in the separate generation of each set of events.}. 

As is evident from Fig.~\ref{fig:dsigmadxpdf}, the differences with the
central values of other PDF sets (CT10 and NNPDF3.0 at NLO in the $n_f = 3$ FFNS 
with their respective default value for $\alpha_s(M_Z)$) 
turn out to be larger than the combined PDF and $\alpha_s$ uncertainty coming
from the 28+1 ABM individual sets. This reflects a feature already observed at
NLO for several examples of LHC cross-sections, where differences arising from
the use of different PDF families turned out to be larger than those from the
variation of $\alpha_S$ and the PDFs through the sets belonging to a same family, see, e.g., Refs.~\cite{Dittmaier:2011ti,Alekhin:2012ig}. 

These differences propagate to the computation of the $Z$-moments, 
although the latter quantities are integrals over all possible $x_E$ values.
As an example, in the right panel of Fig.~\ref{fig:dsigmadxpdf}, the $Z$-moments
for $D^0$ hadroproduction are shown as a function of the $D^0$ energy in the
laboratory frame $E_{lab,\, D^0}$ for the different PDF choices just discussed
above (left panel, Fig.~\ref{fig:dsigmadxpdf}). 
Here, the power-law spectrum as been chosen as input for the CR flux. 
Thus, the change of shape visible around 5 $\cdot 10^6$~GeV 
is due to the change of the spectral index in the power-law spectrum around the knee. 
The largest differences between different PDF sets 
appear at the lowest and at the highest $D^0$ energies. 

It is worth noting that $pp$ collisions with $E_{lab} > E_{lab, D^0}$ 
contribute to the $Z$-moment at any given fixed energy $E_{lab, D^0}$. 
Although, in line of principle, $E_{lab}$ can be very large, 
in practice it turns out that the largest contribution comes from values of $E_{lab}$
within the range $E_{lab, D^0}$ $<$ $E_{lab}$ $<$ $(100-1000) \times E_{lab, D^0}$, due to the fact that the 
distribution in $x_E = E_{D^0}/E_{lab}$ is rapidly suppressed for large $x_E$. 
As a consequence, for energies as those probed by IceCube, the
contributions to the $Z$-moments come mainly from regions with a 
center-of-mass energy $\sqrt{S}$ non too high with respect to the
energy range reached and probed so far at the LHC, 
where perturbative QCD in the standard formalism of collinear factorization has been tested to work. 
Any deviations from this formalism which may exist at the highest energies, e.g., 
in the form of non-linear effects (like gluon recombination as opposed to gluon splittings) 
or due to the dominance  of large logarithms $\ln(S/m^2)$ 
subject to resummation on the basis of a different factorization formalism ($k_T$ factorization)~\cite{Ball:2001pq}, 
are, thus, expected to have only a small impact on the $Z$-moments we are interested in for the aim of understanding the IceCube results.
\begin{figure}[ht!]
\begin{center}
\includegraphics[width=0.75\textwidth]{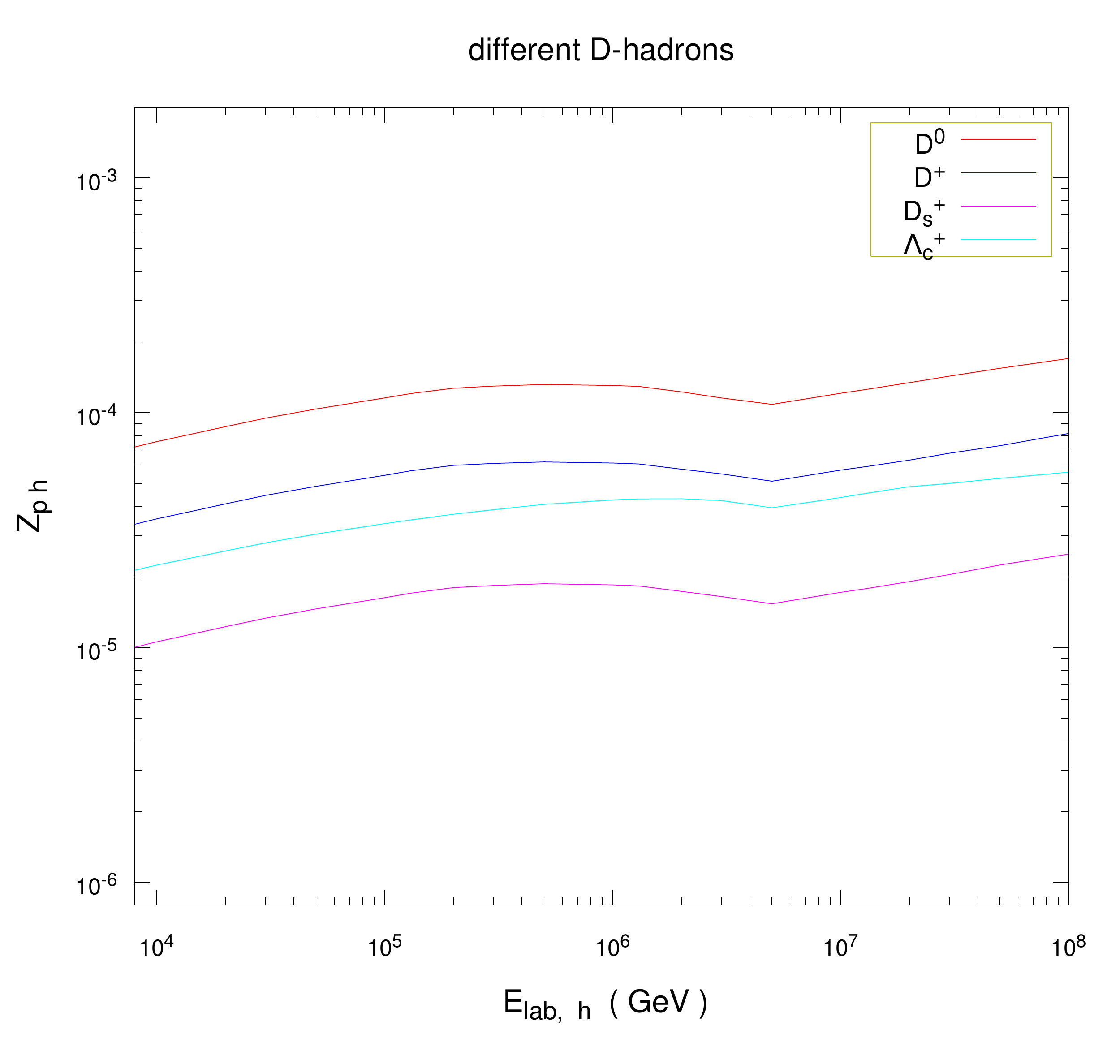}
\caption{\label{fig:zmome} 
  $Z$-moments for the hadroproduction of selected $D$-hadrons containing 
  a $c$~valence quark (electric charge $Q=+2/3$) ($D^0$, $D^+$, $D^+_s$ and $\Lambda_c^+$). Charm mass, ($\mu_R$, $\mu_F$) scales and PDFs were chosen 
as the central values in Figs.~\ref{fig:dsigmadx} and~\ref{fig:dsigmadxpdf}.
  The power-law CR spectrum has been used as input of our calculation.
}
\end{center}
\end{figure}

The relative importance of $Z$-moments of different $D$-hadron species is shown 
in Fig.~\ref{fig:zmome}, where the $Z$-moments of positively charged
$D$-hadrons and $D^0$ are shown. The $D^0$ contribution is the dominant one at
all hadron energies. 
The different shape of the $\Lambda_c^+$ contribution with respect to
those of other hadrons is partly related to the shape of the differential
distribution $d\sigma/dx_E$ of this hadron, which turns out to be less steep at large $x_E$ 
than those of the $D$-meson distributions.

The expressions for $Z_{p\,h}$ enter directly into those for the fluxes eqs.~(\ref{philow}) and (\ref{phihigh}), obtained after solving the system of coupled differential equations descri\-bing the linear development of the hadronic cascade in the atmosphere, under the ap\-proxi\-mations outlined in Sec.~\ref{sec:method}.

\subsubsection{$Z_{h\,l}$, $Z_{p\,p}$ and $Z_{h\,h}$}

In the following we briefly summarize our treatment of the other
$Z$-moments $Z_{h\,l}$, $Z_{p\,p}$ and $Z_{h\,h}$ entering eqs.~(\ref{philow}) and/or~(\ref{phihigh}). 

For $Z_{h\,l}$, our treatment of the semileptonic decay of $D$-hadrons follows   
closely Ref.~\cite{Bhattacharya:2015jpa}. 
Form factors for analytical decay distributions $h \rightarrow \mu \nu_\mu X$ were extracted
from Ref.~\cite{Bugaev:1998bi} and for the decay branching ratios the most recent values reported by the
PDG~\cite{Agashe:2014kda} were taken.

In order to evaluate the proton regeneration $Z$-moment, $Z_{p\,p}$, we have ap\-proxi\-mated
the inelastic $x_E$ distribution for the process $pA \rightarrow pX$ 
by a scaling form
$d\sigma/dx_E \sim \sigma_{pA}^{inel}(E_{lab}) (1-x_E)^n (1+n)$
with $n$ = 0.51, 
as already done in Ref.~\cite{Bhattacharya:2015jpa}.
Here, for $\sigma_{pA}^{inel}$ we have considered the three different models already described in Sec.~\ref{subsec:pair}, which also enter the generation
of the production moments $Z_{p\,h}$. 

Due to the obvious difficulties in measuring $hA$ cross-sections with $h$
being a $D$ or $B$ hadron, caused by the short lifetime of these particles, 
the moments $Z_{h\,h}$ are approximated by considering the available estimates 
for $KA$ cross-sections, on the basis of analogies between $K$ and $D$ mesons. 
Both include quarks belonging to the same flavor family (charm quarks in case of $D$'s  
are replaced by strange quarks in case of $K$'s).
In particular, following Ref.~\cite{Bhattacharya:2015jpa},
the inelastic $x_E$ distribution for the reaction $hA \rightarrow hX$ 
is estimated by the ansatz 
${d\sigma/dx_E} \sim A^{0.75}\, \sigma_{KN}^{inel} (E_{lab}) \,(1 - x_E)^n \,(1 + n)$, 
where $n$~=~1 and $\sigma_{K^\pm N}^{inel}$ is the total inelastic cross-section
for $K^\pm$-nucleon interactions. 
To estimate the latter one, total and elastic cross-sections were extracted
from the latest version of the PDG. 
However, the behavior of the $K^\pm p$ elastic cross-section at high energies 
is uncertain because no data above $E_{lab} \sim$ 300~GeV exist. 
We have thus assumed that the slope of the $K^\pm$$p$ elastic
cross-section at high energies is similar to the one of the $pp$ elastic
cross-section, which was recently constrained at LHC
energies by TOTEM data~\cite{Antchev:2013iaa}. 

The regeneration $Z$-moments $Z_{p\,p}$ and $Z_{h\,h}$ enter the expression for
the atte\-nuation lengths $\Lambda_p$ and $\Lambda_h$, respectively, as defined
below eq.~(\ref{phihigh}).
It is worth noting that estimates of the $Z_{hh}$-moments 
and of the related uncertainties
only affect the high-energy approximated solution of the cascade equations,
eq.~(\ref{phihigh}).

\section{Neutrino fluxes and their uncertainties}
\label{sec:predictions}

The IceCube experiment is looking for a diffuse flux of neutrinos
at high-energies, including both downward and upward going neutrinos,
by trying to establish the nature of the observed events as due to astrophysical
signals or to the atmospheric (conventional + prompt) background. 
In this Section we focus on ($\nu_\mu$ + $\bar{\nu}_\mu$)-fluxes, by taking into
account that the largest contribution to the atmospheric conventional neutrino
flux (to which we will compare our prompt flux) comes from this flavor. 
Predictions for other leptons can be obtained with the same method as well. 
The qualitative/quantitative difference between the results for 
($\nu_\mu$ + $\bar{\nu}_\mu$) fluxes and those for other leptons depends on
the specific decay modes and branching fractions of $D$ hadrons in each
species. 
\begin{figure}[ht!]
\begin{center}
\includegraphics[width=0.75\textwidth]{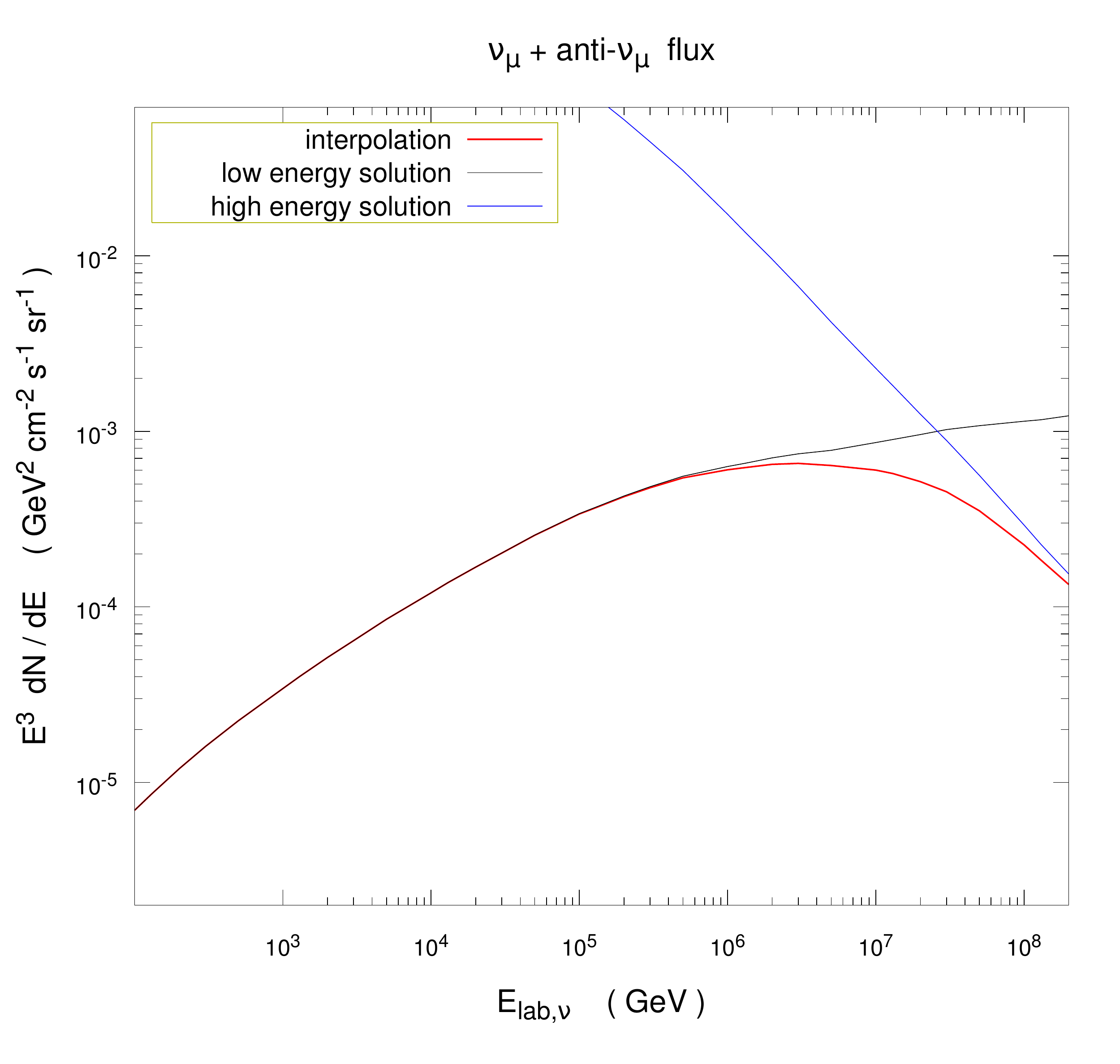}
\end{center}
\caption{\label{fig:interpola} 
  The ($\nu_\mu$ + $\bar{\nu}_\mu$)-flux
  as a function of the neutrino energy $E_{lab,\, \nu}$  
  illustrating the geometric interpolation between low-energy and high-energy solution to the cascade
  equations, in case of a power-law primary cosmic-ray spectrum.}
\end{figure}

The lepton fluxes are derived after evaluating all quantities entering
eqs.~(\ref{philow}) and ~(\ref{phihigh}), already described in previous Sections,
and by interpolating between the high energy and the low energy solutions
according to eq.~(\ref{geometric}). 
An example of the typical behavior of the
two solutions and of their interpolation is shown in Fig.~\ref{fig:interpola}
for the case of a power-law primary CR spectrum as input of the whole calculation.  
In the following we will present the central values of our fluxes, together
with the uncertainty bands arising from the different source of uncertainties, 
both of QCD and of astrophysical origin. 

\subsection{Main uncertainties from QCD and astrophysics}
\label{sec:uncQCD}

Uncertainties on the fluxes whose origin can be ascribed to perturbative QCD mainly
reflect those uncertainties already found in the differential distributions
$d\sigma/dx_E$ and in the $Z$-moments. 
In particular, we discuss in the following the scale, charm mass and PDF variation, 
as well as matching uncertainties, related to the NLO matching to the parton shower. 
\begin{figure}[ht!]
\begin{center}
\includegraphics[width=0.75\textwidth]{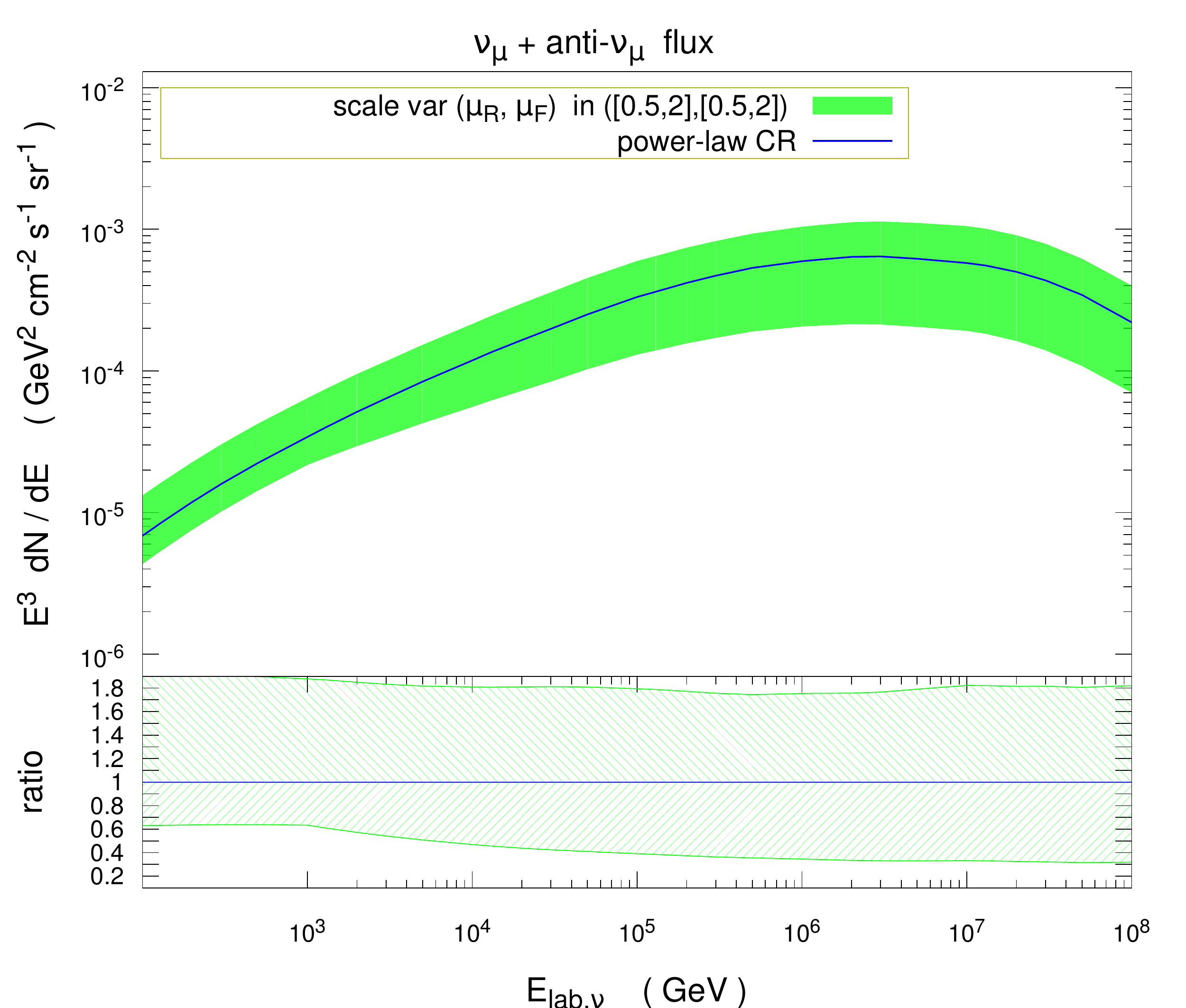}
\end{center}
\caption{\label{fig:scale0}
  The ($\nu_\mu + \bar{\nu}_\mu$)-fluxes 
  as a function of the neutrino energy $E_{lab,\, \nu}$  with 
  uncertainties due to renormalization and factorization scale variation.
  Charm mass, PDF and ($\mu_R$,$\mu_F$) scales
  were fixed as in the left panel of Fig.~\ref{fig:dsigmadx}.} 
\end{figure}
\begin{figure}[ht!]
\begin{center}
\includegraphics[width=0.475\textwidth]{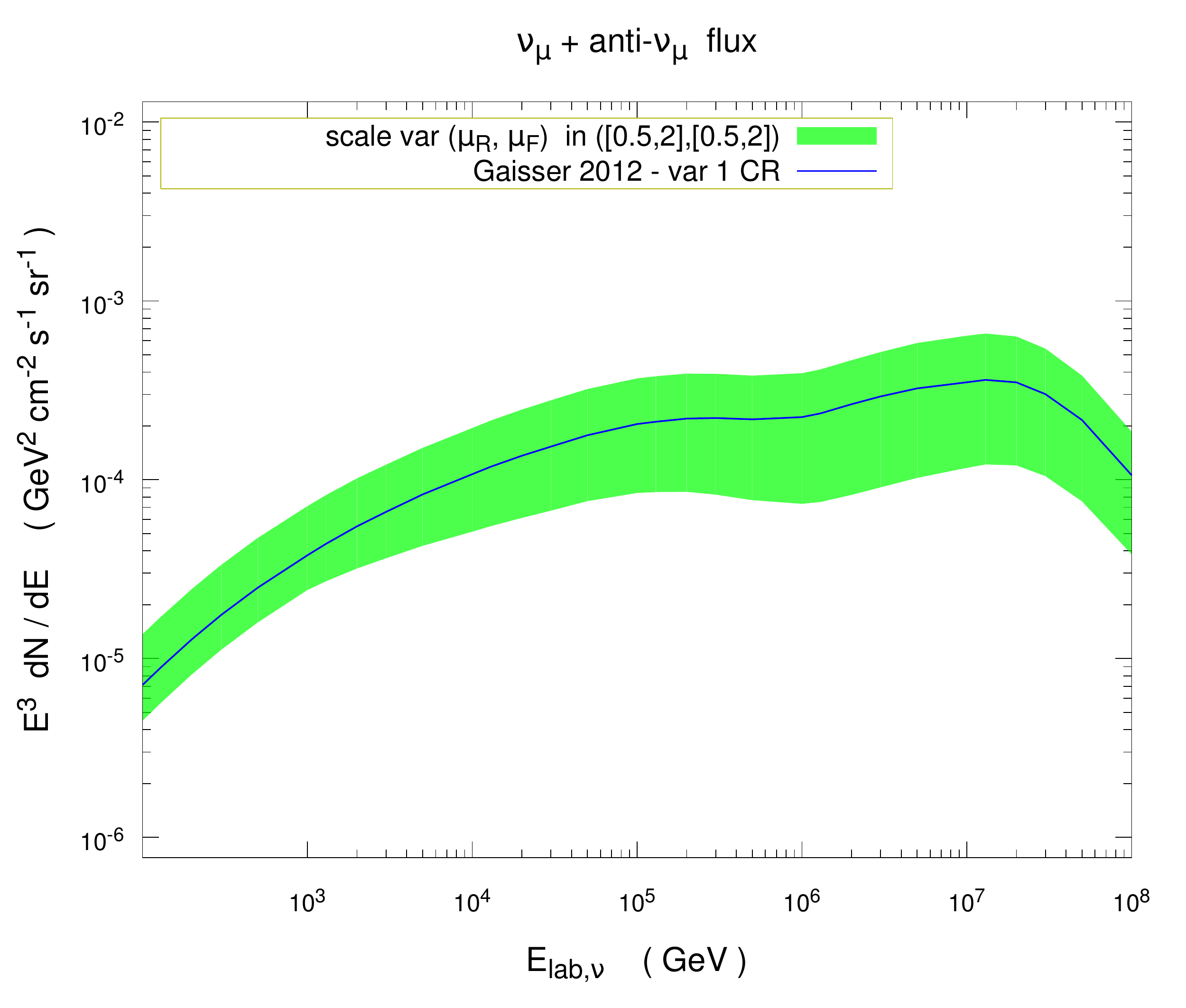}
\includegraphics[width=0.475\textwidth]{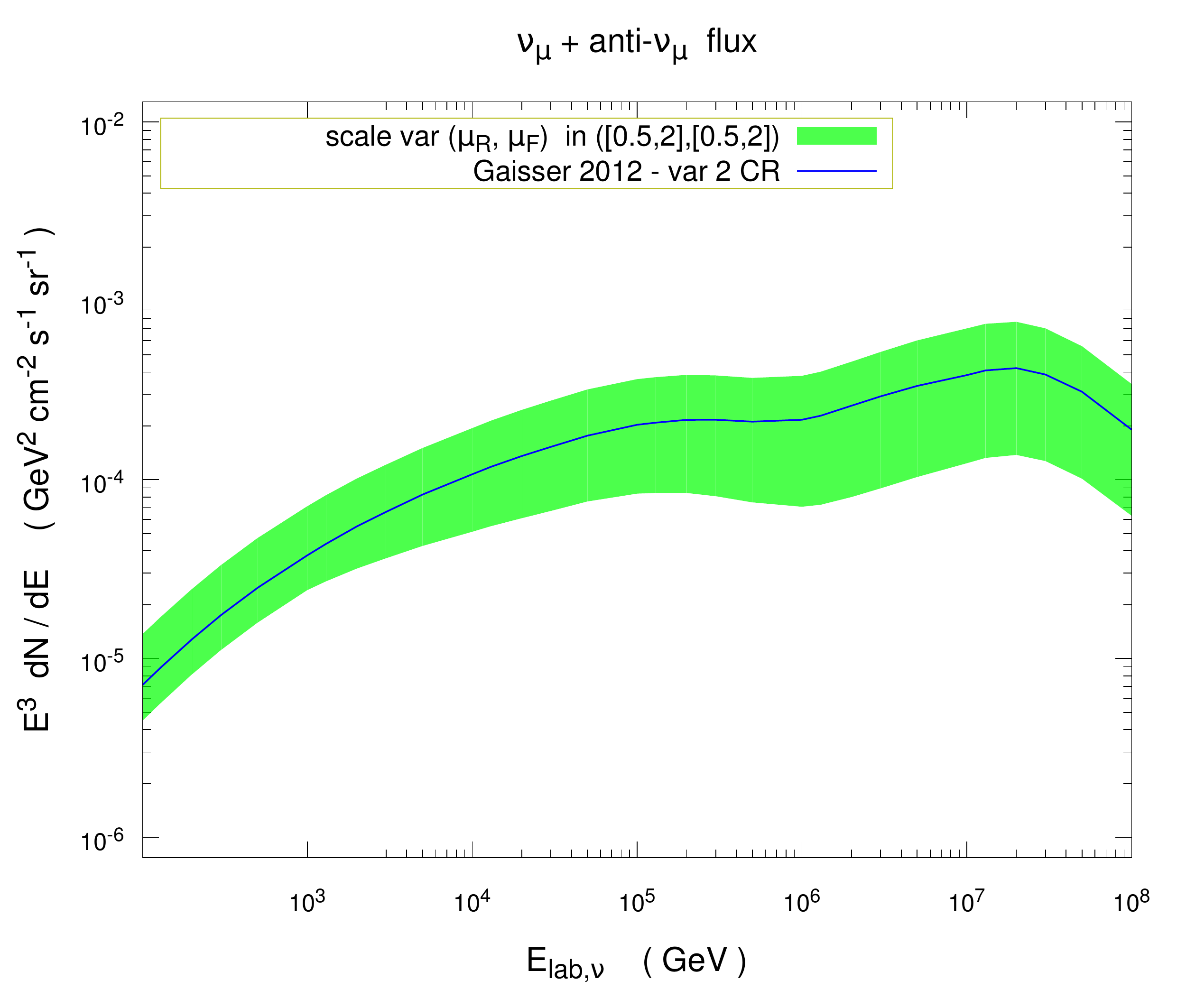}\\
\includegraphics[width=0.475\textwidth]{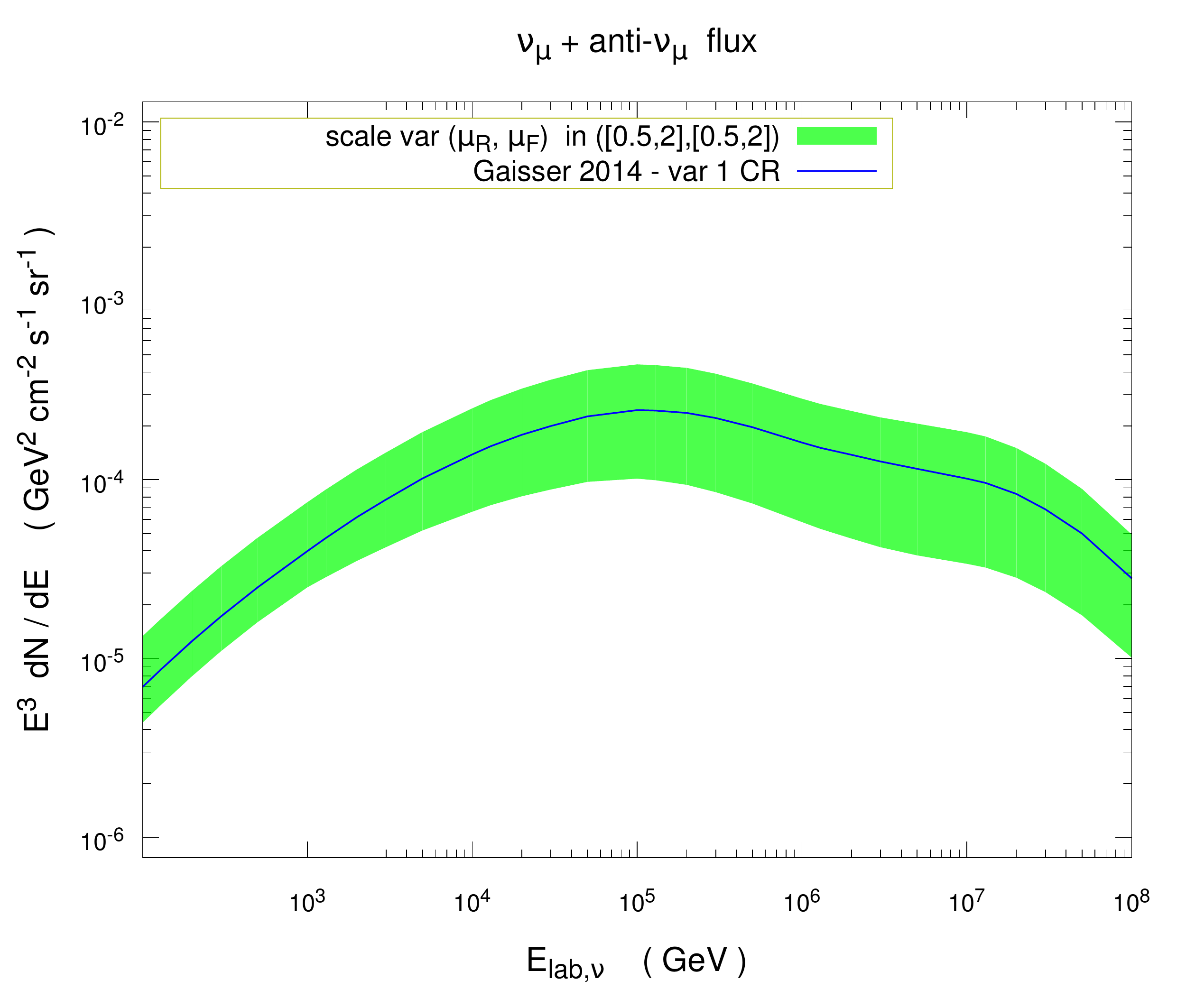}
\includegraphics[width=0.475\textwidth]{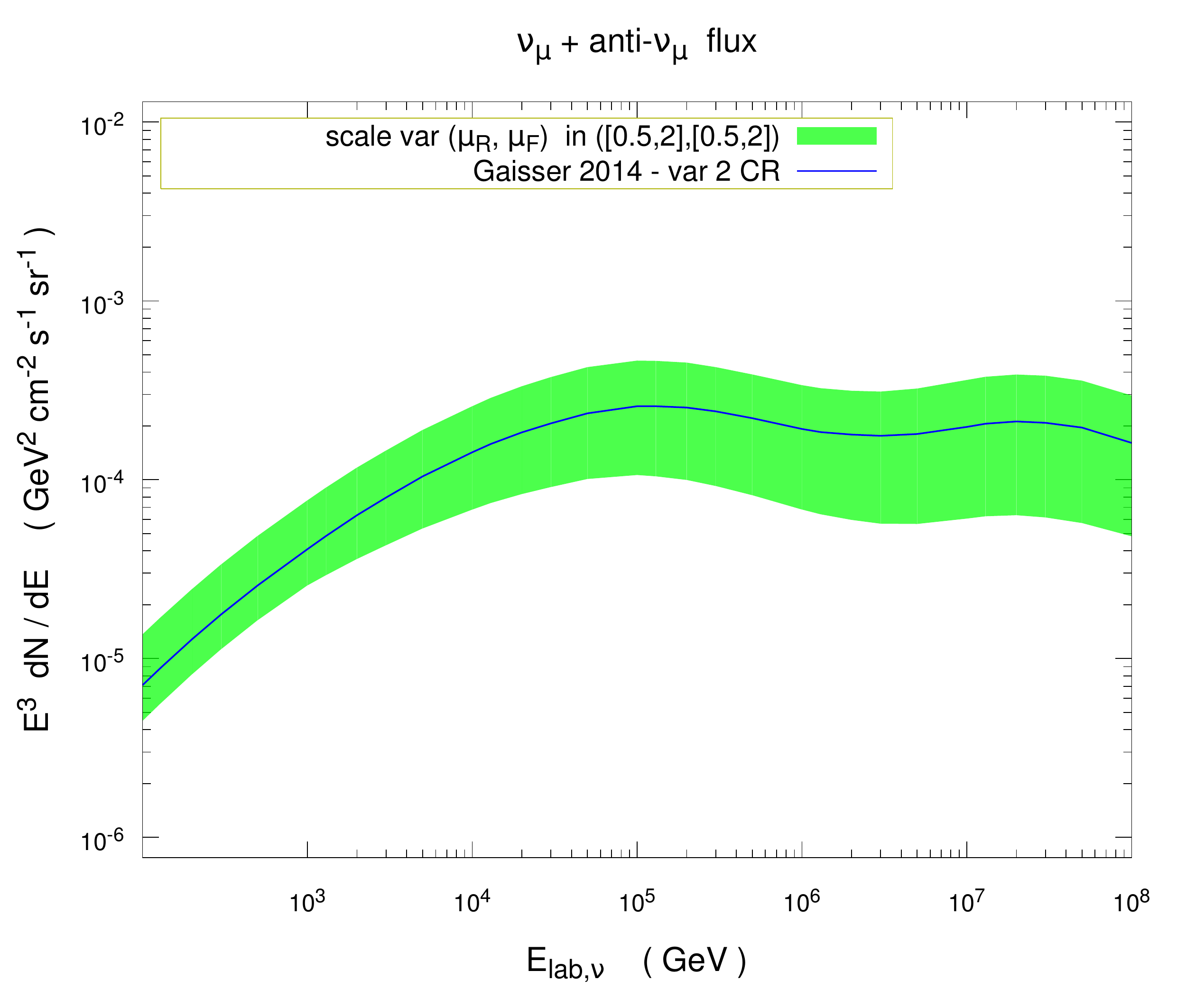}
\end{center}
\caption{\label{fig:scale1234}
Same as in Fig.~\ref{fig:scale0} for different
primary CR spectra, where each panel corresponds to a variant of the 
Gaisser primary spectrum, cf. Sec.~\ref{subsec:cosmic}.
} 
\end{figure}

Uncertainties in the ($\nu_\mu + \bar{\nu}_\mu$)-fluxes due to $\mu_R$ and
$\mu_F$ scale variation for a power-law CR spectrum as input, are reported in Fig.~\ref{fig:scale0}, 
while the corresponding uncertainties for other CR spectra are shown in Fig.~\ref{fig:scale1234}.
The scale variation turns out to be the largest source of uncertainties. 
Including cases with $\mu_R \ne \mu_F$, i.e., the independent variation of $\mu_R$ and $\mu_F$, leads to an uncertainty band
which is almost uniform on a wide interval of energies $E_{lab}$.
In this respect our findings in Figs.~\ref{fig:scale0} and~\ref{fig:scale1234}
are different from the result of Ref.~\cite{Bhattacharya:2015jpa}, 
where the non-diagonal choices with $\mu_R \ne \mu_F$ were neglected 
and the scale uncertainty is underestimated, especially at low energies.  

\begin{figure}[th!]
\begin{center}
\includegraphics[width=0.75\textwidth]{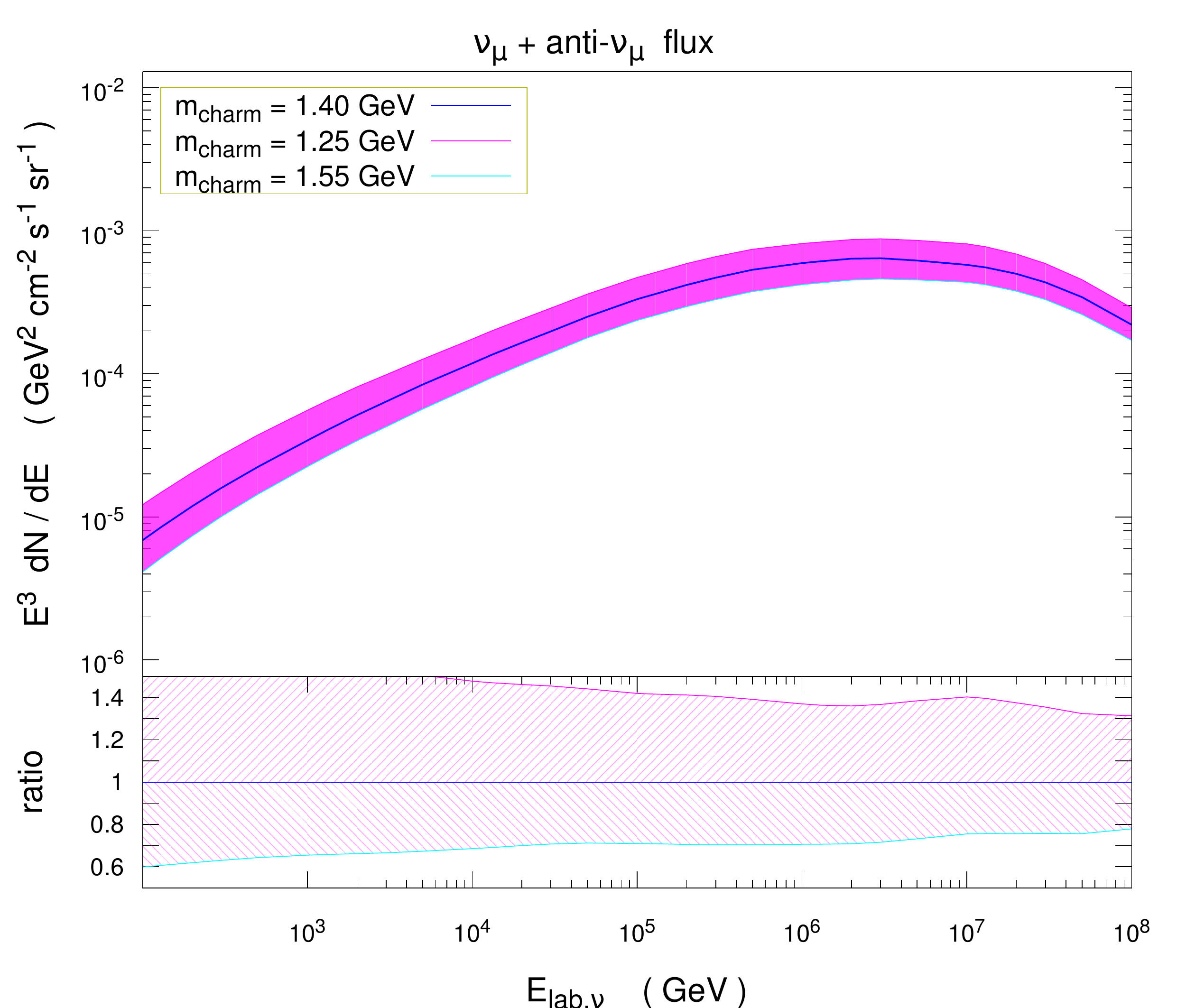}
\end{center}
\caption{\label{fig:mass0}
  The ($\nu_\mu + \bar{\nu}_\mu$)-fluxes 
  as a function of the neutrino energy $E_{lab,\, \nu}$  with 
  uncertainties due to the variation of the pole mass $m_c^{\rm pole}=1.40 \pm 0.15$. PDF, ($\mu_R$,$\mu_F$) scales, charm mass 
  were fixed as in the right panel of Fig.~\ref{fig:dsigmadx}.
} 
\end{figure}

\begin{figure}[ht!]
\begin{center}
\includegraphics[width=0.475\textwidth]{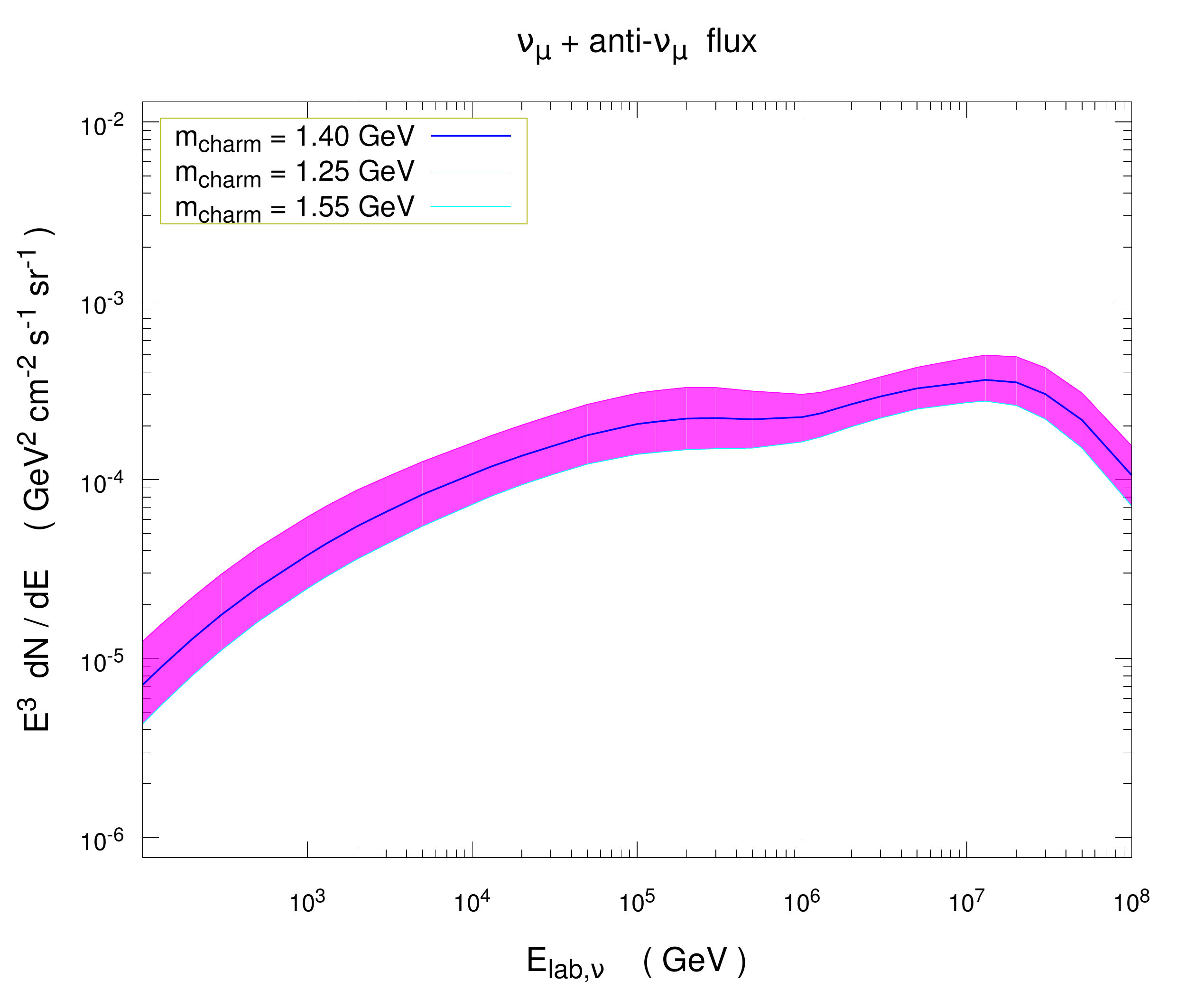}
\includegraphics[width=0.475\textwidth]{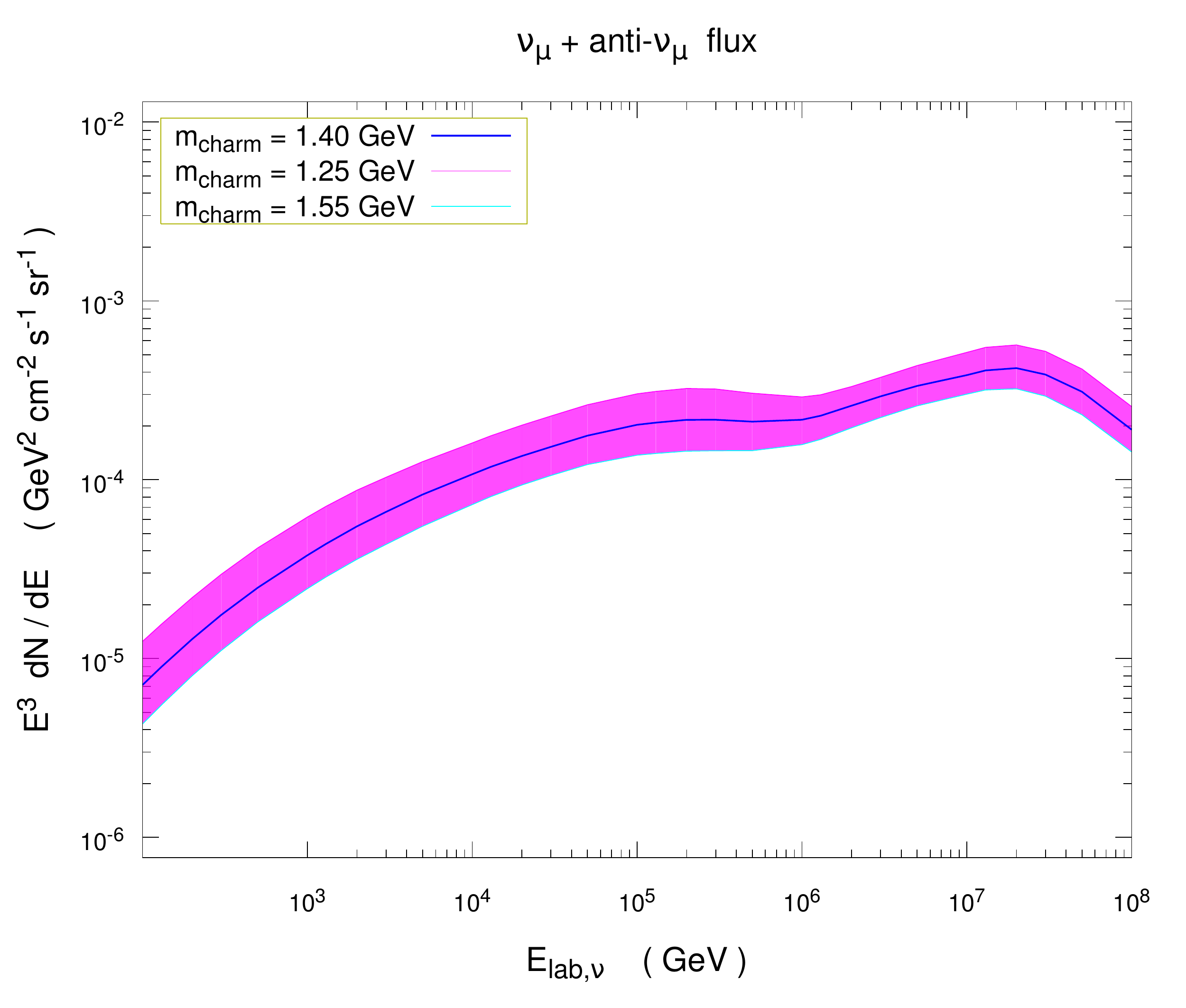}\\
\includegraphics[width=0.475\textwidth]{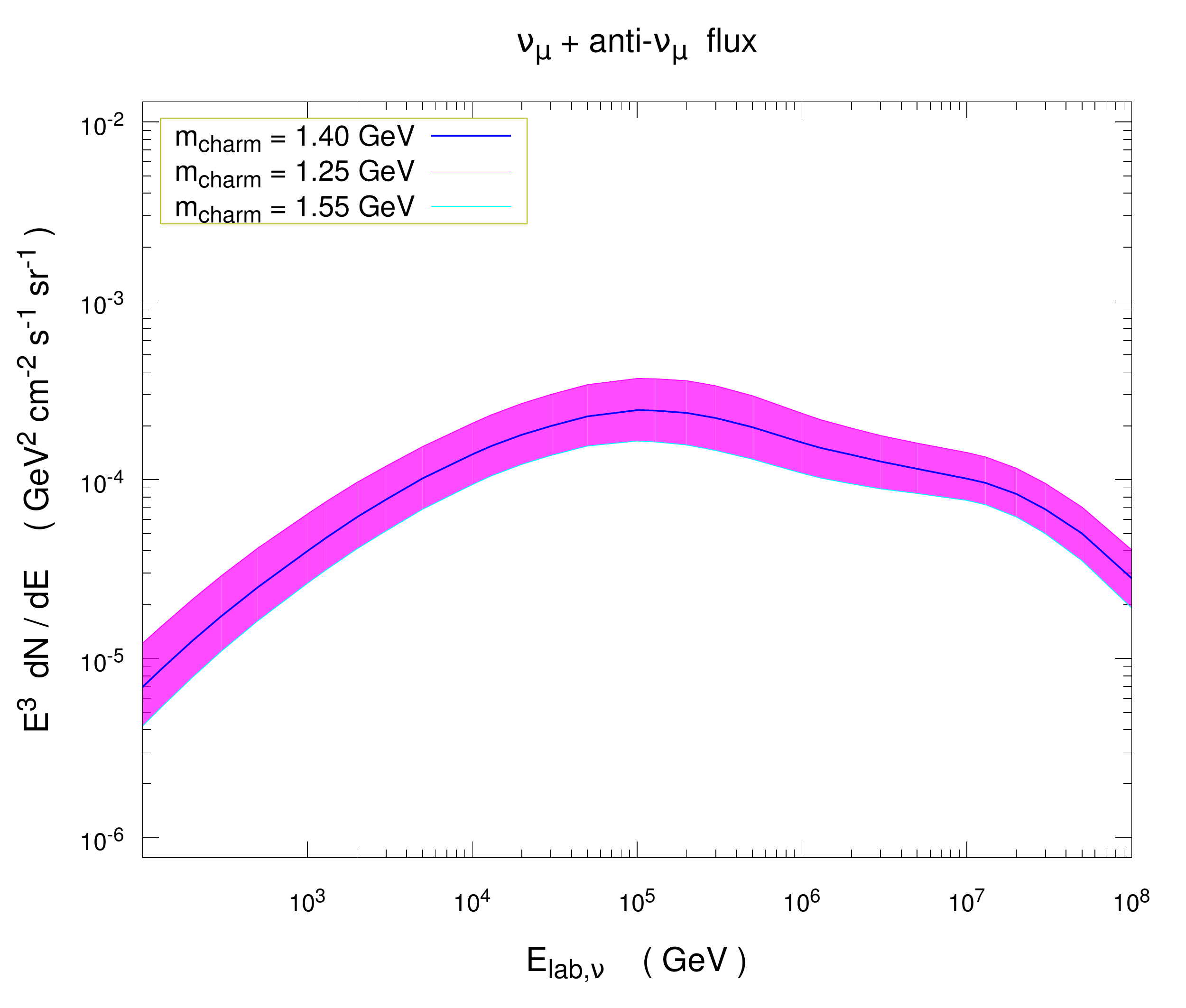}
\includegraphics[width=0.475\textwidth]{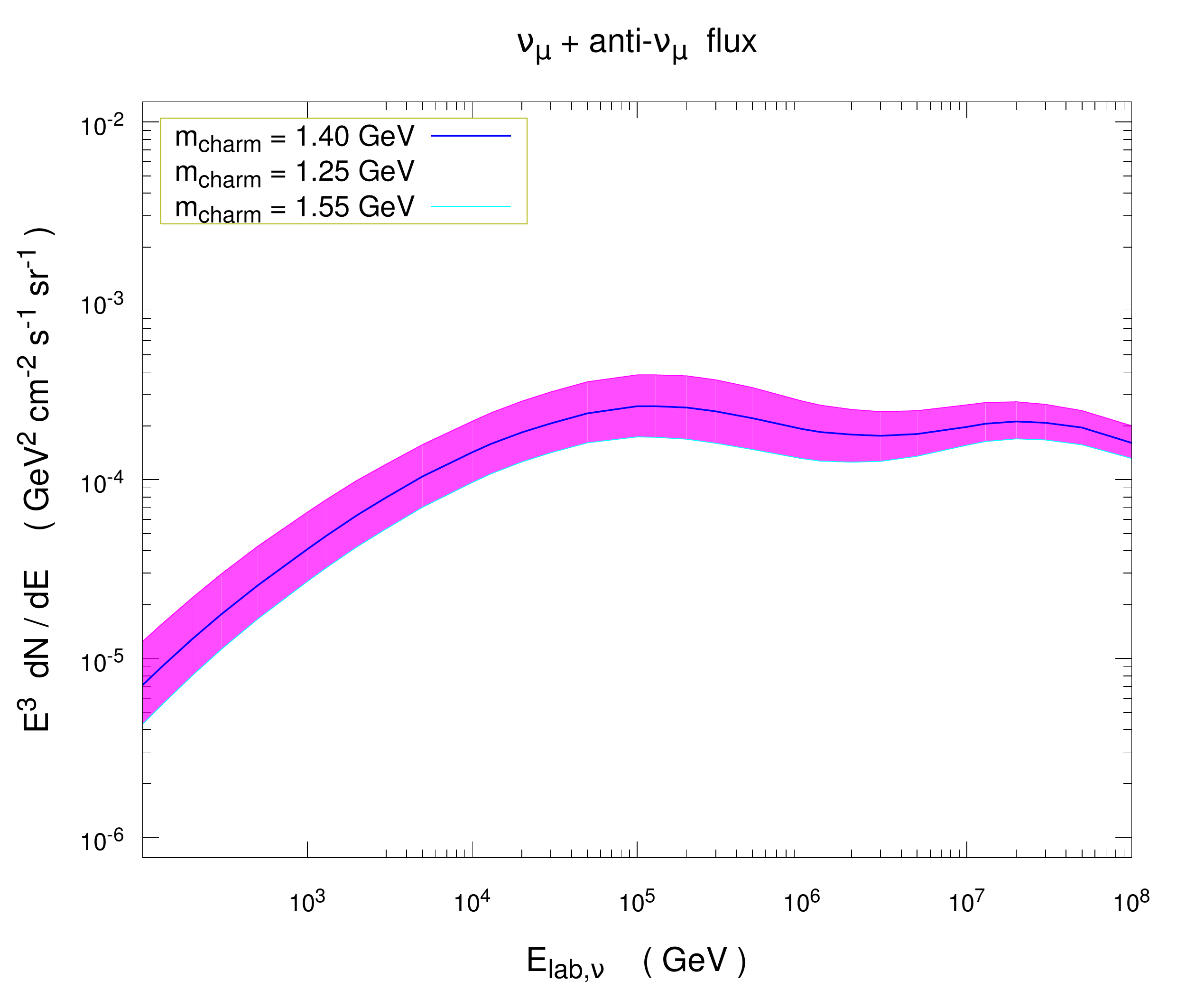}
\end{center}
\caption{\label{fig:mass1234}
Same as in Fig.~\ref{fig:mass0} for different
primary CR spectra, where each panel corresponds to a variant of the 
Gaisser primary spectrum, cf. Sec.~\ref{subsec:cosmic}.
}
\end{figure}

Uncertainties arising from the variation of the charm mass $m_c$ 
within the range motivated in Sec.~\ref{subsec:charm} 
are illustrated in Fig.~\ref{fig:mass0} for a power-law CR spectrum and  
in Fig.~\ref{fig:mass1234} for other CR spectra.
The mass variation turns out to be the second largest source of QCD uncertainties, 
with an uncertainty band slightly decreasing for increasing laboratory energies $E_{lab}$.

\begin{figure}[ht!]
\begin{center}
\includegraphics[width=0.70\textwidth]{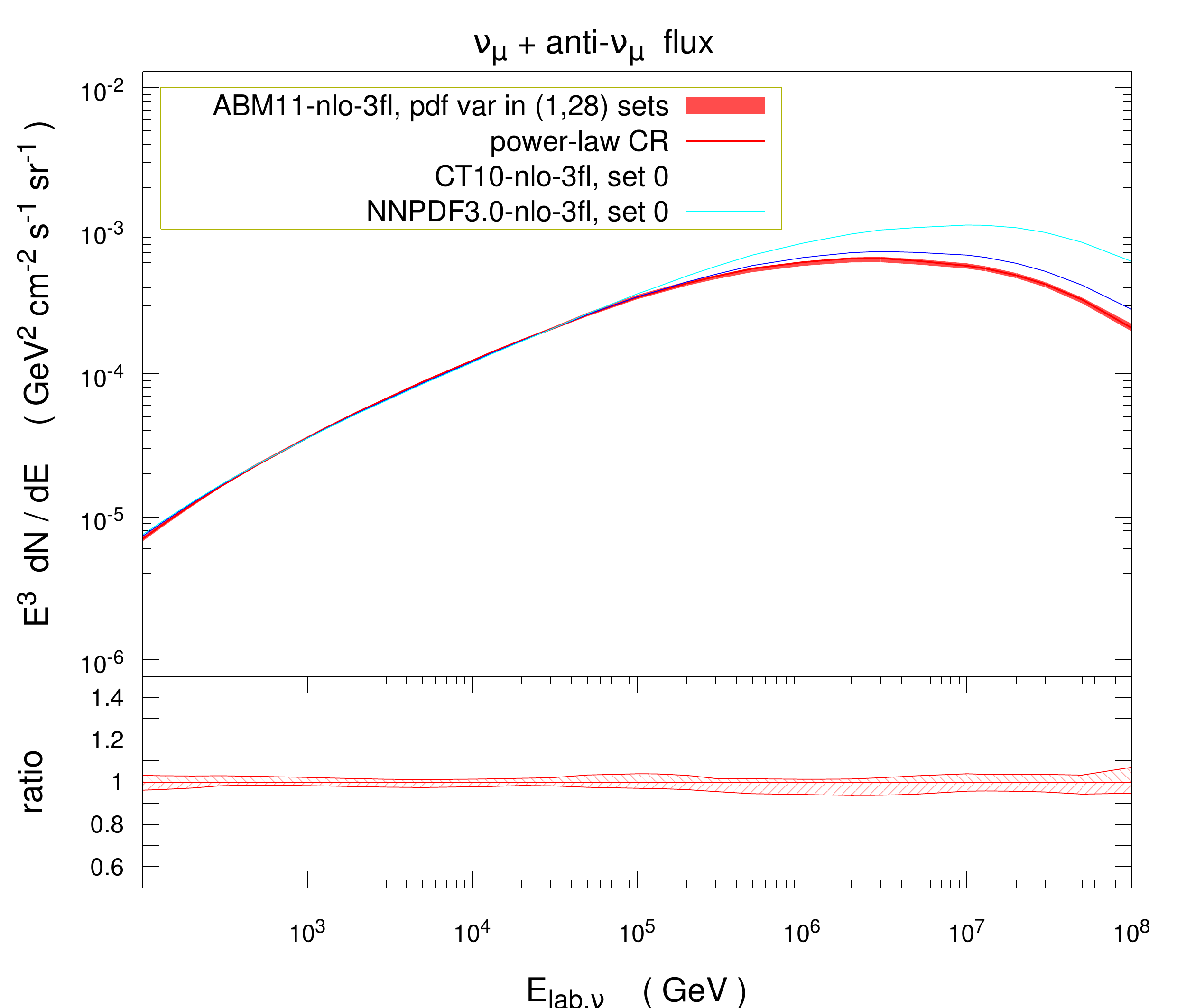}
\end{center}
\caption{\label{fig:pdf0} 
  The ($\nu_\mu + \bar{\nu}_\mu$)-fluxes 
  as a function of the neutrino energy $E_{lab,\, \nu}$ with 
  uncertainties due to PDF variation in the 3-flavor ABM11 PDF set at NLO (red
  band) and predictions for the central set of the
  3-flavor NNPDF3.0 (light-blue line)
  and CT10 (solid blue line) PDFs at NLO, respectively. 
  Charm mass and ($\mu_R$,$\mu_F$) scales 
  were fixed as in Fig.~\ref{fig:dsigmadxpdf}.
  The power-law cosmic ray flux has been used as input in the calculation of $Z$-moments.} 
\end{figure}

\begin{figure}[ht!]
\begin{center}
\includegraphics[width=0.475\textwidth]{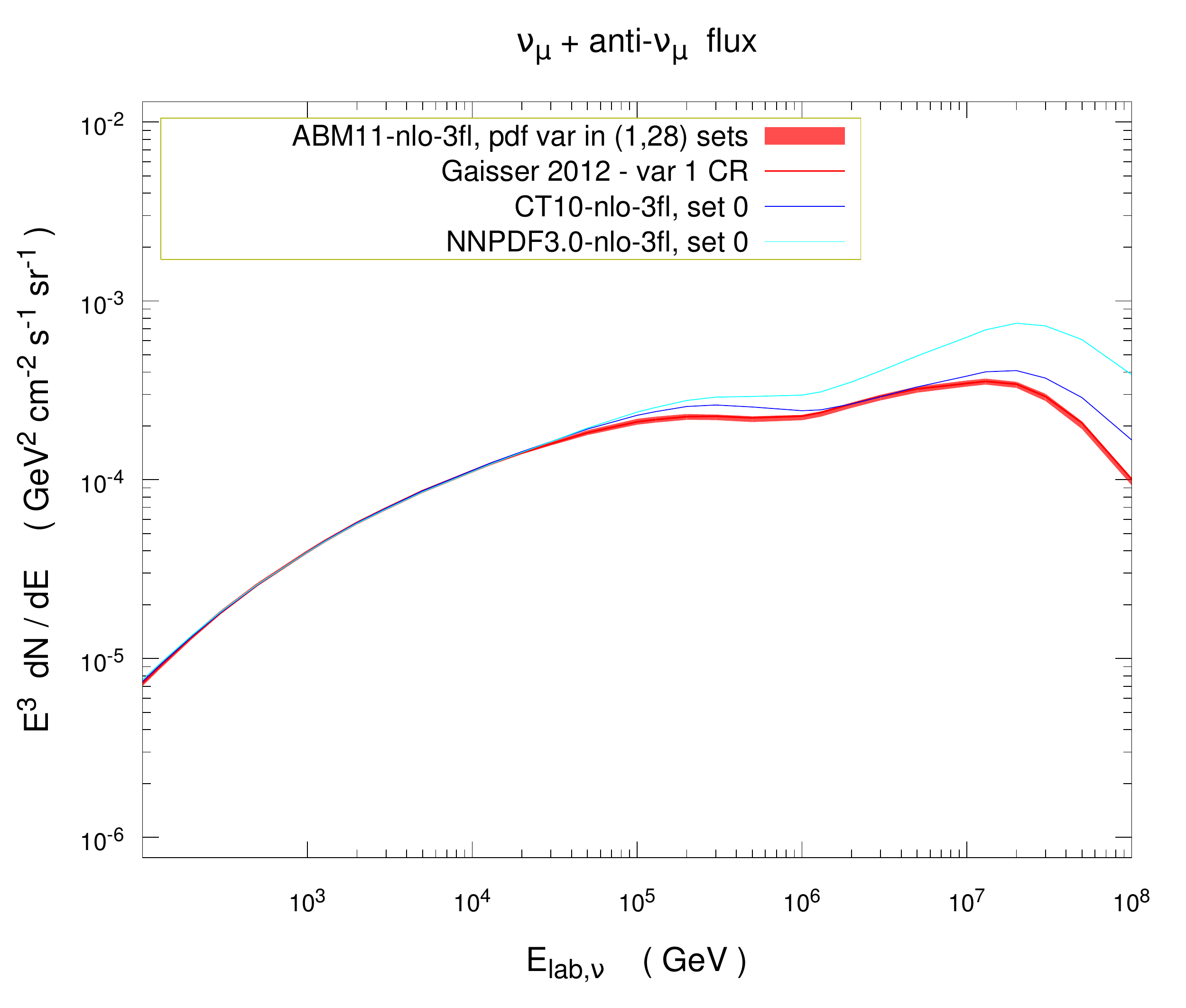}
\includegraphics[width=0.475\textwidth]{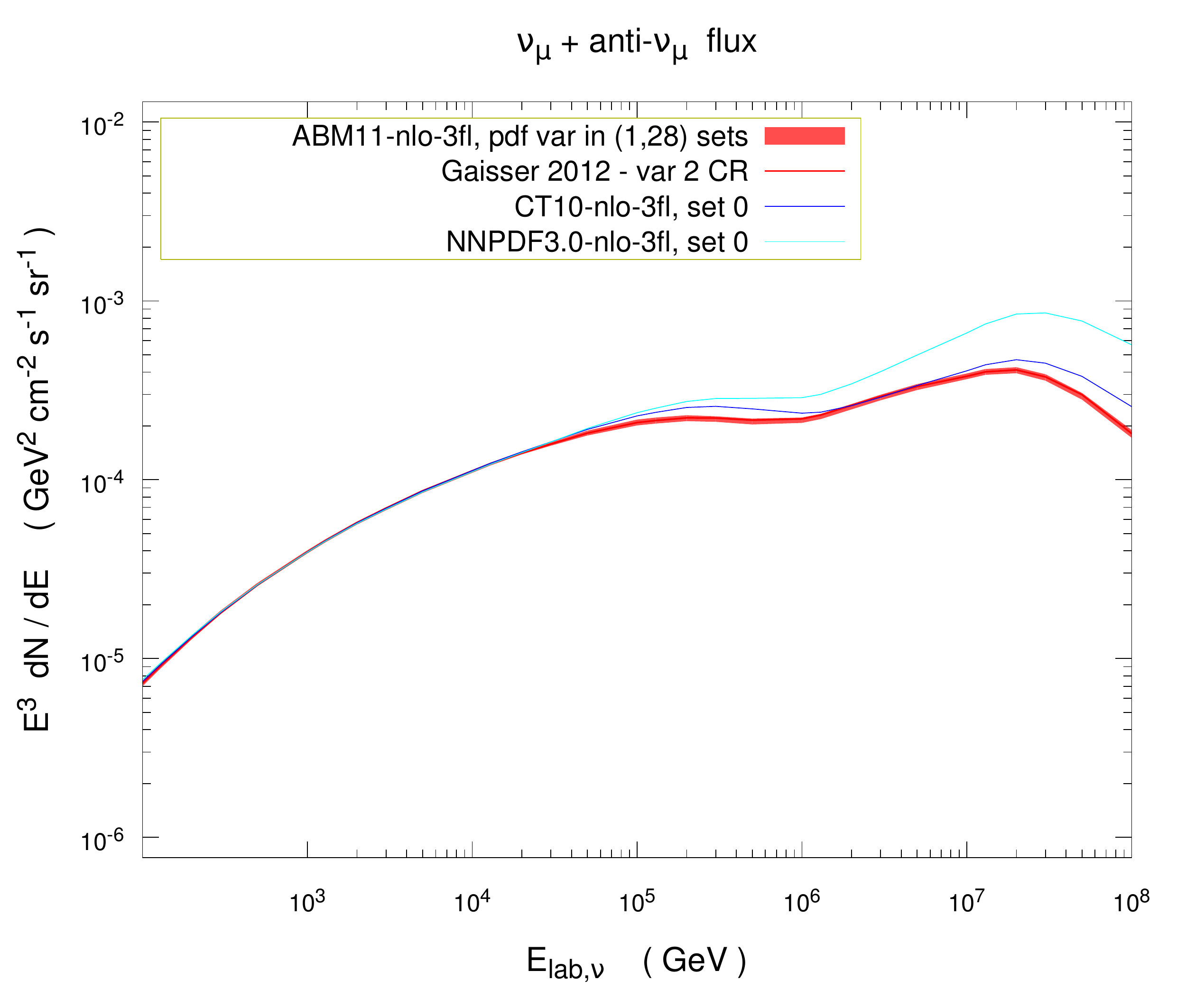}\\
\includegraphics[width=0.475\textwidth]{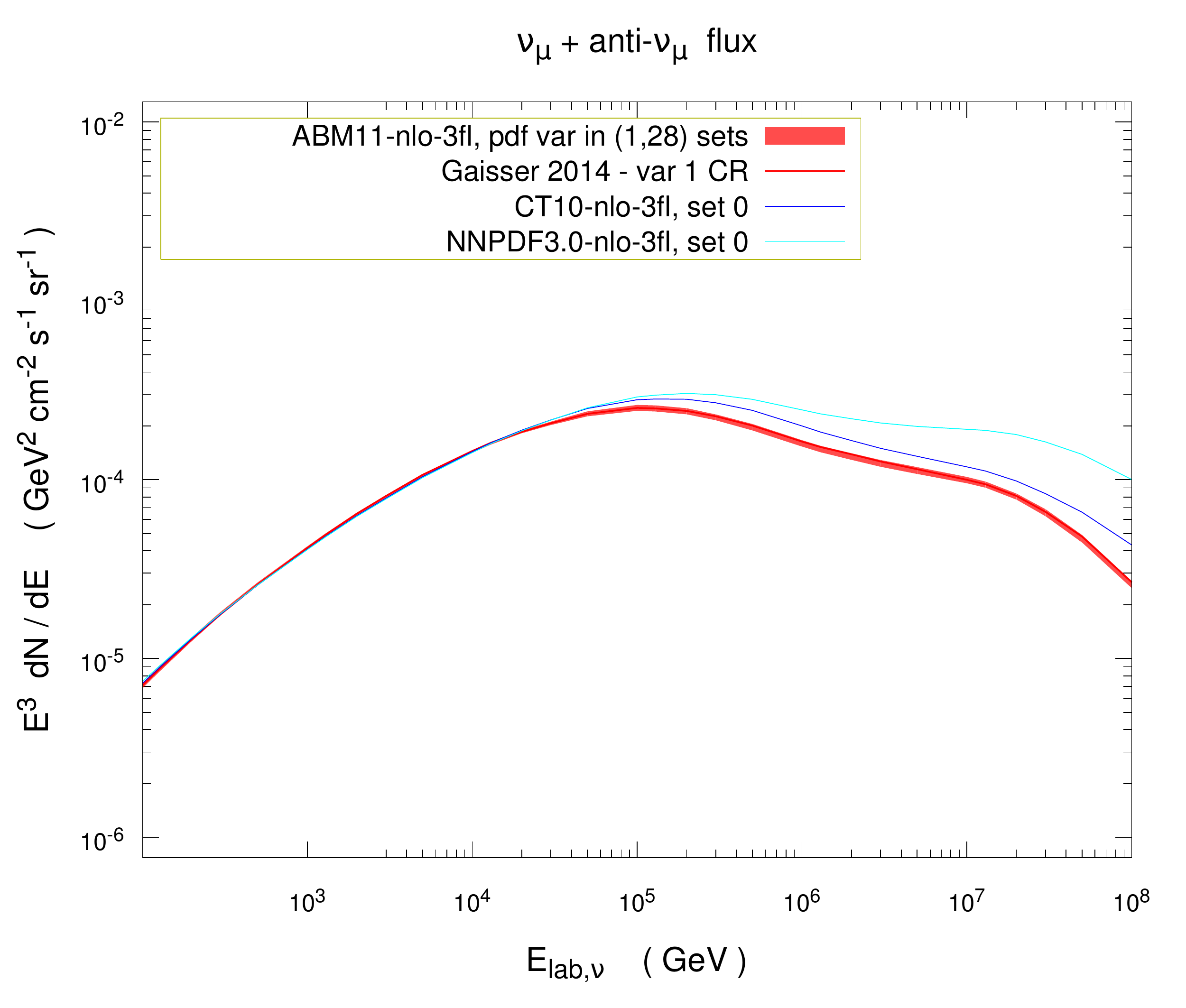}
\includegraphics[width=0.475\textwidth]{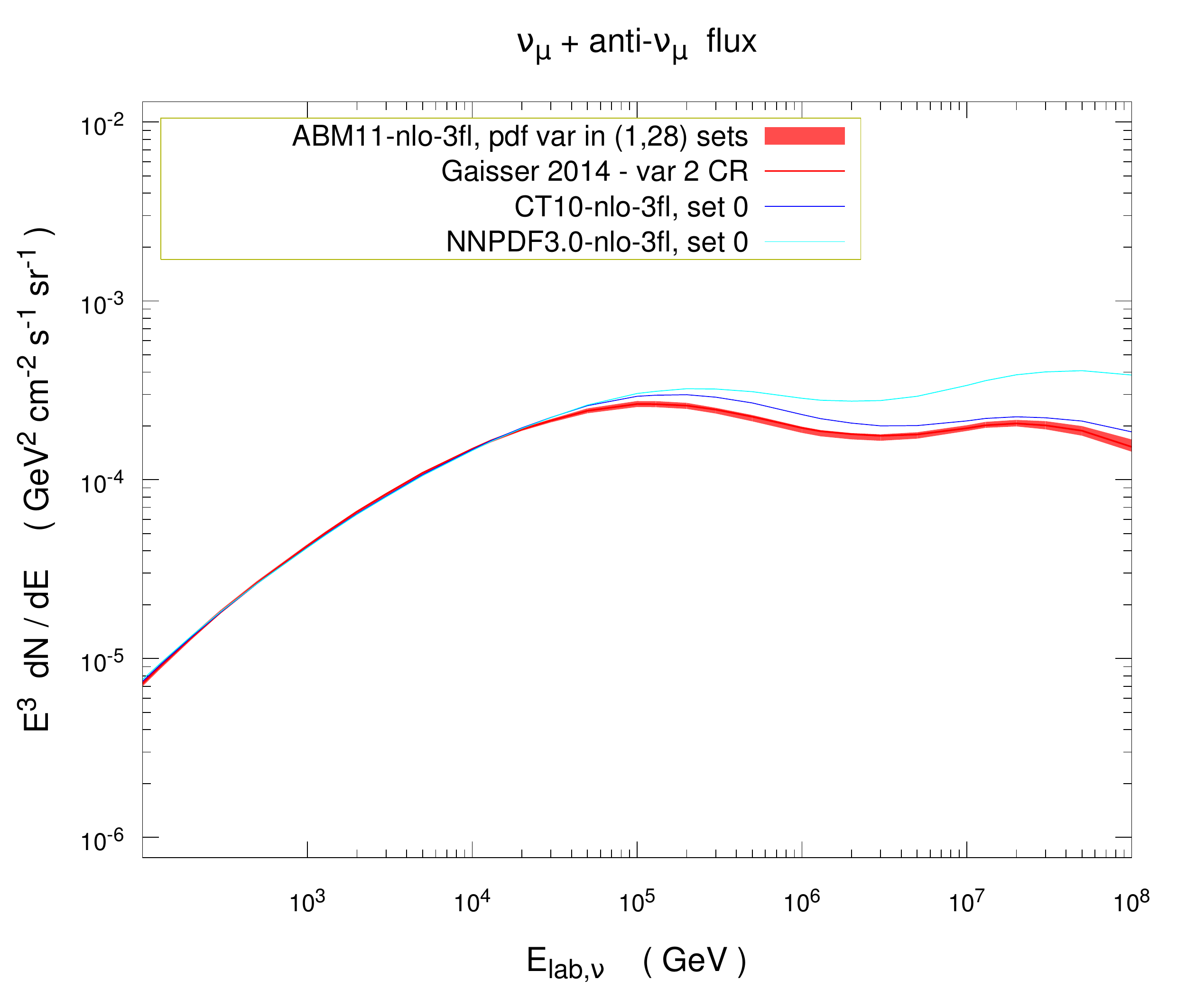}
\end{center}
\caption{\label{fig:pdf1234} Same as in Fig.~\ref{fig:pdf0} for different
primary CR spectra, where each panel corresponds to a variant of the 
Gaisser primary spectrum, cf. Sec.~\ref{subsec:cosmic}.
}
\end{figure}

Uncertainties in the neutrino fluxes related to the PDF variation are displayed 
in Figs.~\ref{fig:pdf0} and ~\ref{fig:pdf1234}
in case of a power-law primary CR spectrum and for the different variants of
Gaisser spectra, respectively.
As already discussed for the case of the $Z_{p\,h}$-moments, the difference of 
the predictions with the ABM11 set (3-flavor FFNS) 
and the central value of other PDF families 
(CT10 and NNPDF3.0 at NLO with $n_f = 3$) turn out to be larger 
than those coming from the 28 sets in the ABM11 fit 
for the combined PDF and $\alpha_s$ uncertainty. 
While the neutrino fluxes from the different PDF fits look quite consistent among each other
for energies in the interval $10^2 < E_{lab,\, \nu} < 4 \cdot 10^4$~GeV, 
visible differences between the fluxes from different PDF families start to appear at higher energies. In this region, 
the predictions based on the ABM11 PDFs are the smallest ones, 
which is related to differences in the shape of the gluon PDF, 
the nominal values of the strong coupling $\alpha_s(M_Z)$ at NLO being largely the same among the ABM11, CT10 and NNPDF3.0 sets used in this work. 
\begin{figure}[ht!]
\begin{center}
\includegraphics[width=0.70\textwidth]{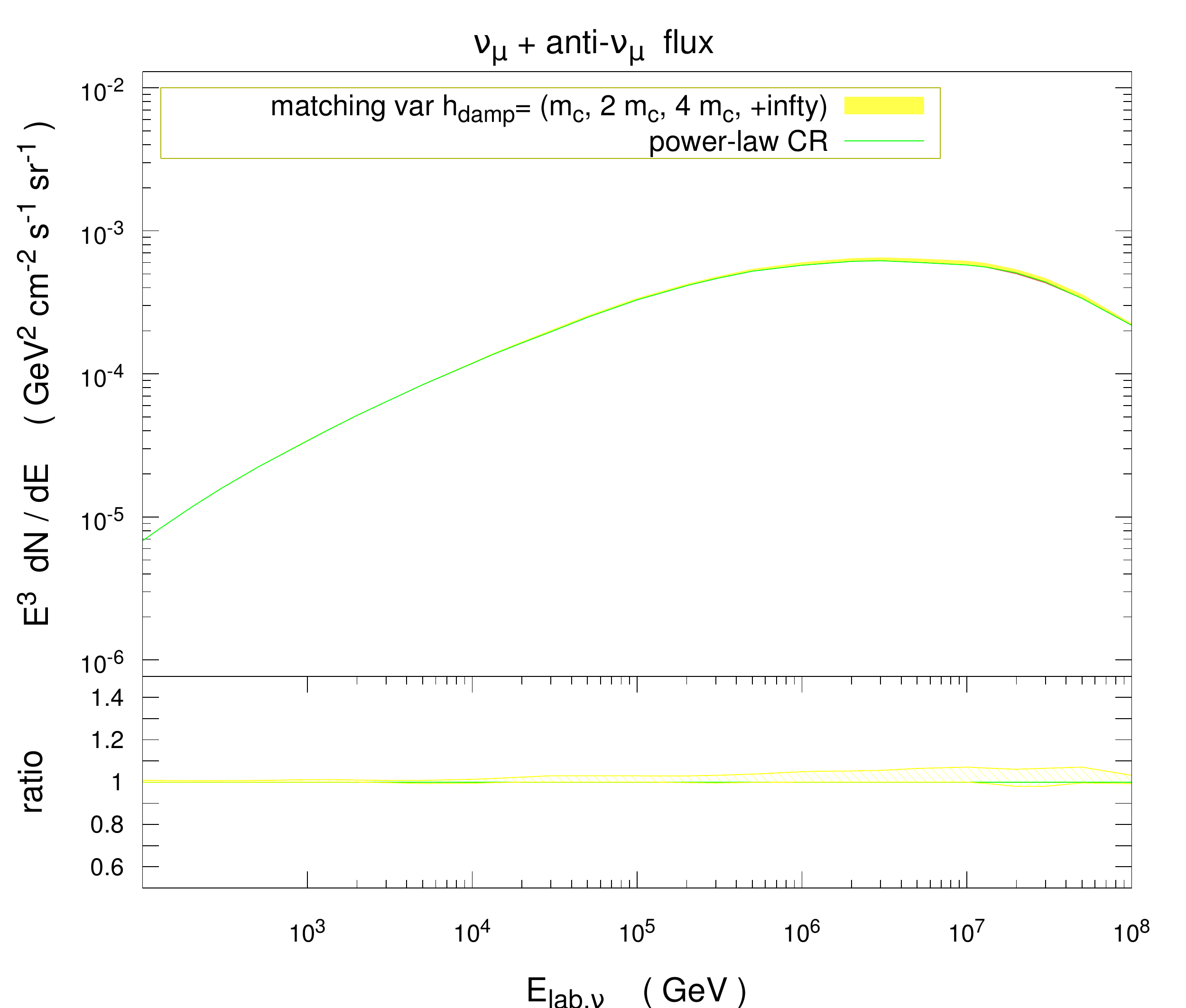}
\end{center}
\caption{\label{fig:match0} 
  The ($\nu_\mu + \bar{\nu}_\mu$)-fluxes 
  as a function of the neutrino energy $E_{lab,\, \nu}$ with 
  the uncertainties from the NLO + PS matching estimated
  in the {\texttt{POWHEG-BOX}} framework through $h_{damp}$ variation.
  See text for more detail. The power-law cosmic ray flux has been used as input in the calculation of $Z$-moments.} 
\end{figure}
\begin{figure}[ht!]
\begin{center}
\includegraphics[width=0.475\textwidth]{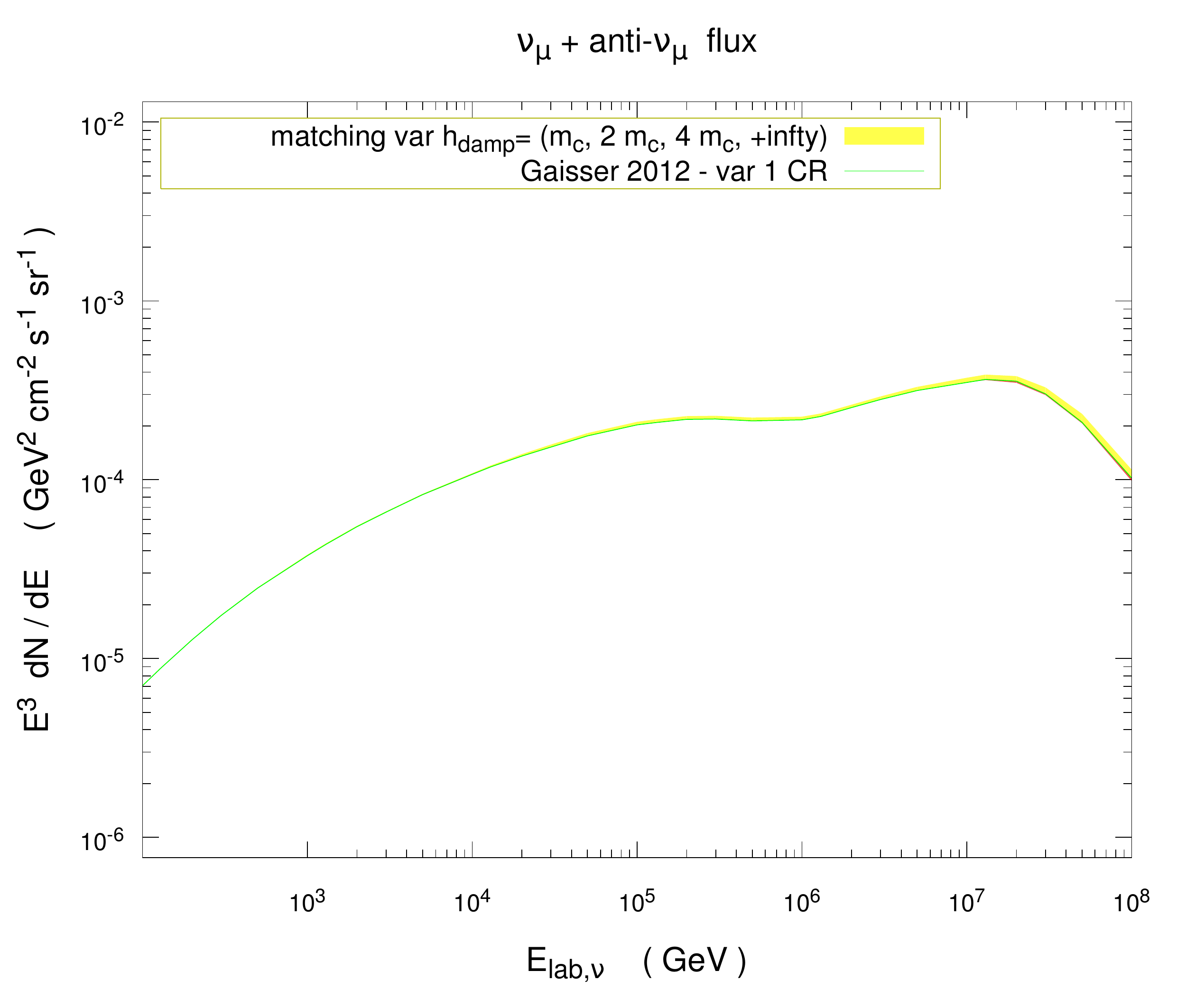}
\includegraphics[width=0.475\textwidth]{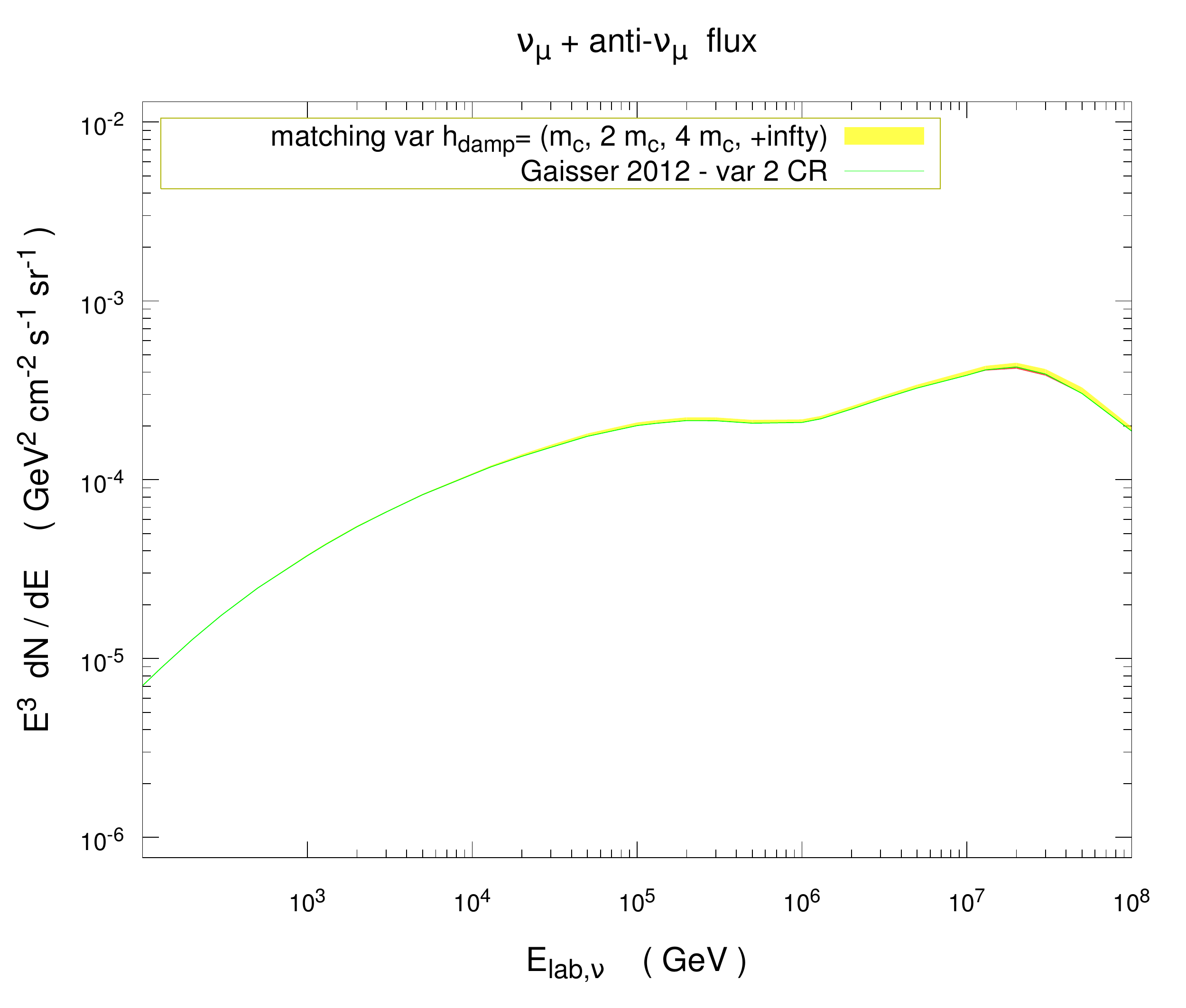}\\
\includegraphics[width=0.475\textwidth]{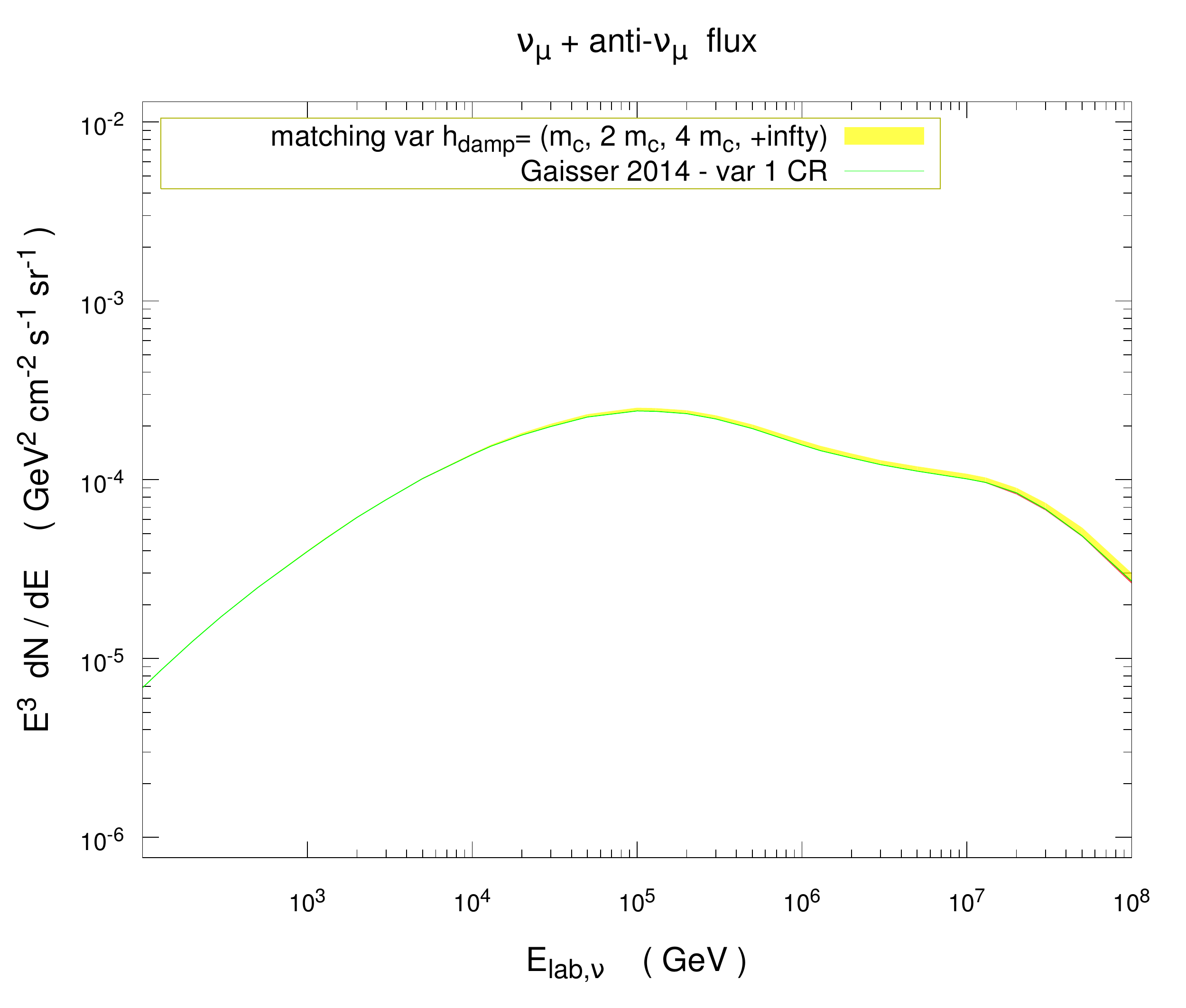}
\includegraphics[width=0.475\textwidth]{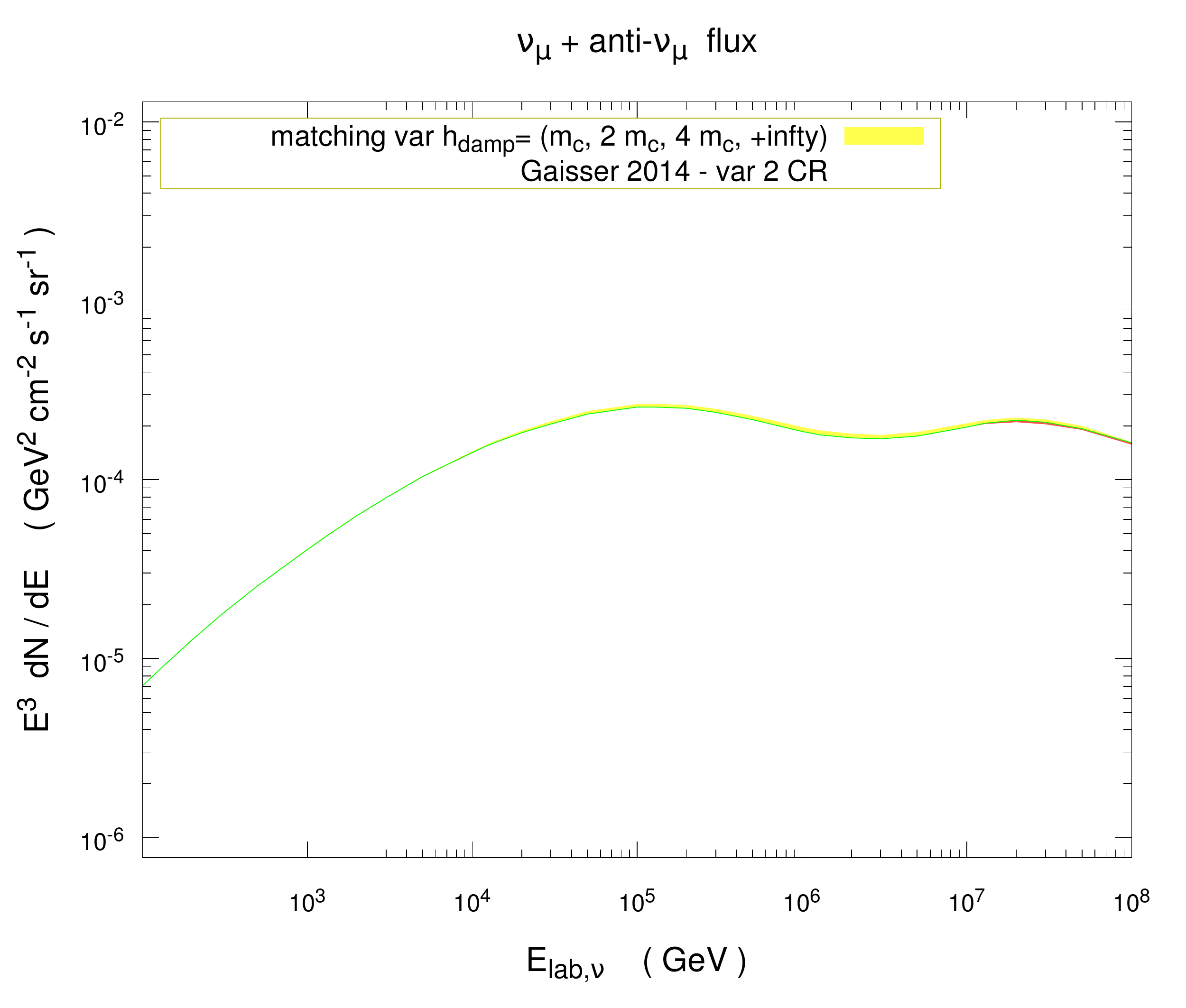}
\end{center}
\caption{\label{fig:match1234} Same as in Fig.~\ref{fig:match0} for different
primary CR spectra, where each panel corresponds to a variant of the 
Gaisser primary spectrum, cf. Sec.~\ref{subsec:cosmic}.
}
\end{figure}

Finally, we provide a first estimate of the uncertainties in the 
NLO matching to the parton shower (NLO~+~PS) by varying the $h_{damp}$ value in the {\texttt{POWHEG-BOX}},
which parameterizes the freedom in 
choosing the form of the separation of the NLO real contribution $R$ into a
singular piece plus a piece damped in the singular region and thus treatable as a finite remainder~\cite{Alioli:2008tz}, $R$
= $R_s$ + $R_f$, with $R_s$~=~$R~\,~h_{damp}^2~/~(h_{damp}^2~+~p_T^2)$ and $R_f$~=~$R~\,~p_T^2~/~(h_{damp}^2 + p_T^2)$~\cite{Dittmaier:2012vm}. 
Only $R_s$ enters the exponent of the Sudakov form factor and 
$h_{damp} = +\infty$ corresponds to the default choice 
in {\texttt{POWHEG-BOX}} so that $R = R_s$, whereas the limit $h_{damp} \rightarrow$ 0 allows to decrease the amount of radiation that is exponentiated and to
recover the $\alpha_s^3$ dependence (pure NLO) in the high-$p_T$ limit. 
In this work we use variations of $h_{damp}$ in the interval \{$m_c$, $2 m_c$, $4 m_c$, $+\infty$\}. 
This is inspired by similar choices performed in experimental studies of
$t\bar{t}$ hadroproduction, see, e.g., the ATLAS note~\cite{ATL-PHYS-PUB-2015-002}. 
The uncertainty is estimated as the envelope of the
predictions corresponding to the different choices above, as shown in
Fig.~\ref{fig:match0}  in case of a power-law primary CR spectrum 
and in Fig.~\ref{fig:match1234} in case of the Gaisser spectra. 
Although several discussions are still on-going about the most meaningful way
of providing NLO~+~PS matching uncertainties which is why we consider our result as a first rough estimate, 
we would like to point out that the uncertainty we got is quite small (less than 10\%)
with respect to other uncertainties of QCD origin. 
This is related to the fact that the key quantities in perturbative QCD to compute 
$Z$-moments, the differential cross sections $d\sigma/d x_E$, are integrated 
over the entire range of transverse momenta $p_T$. 
We thus believe that this conclusion is quite robust, i.e., it does not depend on very specific
details of the way the matching uncertainty is estimated.  
\begin{figure}
\begin{center}
\includegraphics[width=0.475\textwidth]{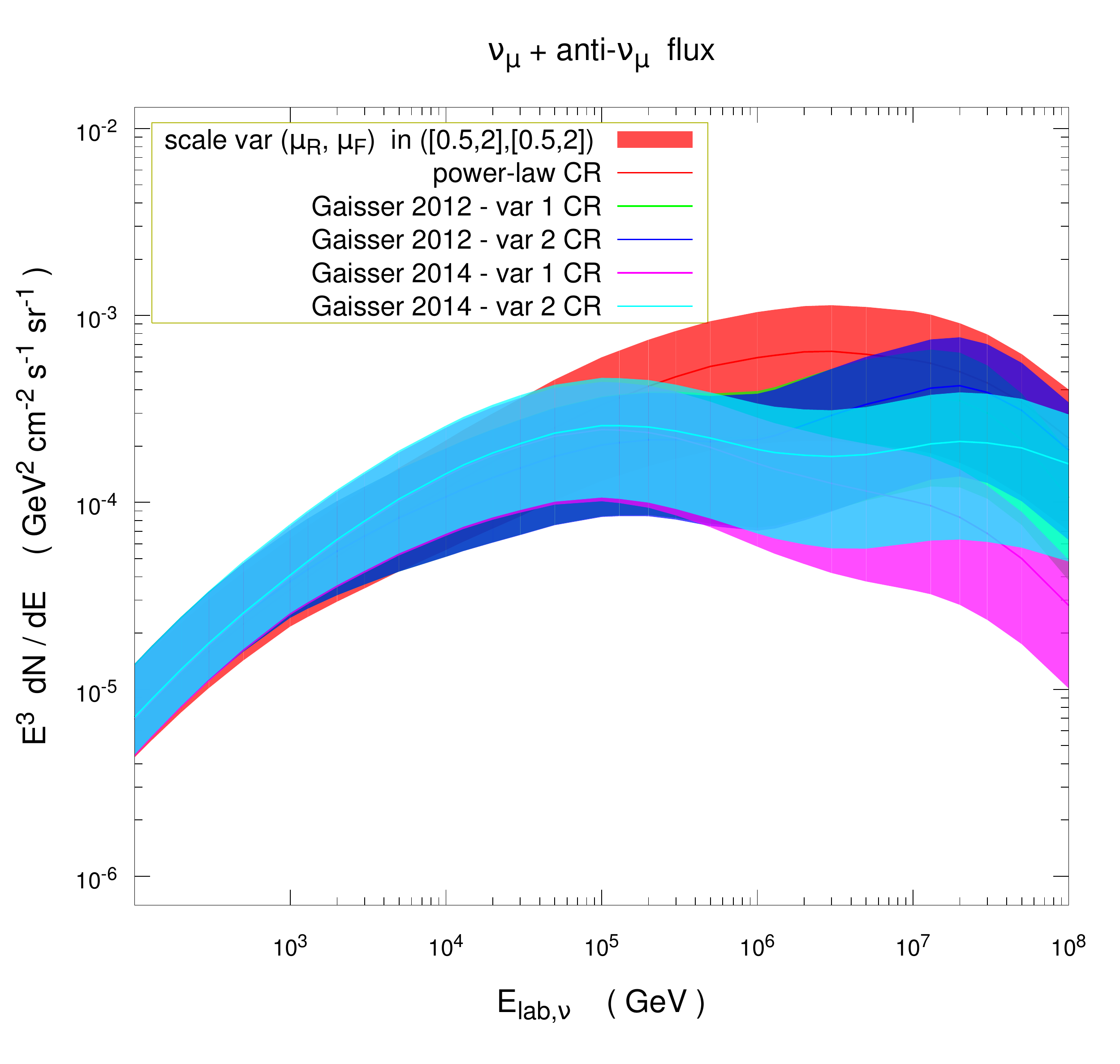}
\includegraphics[width=0.475\textwidth]{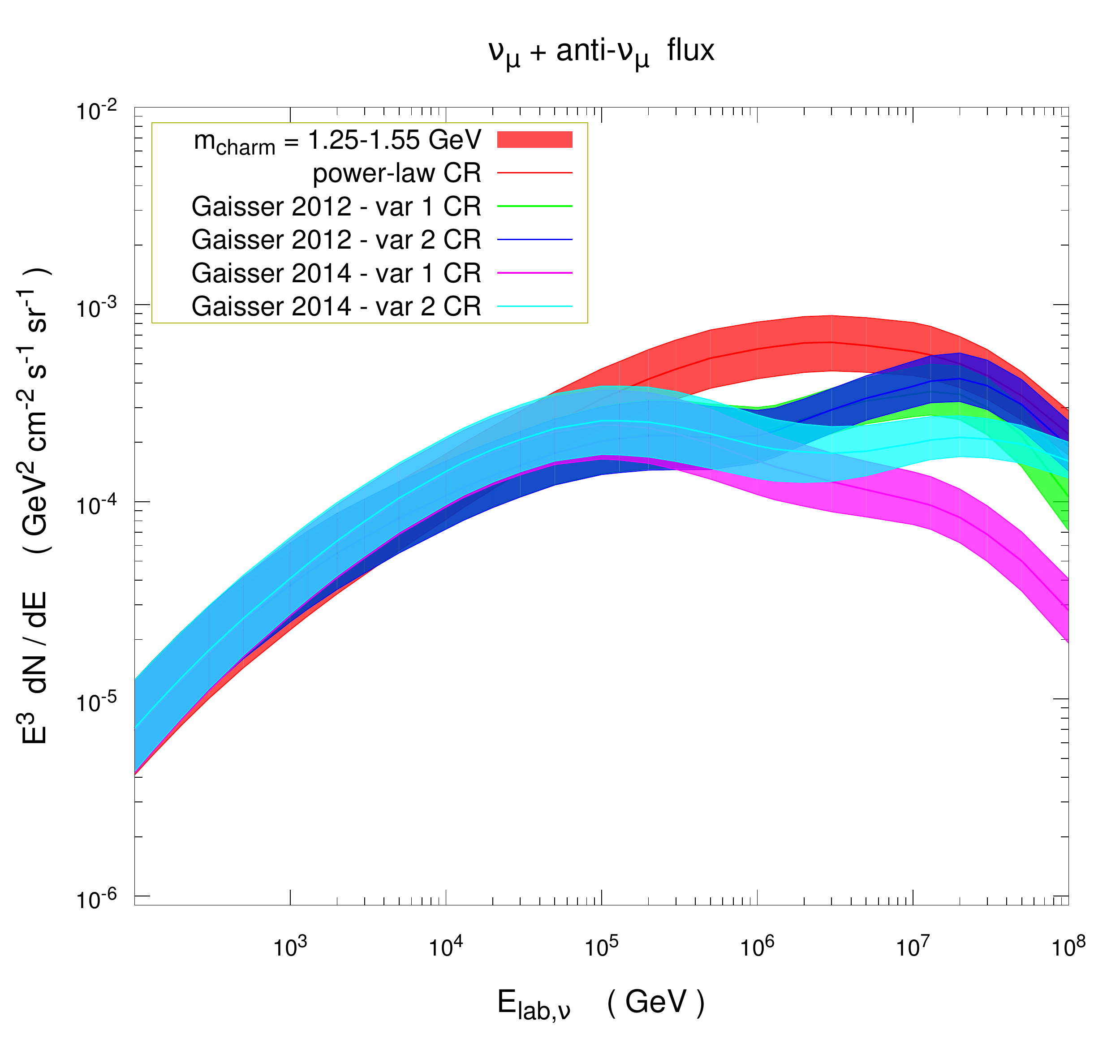}\\
\includegraphics[width=0.475\textwidth]{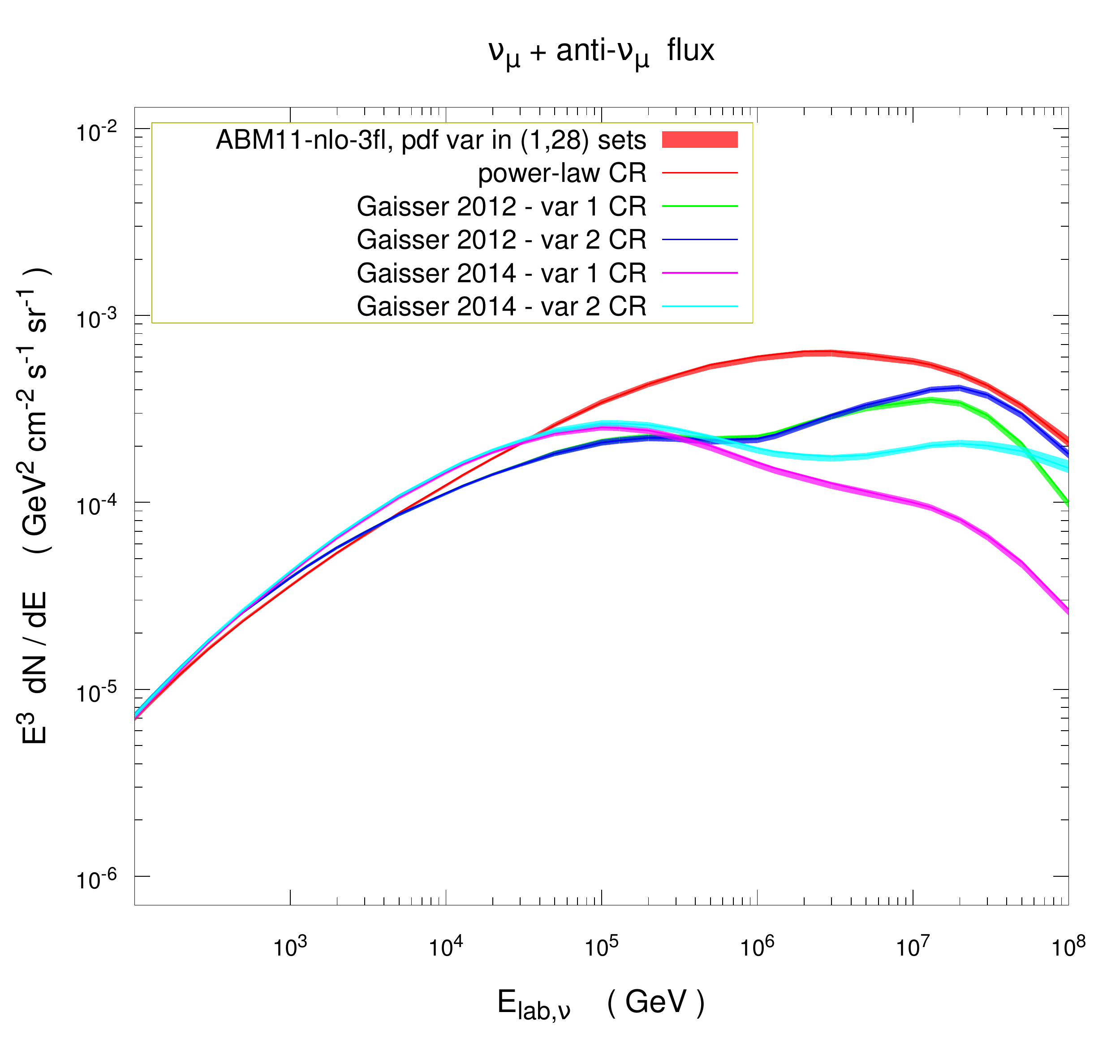}
\includegraphics[width=0.475\textwidth]{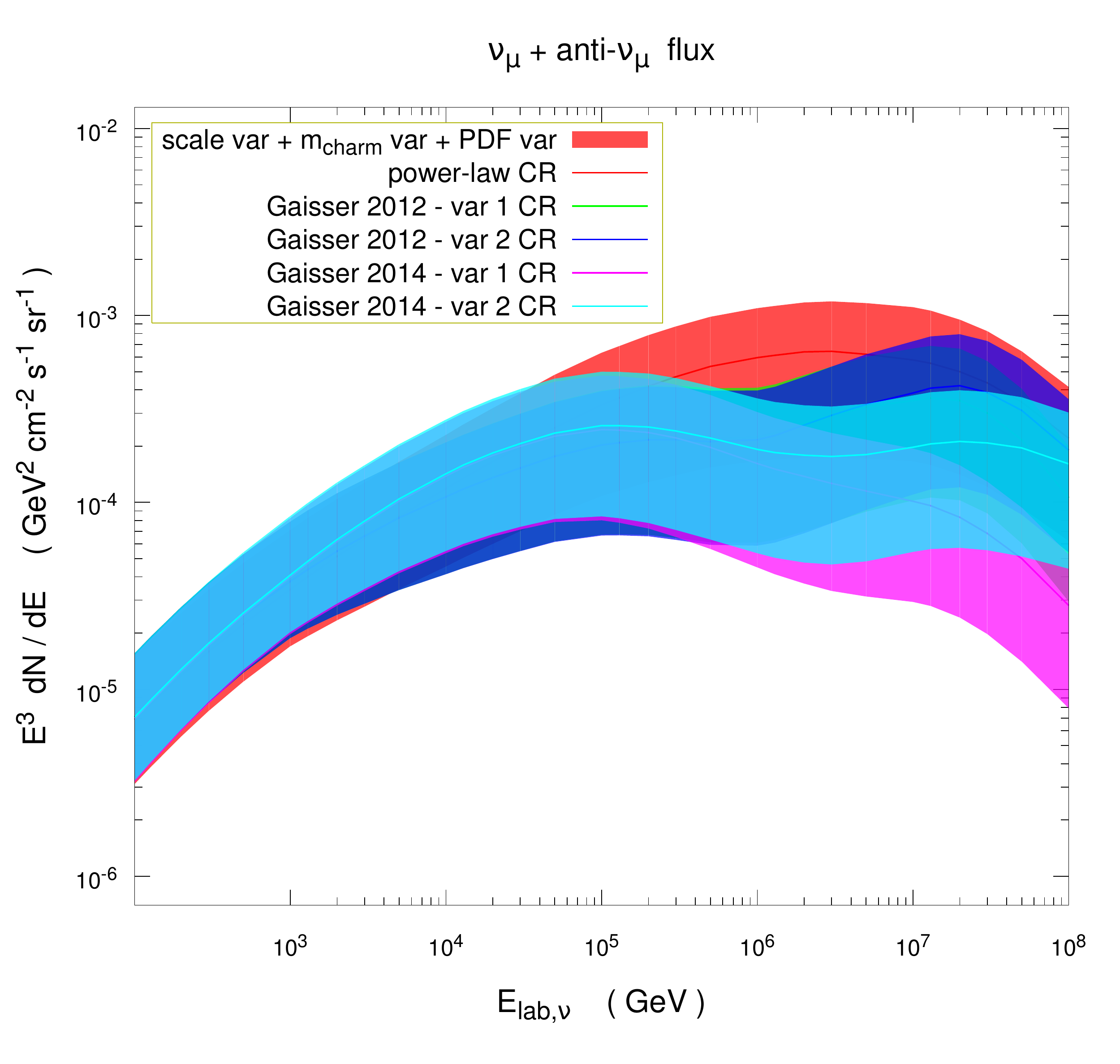}
\caption{\label{fig:summary}
  Summary of the main QCD and astrophysical uncertainties affecting our
  central predictions for ($\nu_\mu$ + $\bar{\nu}_\mu$)-fluxes. 
  Uncertainties due to scale, mass and PDF variation 
  (considering the ABM11 PDF and $\alpha_s$ uncertainty band), 
  are shown separately and combined 
  for each of the five primary CR spectra, cf. Sec.~\ref{subsec:cosmic}.
}
\end{center}
\end{figure}

\begin{figure}
\begin{center}
\includegraphics[width=0.475\textwidth]{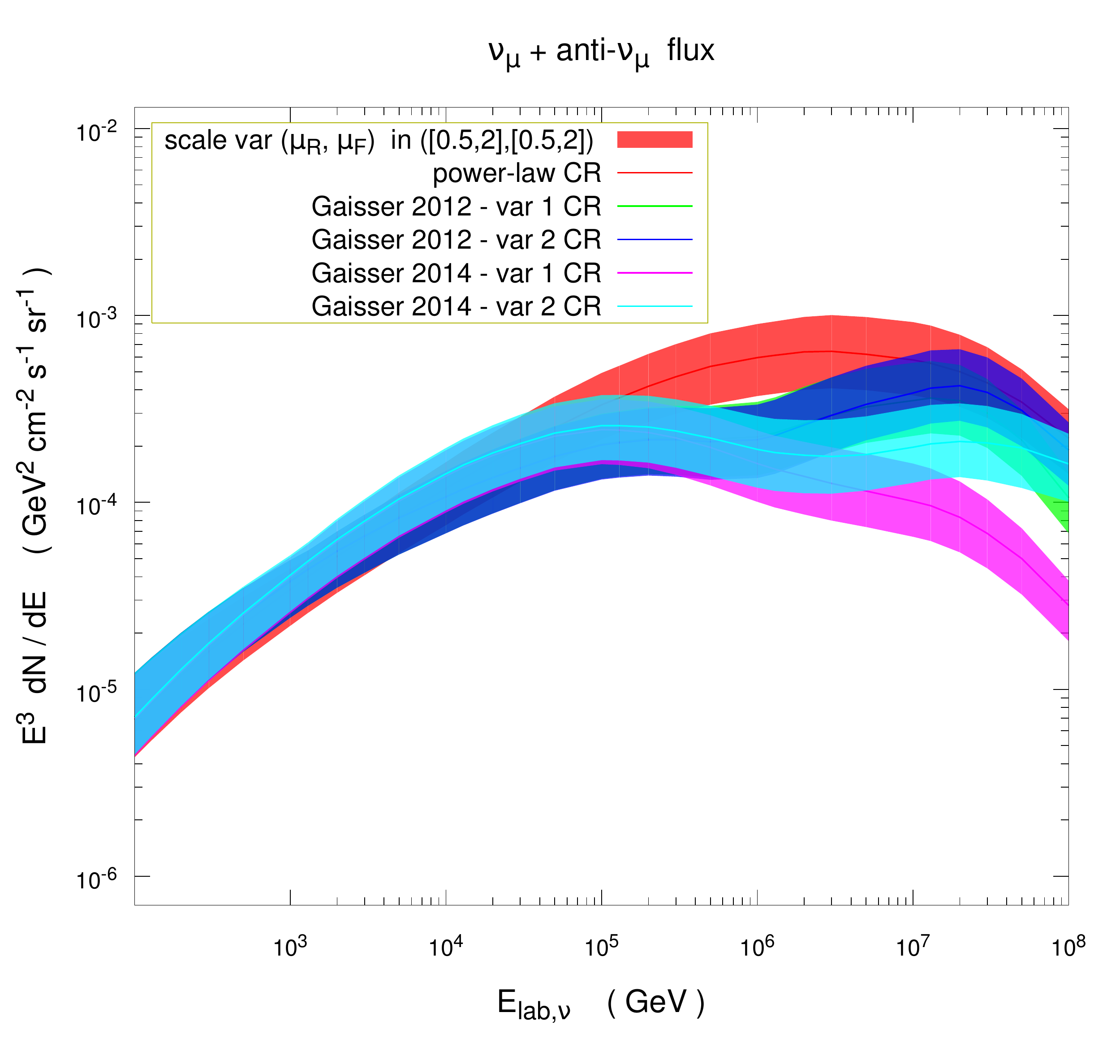}
\includegraphics[width=0.475\textwidth]{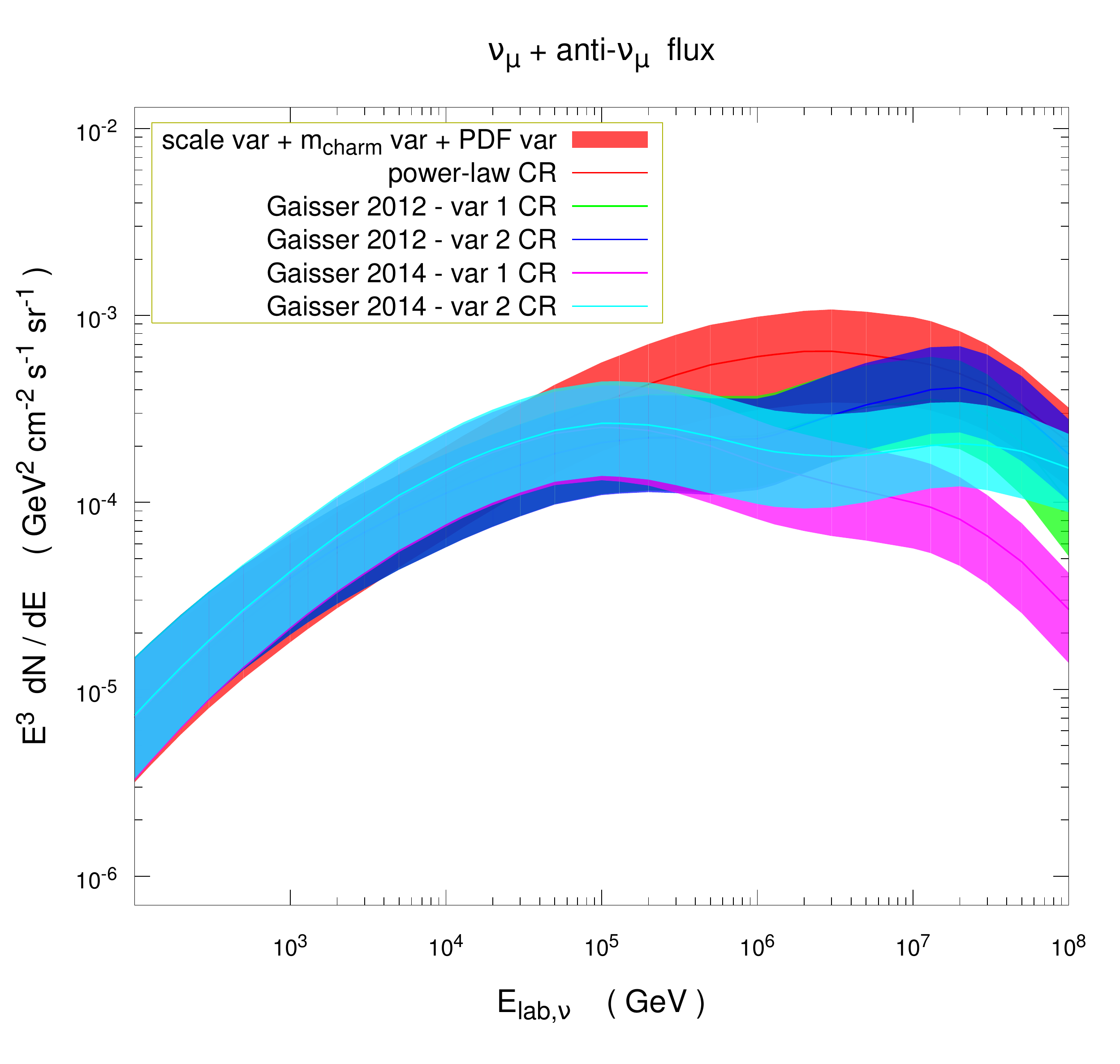}
\caption{\label{fig:summaryrestricted} Same as panel 1 and panel 4 of Fig.~\ref{fig:summary}, but for a restricted choice of $\mu_R$ and $\mu_F$ variations. 
Here the scale uncertainty is obtained as the envelope of the combinations ($\mu_R$, $\mu_F$) = (0.5, 0.5), (1, 1), (2, 2), (1, 2) and (2,1) $\mu_0$, disregarding the cases with ($\mu_R$,~$\mu_F$)~=~(0.5,~1) and (1,~0.5)~$\mu_0$.}
\end{center}
\end{figure}

A summary of the main QCD uncertainties (mass, scale, PDF variation) 
in relation to uncertainties of astrophysical origin, 
in particular those arising from the variations of the primary CR flux used as input
in the ($\nu_\mu + \bar{\nu}_\mu$)-flux calculation, 
is provided in Fig.~\ref{fig:summary}. 
In the first three panels of this figure we show separately the uncertainties
due to scale, mass and PDF variation (by restricting ourselves to 
the ABM11 PDF set) 
for all five primary CR fluxes considered as input in this paper. 
The uncertainties due to scale variation are the dominant component.
Apart from that, it is evident that at energies $\gsim$~10$^{6}$~GeV 
uncertainties due to variations in the CR fluxes dominate over those from 
mass and PDFs.  
On the other hand, uncertainties related to QCD effects always dominate at energies $\lsim 10^5$~GeV, where the primary CR fluxes 
are well constrained by several measurements (see Sec.~\ref{subsec:cosmic}). 
Finally, in the last panel of Fig.~\ref{fig:summary} we show the quadratic combination of the uncertainties above, 
assumed as independent, i.e., $\Delta_{QCD} = \sqrt{\Delta^2_{m_c} + \Delta^2_{(\mu_R, \mu_F)} + \Delta^2_{PDF}}$.
For $E_{lab,\, \nu}~=~10^{6}$~GeV, $- 72\%  \le \Delta_{QCD} \le +84\%$, i.e., 
the uncertainty is slightly asymmetric, and it slightly changes (a few percent) at higher energies. 

We also observe that with a restricted scale variation interval, 
neglecting the combinations ($\mu_R$, $\mu_F$) = (0.5, 1) and (1, 0.5) $\mu_0$
for $\mu_0=\sqrt{p_{T,c}^2 + 4 m_c^2}$, 
scale uncertainties and, as a consequence, the total ones, 
are reduced, as shown in
Fig.~\ref{fig:summaryrestricted}. In particular, for $E_{lab,\, \nu} = 10^{6}$~GeV,  the combined uncertainty amounts to $- 48\%  \le \Delta_{QCD} \le +63\%$, 
and it changes by a few percent at higher energies.

\subsection{Other uncertainties}
\label{subsec:other}

In the previous Sec.~\ref{sec:uncQCD} we have provided a minimal estimate of
the combined QCD and astrophysical uncertainties which affect our predictions for ($\nu_\mu$ + $\bar{\nu}_\mu$)-flux. 
In the following we shortly describe other sources of uncertainties which could be added to the previous ones.

A further QCD contribution arises from heavier hadrons, in
particular $B$-hadrons, which are also a source of neutrinos, 
but whose effect has been neglected here.
Given, that the cross-section for $b\bar{b}$ hadroproduction 
with respect to the one for $c\bar{c}$ hadroproduction 
is smaller by a factor of order $20$ at LHC energies  
and still suppressed by a factor of order 10 at $E_{lab}$ = 100~TeV, 
we expect that the bottom-quark contribution can be neglected 
with respect to the charm one at the energies of interest for IceCube.
However, $b\bar{b}$ hadroproduction may play a larger role at ultra-high-energies.

Other uncertainties can be attributed to the approximate description of the decay of heavy hadrons. 
In particular, a component of secondary neutrinos coming from the decay of the lighter mesons (baryons) produced as decay products of
$D$-mesons (baryons), is missing in our computation as well as in many previous ones (see e.g. Ref.~\cite{Bhattacharya:2015jpa}). 
Furthermore, from the QCD point of view, non-perturbative effects, suppressed by powers of $\Lambda_{QCD}/m$, 
are increasingly important for smaller quark masses $m$. 
In this respect, one should account for a contribution to the uncertainties due to fragmentation, arising from both fragmentation fractions and fragmentation functions~\cite{Behnke:2015qja}. 
The latter can be estimated, for instance, by varying the choice of the
functional form of fragmentation functions for heavy flavors 
in {\texttt{PYTHIA}} together with the parameters involved. 
The corresponding uncertainty estimates have been discussed in the literature,
see, e.g., Ref.~\cite{Cacciari:2012ny}.
Potential uncertainties related to the variation 
of the partonic intrinsic transverse momentum $\langle k_T \rangle \sim
\Lambda_{QCD}$ are indeed smaller since they mostly affect the $p_T$ distributions 
to which our work is not particularly sensitive. 
\begin{figure}[ht!]
\begin{center}
\includegraphics[width=0.75\textwidth]{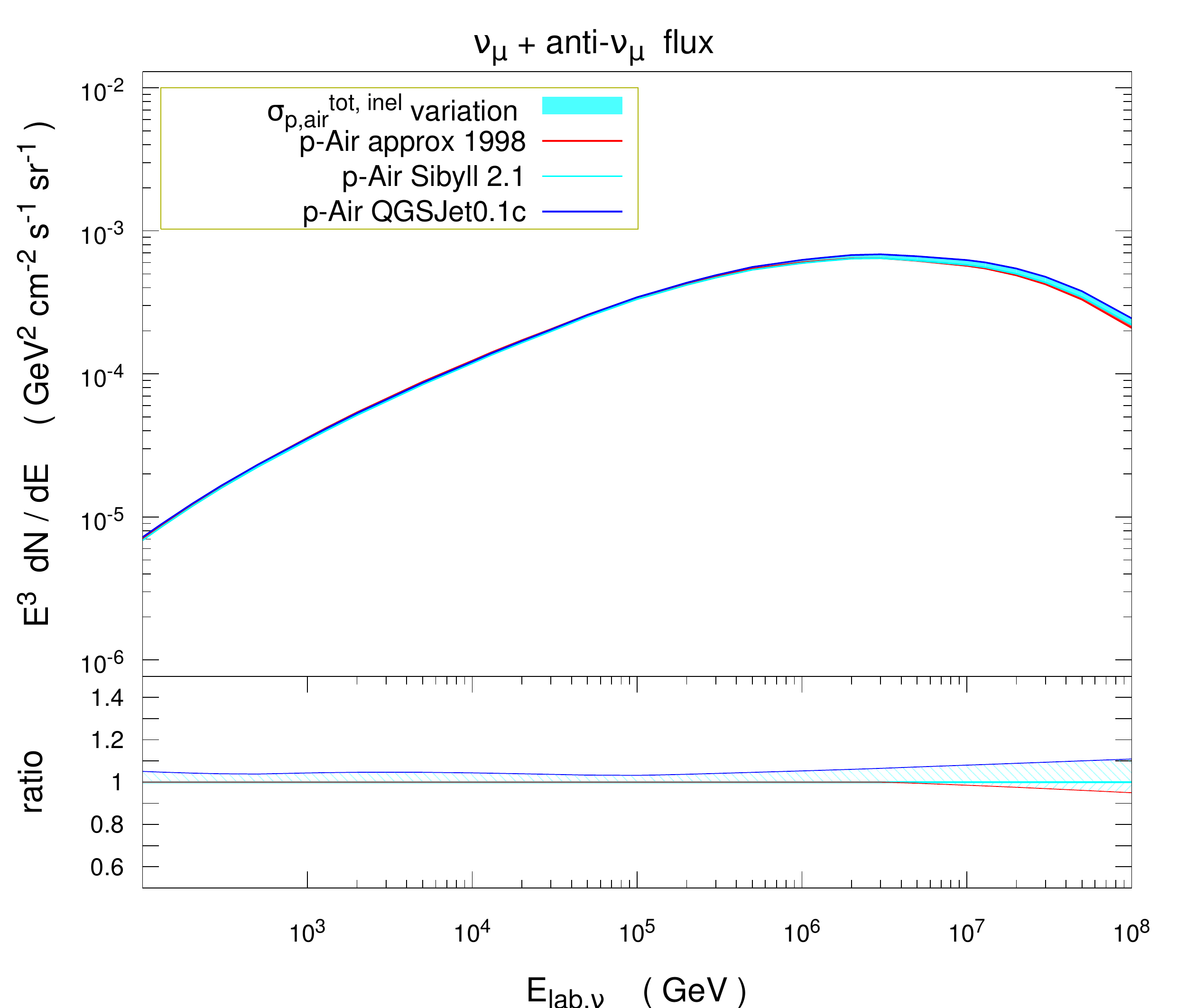} 
\caption{\label{fig:pair-b}
The ($\nu_\mu + \bar{\nu}_\mu$)-fluxes as a function of the neutrino energy $E_{lab,\, \nu}$ 
obtained by considering different models for the $p$-Air total inelastic
cross-section shown in Fig.~\ref{fig:pair-a}
for a power-law primary CR flux. Charm mass, PDFs and scale were fixed
to our central values (see Figs.~\ref{fig:dsigmadx} and~\ref{fig:dsigmadxpdf}).}
\end{center}
\end{figure}
\begin{figure}[ht!]
\begin{center}
\includegraphics[width=0.75\textwidth] {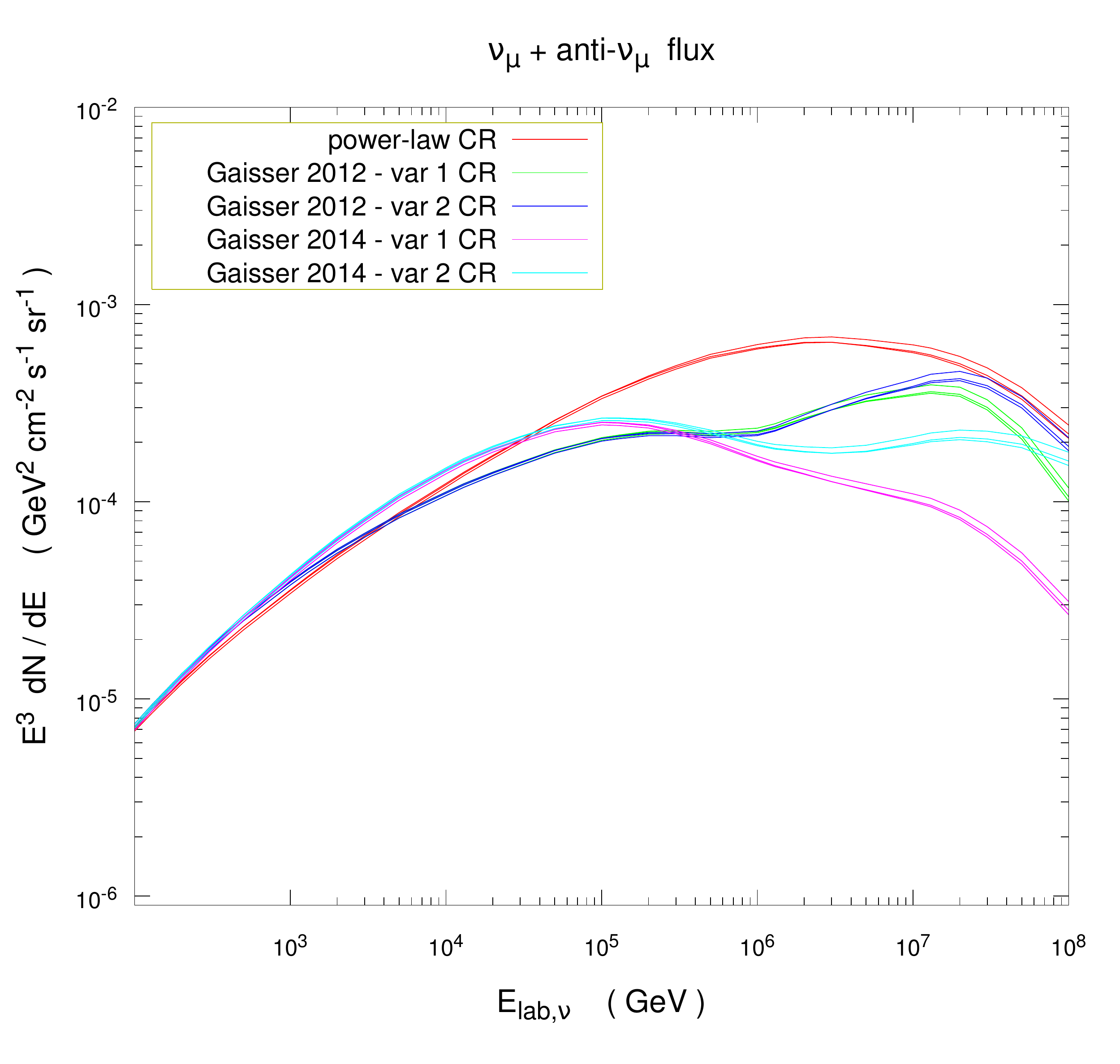} 
\caption{\label{fig:pair-c} 
Same as in Fig.~\ref{fig:pair-b} for the five different primary CR spectra
considered in this work, cf. Sec.~\ref{subsec:cosmic}.
}
\end{center}
\end{figure}

From an astrophysical point-of-view, further uncertainties to be included
encompass those related to a change in the $p$-Air cross-section, 
which affect both the $Z_{p\,h}$ production moments and 
the $Z_{p\,p}$ regeneration moments, and, as a consequence, the lepton fluxes. 
To that end, we consider the theoretical predictions coming from the 
three different models described in Sec.~\ref{subsec:pair} 
({\texttt{QGSJet0.1c}}, {\texttt{SYBILL2.1}} and the ana\-ly\-ti\-cal model eq.~(\ref{eq:sigmapAir})).
The larger the inelastic cross section $\sigma^{inel}$($p$-Air) is, 
the smaller are the predictions for our fluxes, 
as is evident when comparing Fig.~\ref{fig:pair-b} with Fig.~\ref{fig:pair-a}. 
However, these global effects on neutrino fluxes turn 
out to be not too relevant, i.e., the uncertainties coming from the use of
different models, amount to less than $10\%$ over the whole $E_{lab,\,\nu}$ energy range considered. 
They are therefore much smaller than those from QCD effects discussed  previously and those
from the choice of different primary CR spectra, as shown in Fig.~\ref{fig:pair-c}.

\section{Comparison with previous results and astrophysical implications} 
\label{sec:astroimpli} 

\begin{figure}
\begin{center}
\includegraphics[width=0.7\textwidth]{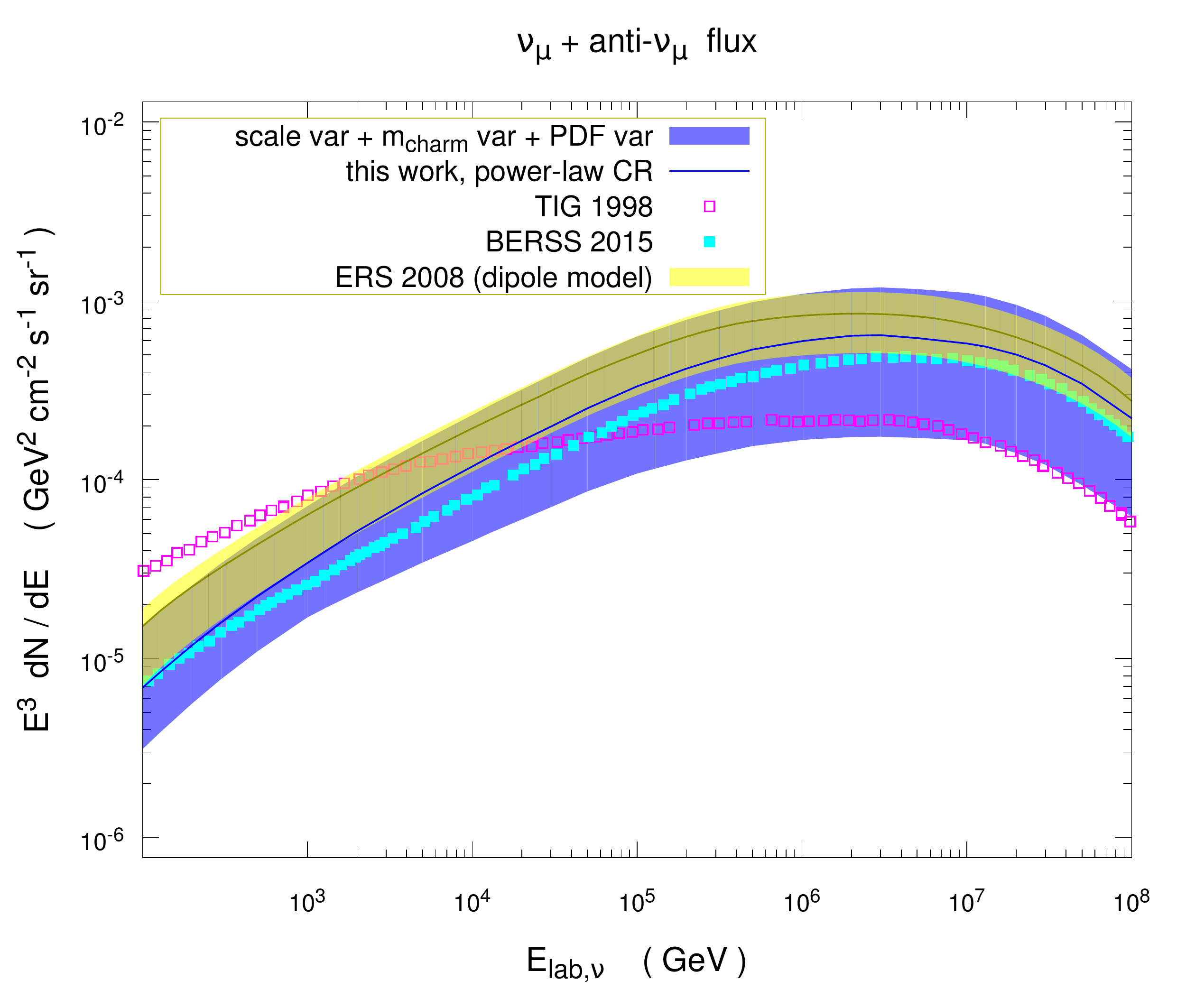}
\caption{\label{fig:comp1} Comparison between the prompt ($\nu_{\mu}$ + $\bar{\nu}_\mu$)-flux obtained 
in this work (blue solid line with blue uncertainty band) with the central
values of those previously obtained by other authors, 
for a power-law primary cosmic ray spectrum. 
The TIG flux (Ref.~\cite{Gondolo:1995fq}) is shown by open magenta squares, 
the ERS central flux (Ref.~\cite{Enberg:2008te}) and its uncertainty is shown in yellow, whereas the more recent BERSS flux (Ref.~\cite{Bhattacharya:2015jpa}) 
is shown by filled light-blue squares.
}
\end{center}
\end{figure}

\begin{figure}
\begin{center}
\includegraphics[width=0.45\textwidth]{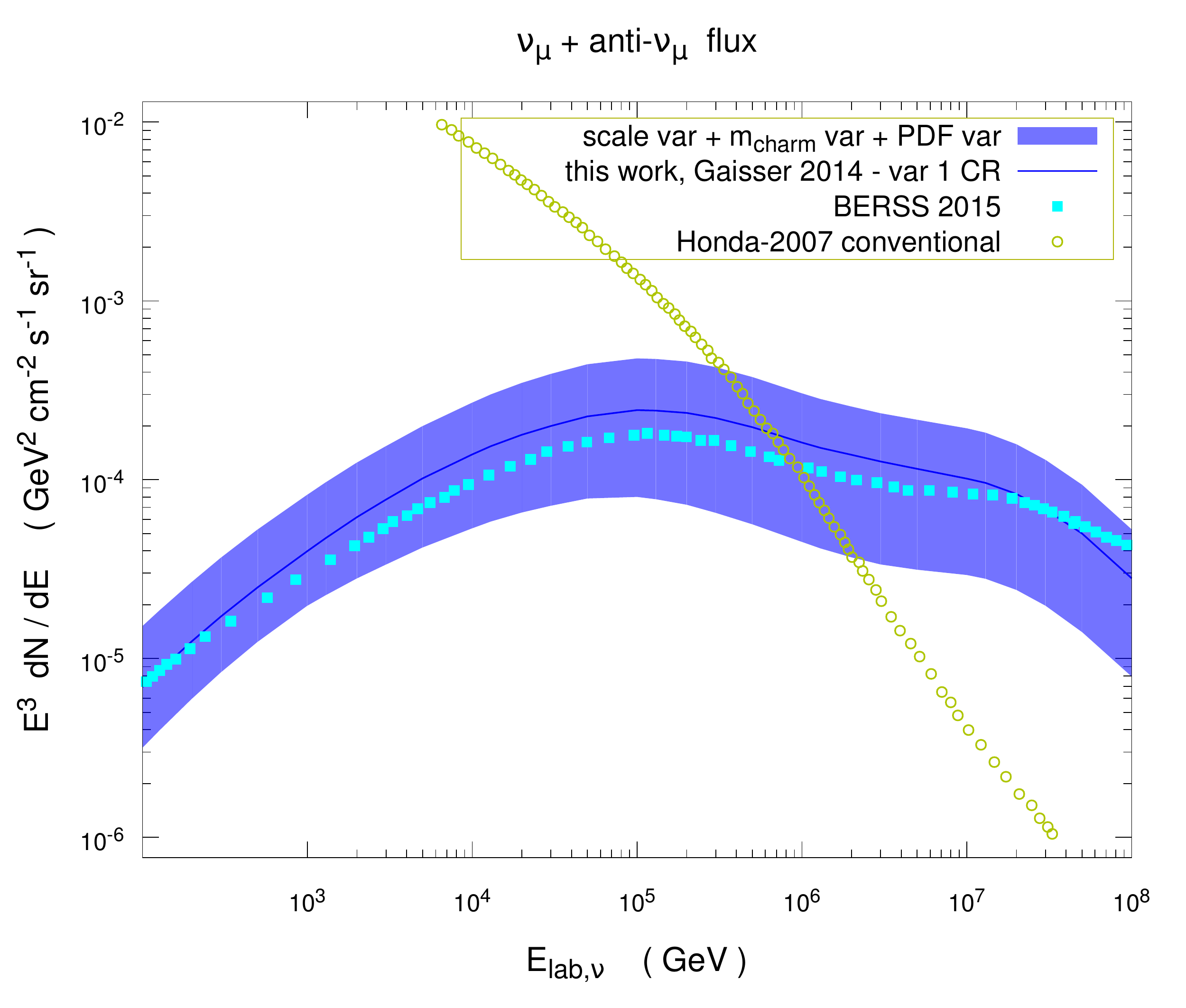}
\includegraphics[width=0.45\textwidth]{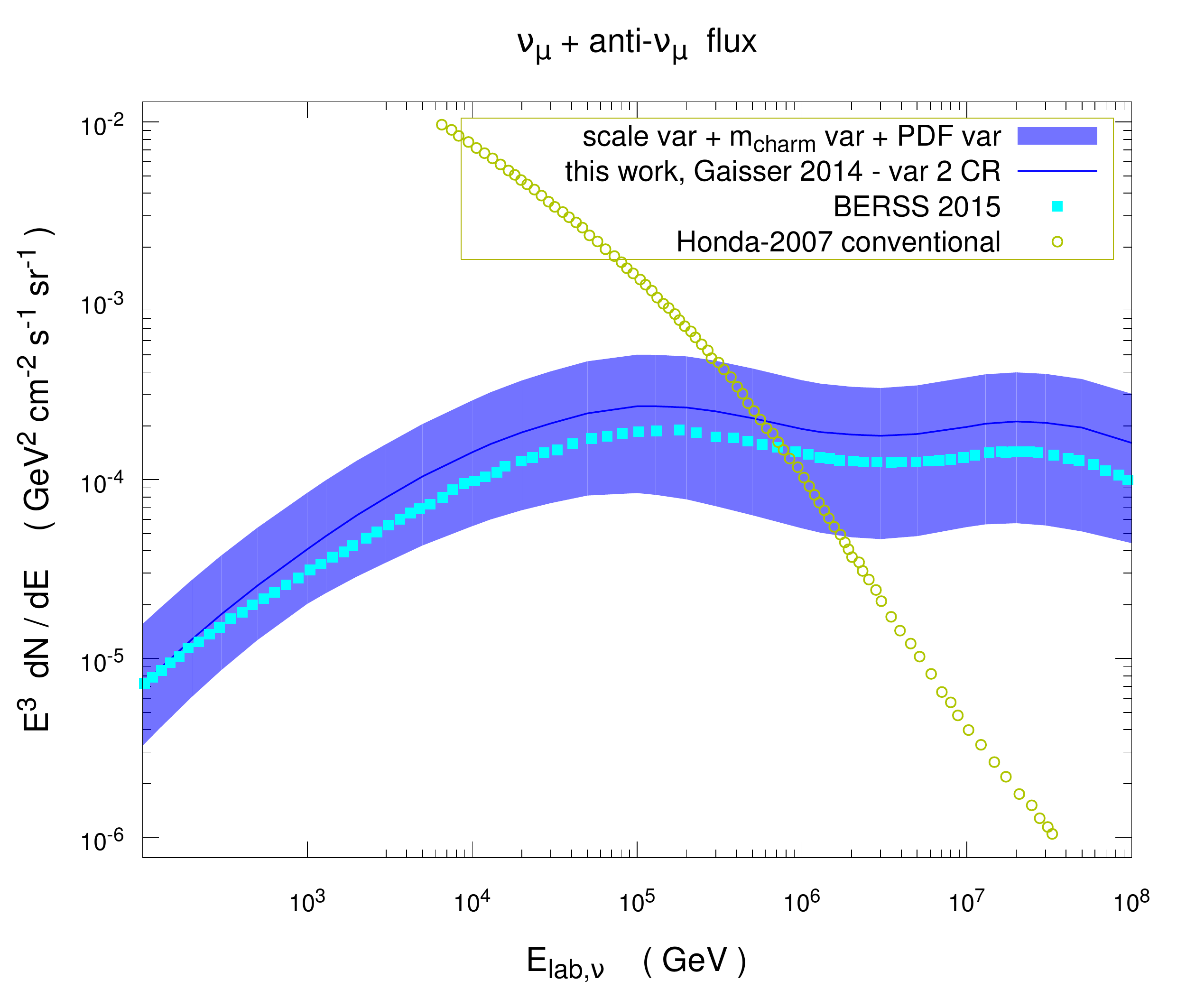}
\end{center}
\caption{\label{fig:comp2} 
Comparison between the prompt ($\nu_{\mu}$ + $\bar{\nu}_\mu$)-flux obtained in this work 
(blue solid line with blue uncertainty band) with the central values 
of the more recent BERSS flux (Ref.~\cite{Bhattacharya:2015jpa}) (light-blue squares) 
in case of recent primary cosmic ray spectra (Gaisser-2014-variant 1 
on the left and Gaisser-2014-variant 2 on the right). 
The conventional neutrino flux computed by Honda (Ref.~\cite{Honda:2006qj}),
after its rescaling to the Gaisser-2014-variant 1 CR primary spectrum 
as presented in Ref~\cite{Bhattacharya:2015jpa}, is shown by open circles.
}
\end{figure}

We compare our prompt ($\nu_{\mu}$ + $\bar{\nu}_\mu$)-flux
with previous results in the literature.
In particular, it turns out that our central fluxes are 
in between central predictions
recently obtained 
by another group using the standard hard-scattering formalism in QCD~\cite{Bhattacharya:2015jpa} and older predictions provided in Ref.~\cite{Enberg:2008te} 
by making use of the dipole picture. 
In particular, our central values are a few ten percent larger  
($\sim$ 40\% at the energies of interest for the IceCube experiment)
than the central values in Ref.~\cite{Bhattacharya:2015jpa}, 
that lie in any case within our quoted uncertainty band, 
for the various primary CR spectra already considered in that
work, as shown in Figs.~\ref{fig:comp1}~and~\ref{fig:comp2}. 
Note, that Ref.~\cite{Bhattacharya:2015jpa} 
is based on a completely independent QCD calculation and on different inputs and methods.

On the other hand, differences with older calculations, like the one in
Ref.~\cite{Gondolo:1995fq} on the basis of {\texttt{PYTHIA}} and including
QCD hard-scattering effects at leading order only, are obvious, 
especially regarding the shape of the distributions. 
This is the case not only for the lepton fluxes, but already for the $Z$-moments 
for $D$-hadron production.

Interestingly, our results are very well compatible with those from
the dipole model of Ref.~\cite{Enberg:2008te}, altough the latter were computed with older sets of PDFs: as shown in Fig.~\ref{fig:comp1}, for $E_{lab,\nu} > 10^3$ GeV, our central predictions are included in the uncertainty band of the latter, whereas the central predictions from the dipole model are included in our uncertainty band.  

In order to infer a value for the transition energy $E_{trans}$ where the prompt
neutrino flux overcomes the conventional one, in Fig.~\ref{fig:comp2} we compare our prompt lepton flux with the conventional neutrino flux originally computed in Ref.~\cite{Honda:2006qj} for a power-law CR primary spectrum and rescaled to one variant of the Gaisser spectra in Ref.~\cite{Bhattacharya:2015jpa}. 
We obtain 
$E_{trans}$~=~$6.0^{+ 12}_{- 3}$~$\cdot 10^5$~GeV. 
Interestingly, the central value lies well within
the interval (4~$\cdot 10^5$~$-$~$10^6$)~GeV where the IceCube experiment did not observe any event 
after the full 988-day analysis~\cite{Aartsen:2014gkd}. 
In fact, the IceCube collaboration has reported an excess of neutrinos in the diffuse flux, all
lying in the neutrino energy regions [0.3~$-$~4]~$10^5$~GeV and [1 $-$ 2] $10^6$~GeV. 
According to our predictions, the ``empty'' region of IceCube corresponds 
to the  ``conventional-prompt'' transition region, i.e., the region where the
contributions of conventional neutrinos and prompt neutrinos to the total
neutrino flux become of the same order of magnitude. We thus believe
that the ``empty'' region seen by IceCube so far, should not be empty, but
actually dominated by prompt neutrinos. 
However, the IceCube error bars in the ``empty'' region are still quite large, 
and we stress that the accumulation of more statistics is necessary before 
judgment can be made, whether this lack of signal is just an artifact due to
poor statistics or due to some other technical issue, or instead has a real physical interpretation.

At higher energies, on the other hand, 
the total observed neutrino flux $E_\nu^2 \, \phi(E_\nu)$ 
for $E_\nu$ in the [1 - 2] $10^6$~GeV energy interval
 looks to be 
slightly suppressed with respect to that 
in the [2 - 3] $10^5$~GeV bins.
However, looking at our central prompt flux distributions and summing them
with the distributions for the conventional flux, as a first rough estimate it turns out that we would expect a much larger suppression in the [1 - 2] $10^6$~GeV region with respect to the [2 - 3] $10^5$~GeV one, disfavoring the interpretation that the events seen by IceCube in the [1 - 2] $10^6$~GeV window are just due to a prompt neutrino component~\footnote{This interpretation is also disfavoured by IceCube observations of the arrival directions of the events with $E$ $>$ 6$\cdot 10^4$ GeV, in presence of a $\mu$ veto (see Fig.~3 of Ref.~\cite{Aartsen:2014gkd}).}.  
The difference between the IceCube (signal + background) observed total yield at high-energy and the yield for prompt neutrinos as predicted by our calculation 
is slightly reduced if we observe that our predictions have a sizable uncertainty band, meaning that even the shapes of the distributions can change in a non-negligible way when a higher-order calculation in QCD will be available, if we consider neutrino flux values corresponding to the upper value of our uncertainty band and if we use as input primary CR fluxes including a population of extra-galactic protons with very-high rigidity (i.e. variants~2 of Gaisser spectra, instead of variants~1 which have a mixed extra-galactic component with a lower global rigidity for the extra-galactic population). 
In fact, the latter give rise to neutrino spectra which are less severely suppressed at the highest energies than those from models with extragalactic mixed components, as is evident when comparing, e.g., the left and right panels of Fig.~\ref{fig:comp2}, obtained with the variants 1 and 2 of the Gaisser 2014 spectrum, respectively.  
In order to go beyond these purely qualitative considerations and
to draw more definite quantitative conclusions, one should definitely
wait for more experimental statistics and, 
also insert our fluxes into the specific experimental analysis software.

In any case, we would like to emphasize that the transition region 
for the prompt ($\nu_{\mu}$ + $\bar{\nu}_\mu$)-flux in our calculation 
turns out to be also a transition region for uncertainties, i.e., 
the QCD uncertainties dominate the total uncertainties at energies below 
the transition region whereas the astrophysical ones start to give
a progressively sizable contribution above it, pointing
out the importance and necessity of pursuing further studies 
of cosmic ray composition at the highest energies~\cite{Aloisio:2013hya},
and, possibly, future measurements independent of Monte Carlo simulations
of hadronic-interactions at the highest energies.

\section{Conclusions}
\label{sec:conclusions}

We have computed the prompt neutrino fluxes from atmospheric charm using 
up-to-date theoretical results and tools for charm hadroproduction in perturbative QCD.
Our results for the neutrino fluxes are several tens percent larger 
over a wide range of neutrino energies than predictions in the recent literature making use of $Z$-moments computed with the standard QCD
hard-scattering formalism, that we also adopt.  
At the energies of interest for the IceCube experiment 
the increase of our prompt ($\nu_{\mu}$ + $\bar{\nu}_\mu$)-flux amounts to
some 40\%. However, our uncertainties on the fluxes both of QCD and
astrophysical origin are dramatically larger. Partly as an effect of this
fact, even predictions obtained by making use of 
the dipole picture, representing an alternative description to the undelying hard-scattering, lie within our uncertainty band over a wide energy range.

We have discussed extensively the different sources of uncertainties which affect the fluxes. 
The main sources come from 
(i) the renormalization and factorization scale variation allowing for independent variations of $\mu_R \ne \mu_F$, 
(ii) the charm mass uncertainties for the pole mass choice,
and
(iii) PDF uncertainties evaluated for the ABM11 set and studied by comparing its predictions to the central predictions of different PDF sets (CT10, ABM11, NNPDF3.0) at NLO.
Further uncertainties due to hadronization and hadron decay have been
discussed as well.
In particular (i) and (ii) had not been included in a systematic way in studies in literature before, so we conclude that previous uncertainties on prompt neutrino fluxes are underestimated.

The uncertainties of QCD origin dominate at low neutrino energies, whereas for
increasing energies $E_{lab,\, \nu} \gsim 10^{5} - 10^{6}$~GeV 
the uncertainties in the astrophysical input, in
particular the primary CR flux and its composition in terms of different
populations, turn out to add a progressively important contribution
to those from QCD.

The results presented may benefit from a number of future developments.
On the QCD side, a fully differential NNLO computation of charm hadroproduction, when available, will be of great help in reducing the theoretical uncertainties from scale, mass and PDF variation. In this respect, the role of resummation of different kinds of logarithms deserves further exploration as well. 
Furthermore, a dedicated systematic survey of the uncertainties related to
both the fragmentation functions in the Monte Carlo parton shower matched
to NLO predictions and the fragmentation fractions, would allow to quantify 
those effects. This could be a step towards the optimization of  
Monte Carlo tunes, to make them especially tailored to studies like those
performed in this work. This optimization also concerns the search for the best 
parameter values for the description of dual and multiple particle 
interactions. On the experimental side, measurements of the $c\bar{c}$ and $b\bar{b}$ production cross-section at the LHC, looking not only at central rapidities but also in the forward rapidity regions, can be of importance especially at the highest energies, where the contribution of low $x$ events becomes increasingly important. 

Finally, from the astrophysical point of view, one could obtain a substantial reduction of the uncertainties on prompt neutrino fluxes at the highest energies 
once issues related to the transition between a galactic and an extragalactic
component in the CR primary spectrum and to the composition of the latter
will be understood better.  

\bigskip

Our lepton fluxes will be made available as numerical tables for download 
at {\texttt{http://www.desy.de/$\sim$promptfluxes}}. Further predictions 
can be requested to the authors of this paper by e-mail.

\bigskip
\bigskip
\noindent{\bf Acknowledgments}\\
We acknowledge encouragement and/or hints from many of our colleagues or col\-la\-bo\-ra\-tors at DESY and University of Hamburg, in particular G.~Kramer, M.~Benzke, B.~Kniehl, O.~Zenaiev, K.~Lipka, A.~Geiser, E.~Bagnaschi, M.~Vollmann, A.~Mirizzi, C.~Evoli, at various stages of this work. We are additionally grateful to C.~Giunti, V.~Niro, V.~de~Souza, P.~Monni and A.~Kulesza for interesting comments. Finally, we are grateful to M.~Cacciari and P.~Nason for useful discussions at the latest stage of this work. M.V.G. is grateful to the Mainz Institute for Theoretical Physics (MITP) for its hospitality and its partial support during the completion of this work.  This work has been supported by Bundesministerium f\"ur Bildung und Forschung through contract (05H12GU8) and  by Deutsche Forschungsgemeinschaft in Sonder\-for\-schungs\-be\-reich~676.


{\footnotesize

\providecommand{\href}[2]{#2}\begingroup\raggedright\endgroup

}
\end{document}